\documentclass[
reprint,
amsmath,amssymb,
floatfix,
]{revtex4-2}
\usepackage{graphicx} 
\usepackage{dcolumn}  
\usepackage{tabularx} 
\usepackage[svgnames]{xcolor}
\usepackage{hyperref} 
\hypersetup
{
    colorlinks=true,
    linktocpage=true,    
    linkcolor=blue,
    citecolor=blue,
    urlcolor=blue,
    filecolor=ForestGreen,
    pdftitle={InvarianceTheoremCJdetonation},
}
\usepackage{chemfig}  
\usepackage[version=4]{mhchem}
\usepackage{bm}       

\begin{document}

\preprint{APS/123-QED}

\title{A velocity-entropy invariance theorem for the Chapman-Jouguet detonation}

\author{Pierre Vidal}
\email{pierre.vidal@ensma.fr, \ @cnrs.pprime.fr\hfill}
\homepage{https://pprime.fr/vidal-pierre/}
\author{Ratiba Zitoun}%
\email{ratiba.zitoun@ensma.fr, \ @univ-poitiers.fr}
\affiliation{%
Institut Pprime, UPR 3346 CNRS, ENSMA, BP40109, 86961 Futuroscope-Chasseneuil, FRANCE
}%




\date{\today\; - Update of the $1^{\text{st}}$ version, June 20, 2020, \href{https://arxiv.org/abs/2006.12533}{arXiv:2006.12533}}

\begin{abstract}
The velocity and specific entropy of the Chapman-Jouguet (CJ) equilibrium detonation are shown to be invariant under the same variations of initial temperature with initial pressure. This leads to additional CJ relations, for example, for calculating the CJ  state -- including the adiabatic exponent -- from the only CJ velocity, without using an equation of state for the detonation products. For gaseous stoichiometric explosives with ideal products, numerical calculations with detailed chemical equilibrium confirm the invariance theorem to $\mathcal{O}(10^{-2})$\% and the additional CJ  properties to $\mathcal{O}(10^{-1})$\%. However, for four liquid carbon explosives, the predicted CJ pressures are about 20\% higher than the measurements. The analysis emphasizes the limited physical representativeness of the hydrodynamic framework of the modelling, i.e. single-phase inviscid fluids at equilibrium for the initial and final states of the explosive. This invariance may illustrate a general feature of hyperbolic systems and their characteristic surfaces.
\end{abstract}


\maketitle

\section{\label{sec:Intro}Introduction}
The Chapman-Jouguet (CJ) detonation \cite{Jouguet1901} is a staple of detonation modelling, defined as the fully reactive, planar and compressive discontinuity with a constant velocity supersonic relative to the initial state and sonic relative to the final burnt state. The CJ state and velocity are calculated from the Rankine-Hugoniot (RH) relations and an equation of state for the detonation products at chemical equilibrium. However, detonation reaction zones have a cellular structure and are highly sensitive to losses, which the RH relations cannot describe. The CJ model can only give reference velocities and reaction-end states, and no conditions for detonation existence. The limited purpose of this study is to bring out two unnoticed CJ properties, which may be useful for improving modelling and interpreting experiments. They apply only to explosives whose fresh and burnt states are single-phase inviscid fluids at thermal equilibrium.
\newline

\begin{figure}[t]
\includegraphics{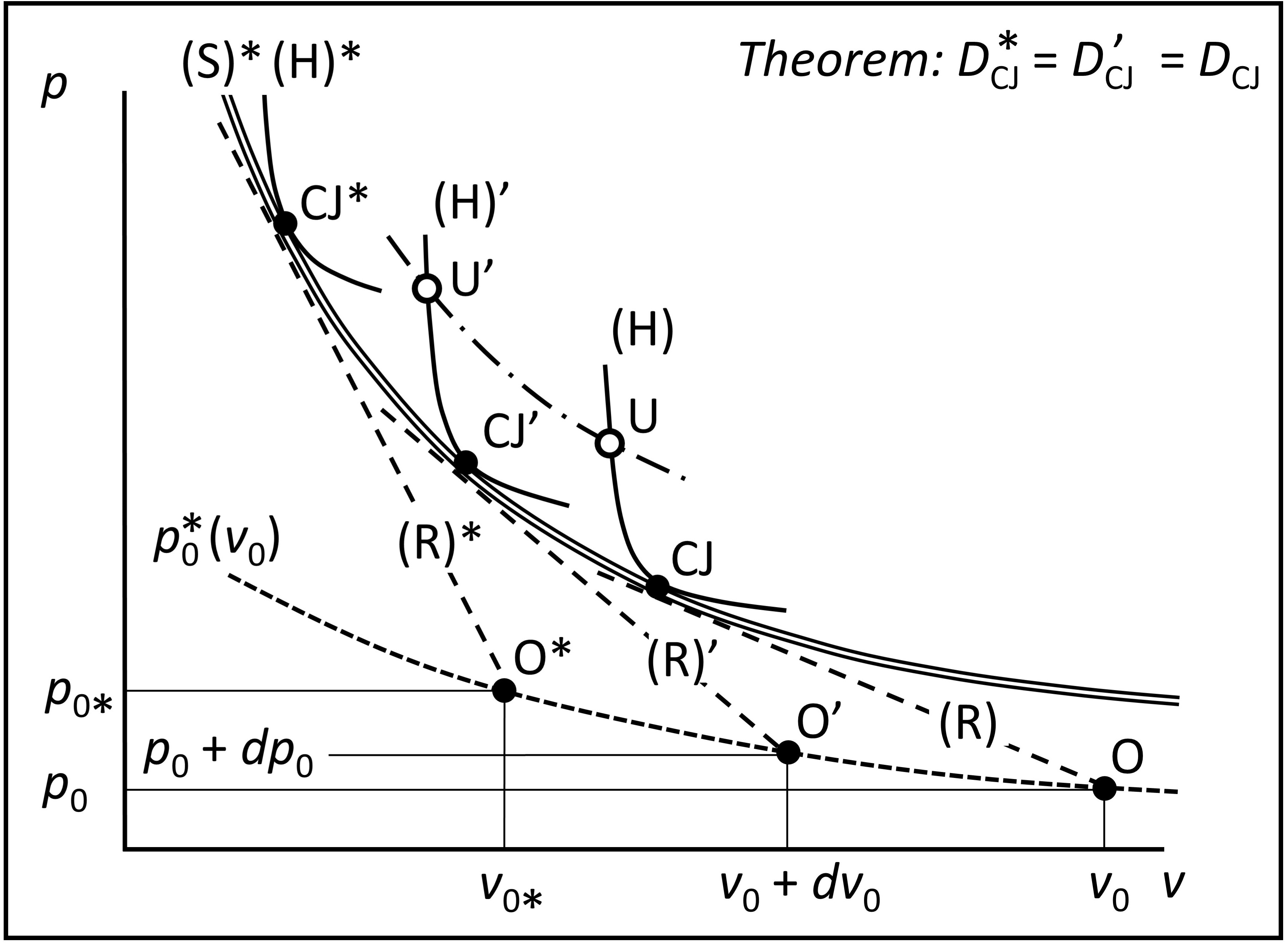}
\caption{An equilibrium isentrope of detonation products (S)* is the common envelope of equilibrium Hugoniot curves (H)*, (H)' and (H) and Rayleigh-Michelson lines (R)*, (R)' and (R) if their poles O*, O' and O lie on a particular $p_0^{\ast}(v_0)$ line through a reference initial state O*$(p_0\ast,v_0\ast)$. The slopes $-(D_{\text{CJ}}/v_0)^{2}$ of these (R) lines increase with increasing initial volume $v_0$, but the theorem ensures they have the same CJ velocity $D_{\text{CJ}}^{\ast}$.}
\label{fig:Fig1}
\end{figure}

The first property is that the CJ detonation velocity $D_{\text{CJ}}$ and the CJ specific entropy $s_{\text{CJ}}$ are invariant under the same dependence of the initial temperature $T_0$ on the initial pressure $p_0$. If $D_{\text{CJ}}$ is invariant, so is $s_{\text{CJ}}$, and vice versa, i.e. different initial temperatures and pressures producing the same $D_{\text{CJ}}$ produce different CJ states on the same isentrope. Figure \ref{fig:Fig1} depicts this velocity-entropy invariance theorem in the pressure ($p$)-volume ($v$) plane \hyperref[sec:Remind]{(Sect.~\ref{sec:Remind})}.
\vspace{1mm}

The second property is a set of additional CJ relations, for example, for calculating the CJ state -- including the adiabatic exponent and the isentrope -- from the only value of $D_{\text{CJ}}$ without using an equation of state for the detonation products. Conversely, $D_{\text{CJ}}$ can be obtained from any one of the CJ-state variables.
\newpage

Fundamental detonation studies today focus less on the CJ model and more on the identification and modelling of processes in the detonation reaction zone. These include cellular structure, high-pressure chemical kinetics, adiabatic or non-adiabatic losses in homogeneous explosives, and carbon condensation and intergranular heat exchange in heterogeneous explosives. Most prevent the CJ equilibrium from being reached. The CJ model is essentially an ideal limit used to calibrate equations of state for detonation products at chemical equilibrium.
\vspace{1mm}

Can the detonation regime be identified from experimental detonation velocities and pressures? Models are usually rejected if they do not reproduce observations. However, disagreement may be due to inaccurate measurements, non-physical parameters or assumptions not relevant to the experiment, and agreement should not preclude that fewer assumptions may be sufficient.

Equations of state for detonation products are calibrated by fitting calculated CJ properties to experimental values, although no criterion ensures that they represent the CJ equilibrium state. This study proposes that they do not if they do not satisfy the additional properties.
\vspace{1mm}


The basis of the analysis has similarities with the semi-empirical Inverse Method of Jones \cite{Jones1949}, Stanyukovich \cite{Stanyuk1955} and Manson \cite{Manson1958a}. This method gives the CJ hydrodynamic variables from experimental values of $D_{\text{CJ}}$ and its derivatives with respect to two independent initial-state variables, such as $p_0$ and $T_0$ \hyperlink{addIM}{(Subsect. \ref{subsec:CJsuppl}, §5)}. The theorem shows that the value of $D_{\text{CJ}}$ is sufficient.
\vspace{1mm}

\hyperref[sec:CJPos]{Section \ref{sec:CJPos}} is an overview of detonation dynamics and physics to indicate what the CJ model is not good at. \hyperref[sec:Remind]{Section \ref{sec:Remind}} is a reminder of basic but all necessary elements, also introducing the main notations and what this work contributes. The informed reader can go directly to \hyperref[sec:DSITh]{Section \ref{sec:DSITh}}, which presents the theorem and its consequences. \hyperref[sec:Applic-Gas]{Sections \ref{sec:Applic-Gas}} and \hyperref[sec:Applic-Liq]{\ref{sec:Applic-Liq}} analyze their agreement or disagreement with calculations for gases and measurements for liquids. \hyperref[sec:Disc]{Section \ref{sec:Disc}} is a summary with a discussion of the assumptions, and some conclusions.
\vspace{1mm}

The work is dedicated to the memory of Dr. Michael Cowperthwaite.
\section{\label{sec:CJPos}Chapman-Jouguet limitations}
The CJ postulate is that the flow at the discontinuity front is both sonic and at chemical equilibrium. However, this is more an ideal mathematical situation than physical reality. The Zel'dovich-von Neuman-D\"{o}ring (ZND) detonation model, i.e. a shock sustained by a subsonic, laminar and adiabatic reaction zone is the usual basis for analysis \cite{{Vieille1900},{FickDav2000},{Higgins2012}}.
\vspace{1mm}

Most self-sustained detonations are non-ideal because explosive devices have a finite transverse dimension. As a result, the reaction zones expand transversely and encompass the sonic front of the rear expansion, hence curved leading shocks and lower velocities than for the CJ planar detonation. The flow behind the sonic locus does not sustain the shock, and self-sustained propagation requires a sufficient distance between the sonic locus and the leading shock for the chemical reaction progress to be sufficiently large. At the sonic locus, the reaction rate must balance the loss rate, which includes heat transfer, friction or transverse expansion, for the flow derivatives to remain finite. The dynamics of a self-sustained ZND detonation is thus described by an eigenconstraint between the parameters of the reaction and loss rates \cite{ZeldoKompa1960} and those of the leading shock, namely its normal velocity, acceleration and curvature \cite {{WoodKirkwood1954},{HeClavin1994}, {KasimovStewart2004},{ShortEtAl2020}}.
\vspace{1mm}

The ZND model uses the frozen sound speed, whereas the CJ model uses the equilibrium sound speed. Any reaction process cannot reach CJ equilibrium as the steady planar limit of a sonic-frozen curved ZND detonation \cite{Sharpe2000}. Higgins \cite{Higgins2012} has given several examples of equilibrium-frozen issues and non-ideal detonations. Achieving the CJ balance requires at least wide enough setups so losses are negligible and large enough distances from the ignition position so the gradients of the expanding flow of products are small and the chemical equilibrium can shift.
\vspace{1mm}

The reaction zone for homogeneous explosives is not laminar but has a three-dimensional cellular structure. In gases, the cellular combustion process may be adiabatic or involve some participation of turbulent diffusion, producing, respectively, regular or very irregular cells. Therefore, not all mean cell widths are statistically significant, but, if they are, their magnitudes are one order larger than the calculated characteristic thicknesses of the planar ZND reaction zones \cite{{DenisovTroshin1959},{DesbordesPresles2012},{Monnier2022},{Monnier2023}}. The surface area of the detonation front has to be at least large enough compared to that of the detonation cells for the CJ and ZND properties to be valid averages and for the cellular structure to depend only on the reaction processes, without any effect of the transverse dimensions of the experimental setup. In liquids, the relationship between the cellular structure and the reaction processes requires further investigation \cite{{EdwardsShort2019},{UrtiewKusubov1970},{PerssonBjarnholt1970},{TarverUrtiew2010},{FickDav2000}}. 
\vspace{1mm}

For gases with moderately large equivalence ratios (ER), the prevailing view on reaction processes is that the translational, rotational and vibrational degrees of freedom re-equilibrate faster than chemical kinetics, although this is debated for \ce{H2}:\ce{O2}:\ce{N2} mixtures highly diluted with the mono-atomic inert argon \cite{{Tarver1982a},{Tarver1982b},{Vargas2022},{Vargas2023}}. In contrast, for liquids, molecular bond breaking would make the de-excitation time of vibrations comparable to that of chemical relaxation \cite{{Dremin1999},{Tarver1982c}}. Local thermal equilibrium would be reached before chemical transformation for most gases but not in the detonation products of liquids. Tarver \cite{Tarver2012} has introduced the Non-Equilibrium ZND model. For gases, models of detonation products that include carbon condensation predict lower CJ velocities and pressures with increasing ERs, as opposed to homogeneous, i.e. single-phase, gases \cite{{Kistiakovski1952},{Kistiakovski1955},{Kistiakovski1956},{BatraevEtal2018}}. This condensation is also inherent to the reaction processes of many liquid or solid explosives \cite{{BergerViard1962},{Bastea2017},{EdwardsShort2019}}.
\vspace{1mm}

The work deals with the fully reactive planar discontinuity and the CJ model as such \hyperref[sec:Remind]{(Sect. \ref{sec:Remind})}. The invariance theorem and the additional CJ relations are valid for single-phase detonation products at chemical equilibrium. They also provide a simple criterion for discussing whether the CJ equilibrium model can represent experimental or numerical data but not for indicating which of its hydrodynamic or physical assumptions would be incorrect.
\newpage
\section{\label{sec:Remind}Reminders, notations, remarks}
A single-phase inviscid fluid with composition invariant or at chemical equilibrium obeys equations of state with two independent thermodynamic variables, e.g. temperature $T$ and pressure $p$. Indeed, the minimization of the Gibbs free energy defines an equilibrium chemical composition as a function of $T$ and $p$ \cite{GordonMcBride-I} (\hyperref[sec:Disc]{Sect. \ref{sec:Disc}}). The specific volume $v\left(T,p\right)$ is also a convenient independent variable because it appears explicitly in the balance equations of hydrodynamics, e.g. (\ref{RH_Mass})-(\ref{RH_Energ}). Specific enthalpy $h$ and entropy $s$ are state functions essential in this work, and their differentials write
\begin{align}
dh\left(s,p\right) &=Tds+vdp,  \label{H_SP} \\
dh\left(p,v\right) &=\frac{G+1}{G}vdp+\frac{c^{2}}{G}\frac{dv}{v}, \label{H_PV}\\
dh\left(T,p\right) &=C_{p}dT+\left(1-\frac{T}{v}\left. \frac{\partial v}{\partial T}\right) _{p}\right) vdp,  \label{H_TP} \\
Tds\left(p,v\right) &=\frac{vdp}{G}+\frac{c^{2}}{G}\frac{dv}{v},
\label{S_PV} \\
c^{2} &=Gv\left. \frac{\partial h}{\partial v}\right) _{p}=-v^{2}\left. 
\frac{\partial p}{\partial v}\right) _{s},  \label{SOUND} \\
G &=\frac{v}{\left. \frac{\partial h}{\partial p}\right) _{v}-v}=-\frac{v}{T}\left. \frac{\partial T}{\partial v}\right) _{s},  \label{GRUN}
\end{align}
where $G$ is the Gruneisen coefficient, $C_{p}$ is the heat capacity at constant pressure, and $c$ is the sound speed. In gases, the adiabatic exponent $\gamma$ conveniently defines $c$ by
\begin{equation}
c^{2}=\gamma pv,\quad \gamma =-\frac{v}{p}\left. \frac{\partial p}{\partial v}\right) _{s}.  \label{GAMMA}
\end{equation}
In the $p$-$v$ plane, the isentropes ($ds=0$) have negative slopes ($\gamma>0$), and the fundamental derivative of hydrodynamics
\begin{equation}
\!\!\Gamma=\frac{1}{2}\frac{v^{3}}{c^{2}}\left. \frac{\partial ^{2}p}{\partial
v^{2}}\right) _{s}\!\!=\frac{-v}{2}\left. \frac{\partial ^{2}p}{%
\partial v^{2}}\right) _{s}\!/\left. \frac{\partial p}{\partial v}\right) _{s}%
\!\!=1-\frac{v}{c}\left. \frac{\partial c}{\partial v}\right)
_{s} \label{FUND_HYDRO}
\end{equation}
defines their convexity \cite{{Duhem1909},{Bethe1942},{Weyl1949},{Thomson1971}}. Most fluids have uniformly convex isentropes with slopes that decrease monotonically with increasing volume ($\Gamma >0$).
The initial (fresh, subscript $0$) and final (burnt, no subscript) states of a reactive medium have different chemical compositions and hence different state functions and coefficients. Typically, $\gamma <\gamma _0$ and, if products are brought from a $\left(T,p\right)$ equilibrium state to the $\left(T_0,p_0\right)$ initial state, $v\left(T_0,p_0\right)>v_0=v_0\left(T_0,p_0\right)$ and $h\left(T_0,p_0\right)<h_0=h_0\left(T_0,p_0\right)$. The difference of enthalpies $Q_0=h_0\left(T_0,p_0\right)-h\left(T_0,p_0\right)$ is the heat of reaction at constant pressure.
\newline

The Rankine-Hugoniot relations (RH) express the conservation of mass, momentum and energy surface fluxes across a hydrodynamic discontinuity. Along the normal to the discontinuity, they write

\begin{align}
\rho _0D &=\rho \left(D-u\right),  \label{RH_Mass} \\
p_0+\rho _0D^{2} &=p+\rho \left(D-u\right) ^{2},  \label{RH_Mom} \\
h_0+\frac{1}{2}D^{2} &=h+\frac{1}{2}\left(D-u\right) ^{2}, \label{RH_Energ}
\end{align}%
where $\rho =1/v$ is the density, and $u$ and $D$ the material speed and the discontinuity velocity in a laboratory-fixed frame, with initial state at rest ($u_0=0$). These relations completed with an $h\left(p,v\right) $ equation of state are not a closed system \hyperref[subsec:InVarPb]{(Subsect. \ref{subsec:InVarPb})}, since there are 4 equations for the 5 variables $v$, $p$, $h$, $u$ and $D$, given an initial state $\left(p_0,v_0\right)$ and $h_0\left(p_0,v_0\right)$, hence a one-variable solution, e.g.
\begin{equation}
p,v,h,u,T,s,c,\gamma ,\Gamma ,G,\;...\equiv \eta \left(D;v_0,p_0\right).
\label{CAUS_OVDRV}
\end{equation}

\begin{figure}[h!]
\includegraphics{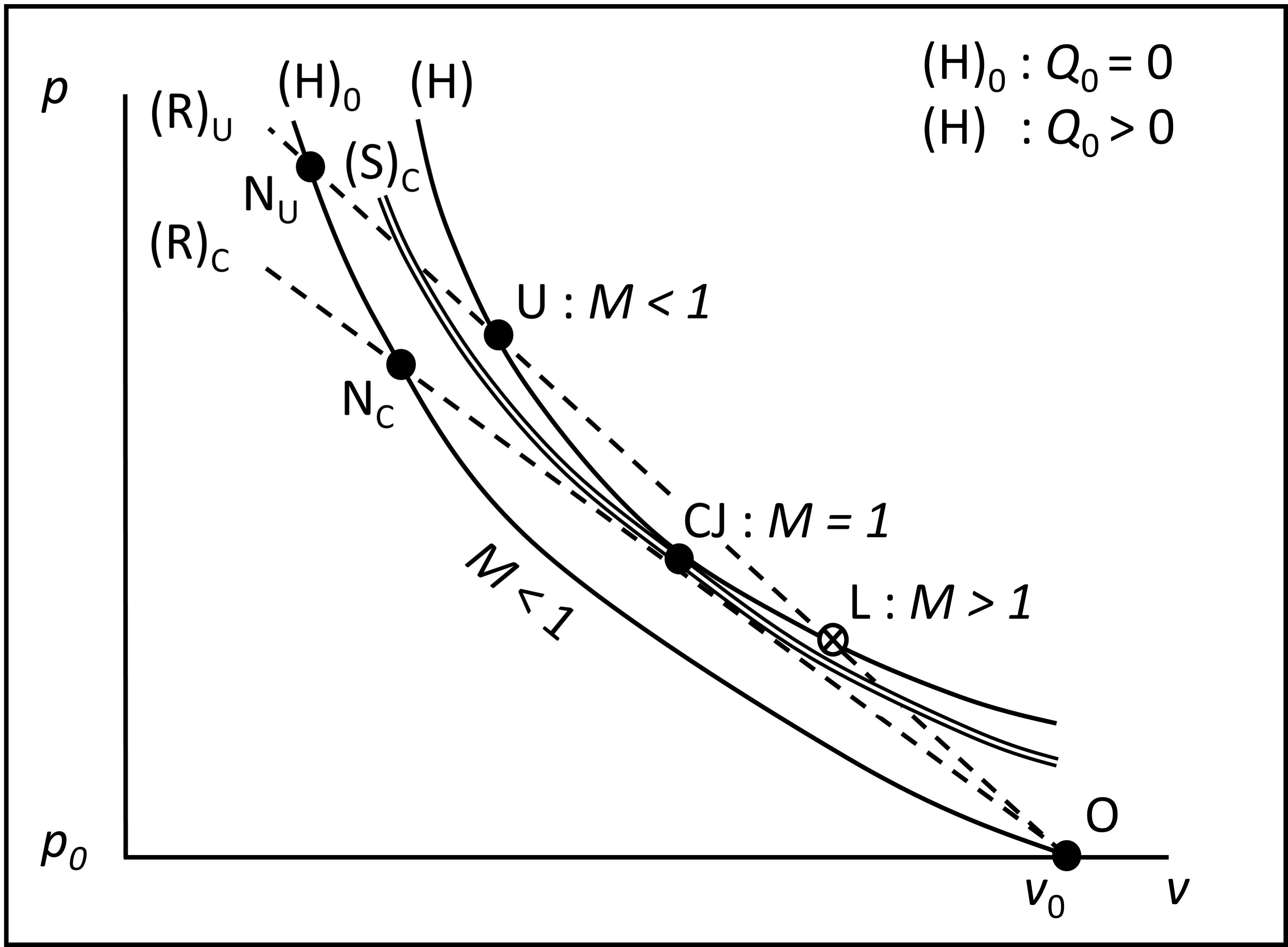}
\caption{Unreacted (H)$_0$ and equilibrium (H) Hugoniot curves, and Rayleigh-Michelson lines with velocities greater than or equal to ${D}_{\text{CJ}}$, i.e. (R)$_{\text{U}}$ and (R)$_{\text{C}}$, resp., for compressive discontinuities ($v/v_0<1$). The physical intersects are the points N, U and CJ ${(M\leqslant 1})$. The CJ isentrope (S)$_{\text{C}}$ is positioned between the (R)$_{\text{C}}$ line and the (H) curve.}
\label{fig:Fig2}
\end{figure}

Its representation in the $p$-$v$ plane (Fig. \ref{fig:Fig2}) is an intersect of a Rayleigh-Michelson line (R) : $p_{\text{R}}\left(v,D;v_0,p_0\right) $ and a Hugoniot curve (H) : $p_{\text{H}}\left(v;v_0,p_0\right)$,
\begin{align}
&\ \ p_{\text{R}}:&p&=p_0+\left(\frac{D}{v_0}\right)
^{2}\left(v_0-v\right) ,  \label{RM} \\
&\ \ p_{\text{H}}:&h\left(p,v\right)&=h_0\left(p_0,v_0\right) +\frac{1}{2}\left(p-p_0\right) \left(v_0+v\right).  \label{HUGO}
\end{align}

Hugoniots for discontinuities with final states at chemical equilibrium ($Q_0>0$) lie above that for shocks ($Q_0=0$). Figure \ref{fig:Fig2} shows the case of compressive discontinuities ($v/v_0<1$). Most fluids have uniformly convex Hugoniots with 1 intersect (N) if $Q_0=0$ regardless of $D$, and 2 (U and L) if $Q_0>0$ and $D$ is large enough.

The observability of states on non-uniformly convex Hugoniots is a debate about the instability criteria derived from stability analyses of discontinuities, e.g. \cite{{Dyakov1954},{Kontorovich1957},{BatesMontgomery2000},{Brun2013},{Clavin2016}}. At least physical admissibility (the discontinuity increases entropy, $s > s_0$) or equivalently mathematical determinacy (uniqueness and continuous dependence of (\ref{CAUS_OVDRV}) on the boundaries) must be satisfied \cite{{Landau1944},{Lax1957},{Fowles1975}}. Denoting the Mach numbers of the discontinuity relative to its initial and final states by $M_0$ and $M$, that is expressed by the subsonic-supersonic evolution condition
\begin{equation}
u+c > D > c_0\ \Leftrightarrow\ \frac{D}{c_0}=M_0 > 1 > M=\frac{D-u}{c}. \label{MACH}
\end{equation}

The tangency of a Rayleigh-Michelson line $p_{\text{R}}\left(v;D\right) $, an equilibrium Hugoniot $p_{\text{H}}\left(v\right)$ and an isentrope $p_{\text{S}}\left(v\right) $ defines CJ points and is
equivalent to the sonic condition
\begin{equation}
M_{\text{CJ}}=\left(\frac{D-u}{c}\right) _{\text{CJ}}=1\ \text{ or }\ D_{\text{%
CJ}}=\left(u+c\right) _{\text{CJ}},  \label{CJ}
\end{equation} as shown by
\begin{align}
\left. \frac{\partial p_{\text{R}}}{\partial v}\right) _{D,p_0,v_0}
&=-\left(\frac{D}{v_0}\right) ^{2}<0,  \label{R_SLP} \\
\left. \frac{\partial p_{\text{S}}}{\partial v}\right) _{s}&=-\left(\frac{D%
}{v_0}\right) ^{2}\times M^{-2}<0,  \label{S_SLP} \\
\left. \frac{\partial p_{\text{H}}}{\partial v}\right) _{p_0,v_0}
&=-\left(\frac{D}{v_0}\right) ^{2}\times \left(1+2\frac{M^{-2}-1}{F}%
\right) ,  \label{H_SLP}\\
F\left(G,v;v_0\right)&=2-G\left(\frac{v_0}{v}-1\right).  \label{F}
\end{align}

\begin{figure}[h!]
\includegraphics{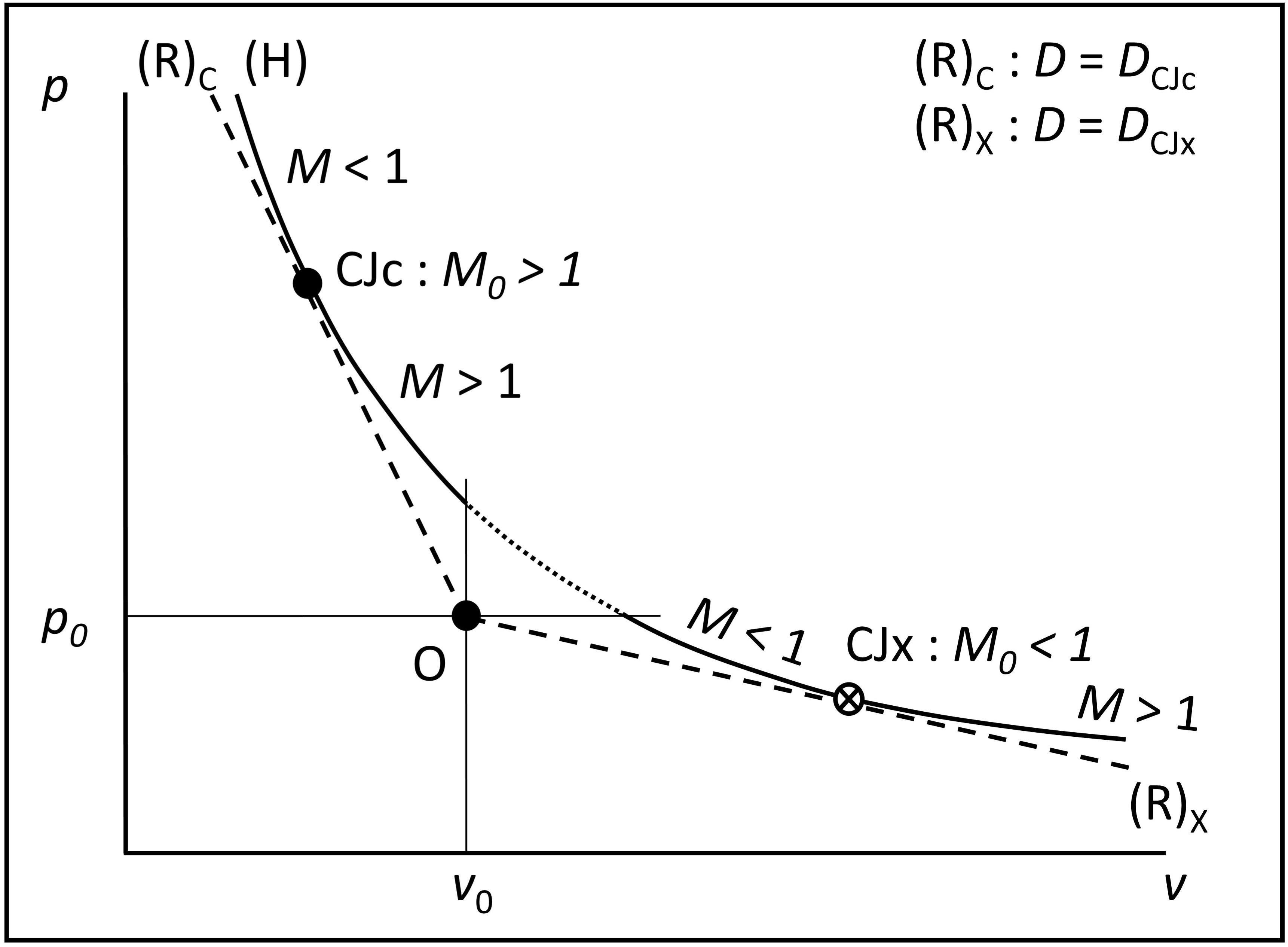}
\caption{Detonation (upper) and deflagration (lower) Hugoniot arcs. The
physical arc is above the compressive CJ point CJc.}
\label{fig:Fig3}
\end{figure}

Uniformly convex equilibrium Hugoniots have 2 CJ points (Fig. \ref{fig:Fig3}). The upper one (CJc) represents the CJ detonation, which is compressive and has a supersonic velocity relative to the initial state: $v_{\text{CJ}}/v_0<1$, $p_{\text{CJ}}/p_0>1$, $D_{\text{CJ}}/c_0>1$. The lower one (CJx) represents the CJ deflagration, which is expansive and has a subsonic velocity: $v_{\text{CJ}}/v_0>1$, $p_{\text{CJ}}/p_0<1$, $D_{\text{CJ}}/c_0<1$.

The admissibility of the CJ detonation \hyperref[sec:CJadm]{(App. \ref{sec:CJadm})} requires $\Gamma_{\text{CJ}}>0$, so $F>0$ around and at a CJ point, the physical arc of an equilibrium Hugoniot is convex and above the CJ point as $M$ decreases from $1$ and $s$ increases with decreasing $v$, and $p_{\text{S}}\left(v\right)$ lies between $p_{\text{H}}\left(v\right)$ and $p_{\text{R}}\left(v\right)$ if $G>0$ (Figs. \ref{fig:Fig1} and \ref{fig:Fig2}). Other properties are especially useful here \hyperref[subsec:Proof]{(Subsect. \ref{subsec:Proof})}, i.e. $0\leqslant \left. \partial s_{\text{H}}/\partial D\right) _{p_0,v_0}<\infty $ regardless of $M$, and,
since $F_{\text{CJ}}\neq 0$, $\left. \partial s_{\text{H}}/\partial
v\right)_{p_0,v_0}^{\text{CJ}}=0$ and $\left. \partial D/\partial
v\right)_{p_0,v_0}^{\text{CJ}}=0$, as shown by
\begin{align}
\frac{v_0T}{D^{2}}\left. \frac{\partial s_{\text{R}}}{\partial v}\right)
_{D,p_0,v_0}&=\frac{v}{v_0}\frac{M^{-2}-1}{G},  \label{DERIV_SR_V} \\
\frac{v_0T}{D^{2}}\left. \frac{\partial s_{\text{H}}}{\partial v}\right)
_{p_0,v_0}&=-\left(1-\frac{v}{v_0}\right) \frac{M^{-2}-1}{F},
\label{DERIV_SH_V}\\
\frac{T}{D}\left. \frac{\partial s_{\text{H}}}{\partial D}\right)
_{p_0,v_0}&=\left(1-\frac{v}{v_0}\right) ^{2}>0,  \label{DERIV_SH_D}\\
\frac{v_0}{D}\left. \frac{\partial D}{\partial v}\right) _{p_0,v_0}
&=-\left(1-\frac{v}{v_0}\right) ^{-1}\frac{M^{-2}-1}{F}.
\label{DERIV_DH_V}
\end{align}
The CJ condition (\ref{CJ}) closes the system (\ref{H_PV}), (\ref{RH_Mass})-(\ref{RH_Energ}).~The one-variable solution (\ref{CAUS_OVDRV}) and (\ref{CJ}) give the CJ velocities $D_{\text{CJ}}$ and variables $\eta _{\text{CJ}}=\left(p,v,h,u,T,s,c,\gamma ,\Gamma ,G,\text{...}\right) _{\text{CJ}}$ as functions of the initial state,
\begin{equation}
D_{\text{CJ}}=D_{\text{CJ}}\left(v_0,p_0\right) ,\quad \eta _{\text{CJ}}=\eta _{\text{CJ}}\left(v_0,p_0\right).  \label{CJ_STATE}
\end{equation}%
Simple $h\left(p,v\right)$ equations of state give explicit relations \hyperref[sec:CJperf]{(App. \ref{sec:CJperf})}, but accurate CJ detonation properties are calculated using thermochemical computer programs, e.g. NASA's CEA \cite{GordonMcBride-I}, which combine detailed chemical equilibria, physical equations of state and thermodynamic properties at high  temperatures and pressures.
\newline

The hydrodynamic variables $y=\left(p,v,u,c,h\right)$ at CJ points have a well-known two-variable representation as exact functions of $D_{\text{CJ}}$ \textit{\textbf{and}} $\gamma _{\text{CJ}}$
\begin{equation}
y_{\text{CJ}}=y_{\text{CJ}}\left(D_{\text{CJ}},\gamma _{\text{CJ}};v_0,p_0\right),  \label{CJ_STATE_2_PARAM}
\end{equation}%
for example,
\begin{align}
\frac{v_{\text{CJ}}}{v_0}&=\frac{c_{\text{CJ}}}{D_{\text{CJ}}}=\frac{%
\gamma _{\text{CJ}}}{\gamma _{\text{CJ}}+1}\left(1+\frac{p_0v_0}{D_{%
\text{CJ}}^{2}}\right) ,  \label{VCJ_PARAM} \\
\frac{v_0p_{\text{CJ}}}{D_{\text{CJ}}^{2}}&=\frac{1+\frac{p_0v_0}{D_{%
\text{CJ}}^{2}}}{\gamma _{\text{CJ}}+1}, \hspace{5mm}\frac{u_{\text{CJ}}}{D_{\text{CJ}}}=\frac{1-\gamma _{\text{CJ}}\frac{%
p_0v_0}{D_{\text{CJ}}^{2}}}{\gamma _{\text{CJ}}+1} \label{PCJ_PARAM}
\end{align}
obtained by combining (\ref{GAMMA}), the mass balance (\ref{RH_Mass}), the (R) relation (\ref{RM}) and the CJ condition (\ref{CJ}). The Hugoniot relation (\ref{HUGO}) rewritten as $(h-h_0)/D^2=(1-(v/v_0)^2)/2$ then gives $h_{\text{CJ}}$. A zero-variable representation (\ref{CJ_STATE}) is obtained from a complete set that includes an explicit equation of state and a detailed chemical equilibrium composition, hence the two-variable representation (\ref{CJ_STATE_2_PARAM}) since it does not use these two constraints.
\vspace{1mm}

The theorem complements (\ref{CJ_STATE_2_PARAM}) by using the equation of state implicitly \hyperref[sec:DSITh]{(Sect. \ref{sec:DSITh})}. The primary consequence is that the $y_{\text{CJ}}$'s \textit{\textbf{and}} $\gamma_{\text{CJ}}$ have an explicit one-variable representation as exact functions of $D_{\text{CJ}}$,%
\begin{equation}
y_{\text{CJ}}=y_{\text{CJ}}\left(D_{\text{CJ}};v_0,p_0\right),\;\gamma
_{\text{CJ}}=\gamma _{\text{CJ}}\left(D_{\text{CJ}};v_0,p_0\right),
\label{CJ_STATE_1_PARAM}
\end{equation}%
or, reciprocally, $D_{\text{CJ}}$ as a function of any one of the CJ variable, for example, $D_{\text{CJ}}\left(\gamma_{\text{CJ}};v_0,p_0\right)$ \hyperref[subsec:CJsuppl]{(Subsect. \ref{subsec:CJsuppl})}.
\section{\label{sec:DSITh}The invariance theorem}
\subsection{\label{subsec:Statemnt}Statement}
Considering a series of experiments on a homogeneous explosive with the same uniform composition, each performed at different but uniform initial pressures $p_0$ and temperatures $T_0$, equivalent statements are:

\begin{enumerate}
\item CJ detonations with the same velocity $D_{\text{CJ}}$ have the same specific entropy $s_{\text{CJ}}$, and reciprocally;
\item $D_{\text{CJ}}$ and $s_{\text{CJ}}$ are invariant under the same dependence of $T_0$ on $p_0$.
\end{enumerate}


\subsection{\label{subsec:InVarPb}The initial-state variation problem}
A semi-infinite tube closed at one end by a constant-speed piston, e.g. \cite{FickDav2000}, provides the visual support for the proof of the theorem \hyperref[subsec:Proof]{(Subsect. \ref{subsec:Proof})}. 
The two independent experimental controls are the initial state $\left(v_0,p_0\right)$ and the speed $u_{\text{p}}$ of the piston that may sustain the wave. The aim is to characterize the velocity of the discontinuity and the flow of detonation products behind it, not only for one initial state but also for several, depending on the constraint placed on the final states. There are three possibilities for the case relevant to the planar discontinuity.
\vspace{1mm}

If $u_{\text{p}}$ is greater than the CJ material speed $u_{\text{CJ}}\left(v_0,p_0\right)$ (\ref{PCJ_PARAM}), the flow is constant-state and subsonic relative to the discontinuity ($D-u_{\text{p}}<c$), and the detonation is overdriven. The velocity $D$ and the final-state variables $\left(\eta\right)=\left(p,v,h,T,s,c,\gamma,\Gamma,G,\text{ ...}\right)$ are one-variable functions \hyperref[sec:Remind]{(Sect. \ref{sec:Remind})}, i.e. $D\left(u_{\text{p}};v_0,p_0\right)$ and $\eta\left(u_{\text{p}};v_0,p_0\right)$, or $\eta\left(D;v_0,p_0\right)$ (\ref{CAUS_OVDRV}), e.g. (\ref{PERF_PV_OVRDRV}).
\vspace{0.5mm}

If $u_{\text{p}}$ is smaller than $u_{\text{CJ}}\left(v_0,p_0\right)$, the flow is expanding and supersonic relative to the discontinuity ($D-u>c$) but sonic just at its front, which defines the CJ self-sustained detonation \hyperref[sec:Remind]{(Sect. \ref{sec:Remind}}, \hyperref[sec:CJadm]{App. \ref{sec:CJadm})}. The CJ condition is a consequence of the Taylor-Zel'dovich-D\"{o}ring (TZD) simple-wave solution $\eta\left( x/t\right)$ to the homentropic (uniform $s$) unsteady flow behind a constant-velocity front: $u+c=x/t\Rightarrow \left( u+c\right)_{\text{CJ}}=x_{\text{CJ}}\left(t\right)/t\equiv D_{\text{CJ}}$, with $t$ the time and $x$ the position in the flow \cite{{Taylor1950}, {DoringBurkhardt1944},{ZeldoKompa1960}}. In contrast to the overdriven case, no disturbance in the flow reaches the front: $x<~x_{\text{CJ}}\Rightarrow x/t~=u+c<~x_{\text{CJ}}/t=\left(u+c\right)_{\text{CJ}}=D_{\text{CJ}}$. The CJ velocity and state are the functions $D_{\text{CJ}}\left(v_0,p_0\right)$ and $\eta _{\text{CJ}}\left(v_0,p_0\right)$ (\ref{CJ_STATE}) of the initial state only, e.g. (\ref{VCJ_PARAM}), (\ref{PCJ_PARAM}) and (\ref{PERF_DCJ}).
\vspace{0.5mm}

If $u_{\text{p}}$ is equal to $u_{\text{CJ}}\left(v_0,p_0\right)$, the flow is both constant-state and sonic for any $x$ and $t$: $u_{\text{p}}+c=\left(u+c\right)_{\text{CJ}}=x_{\text{CJ}}\left(t\right)/t=D_{\text{CJ}}$.
\vspace{1mm}

This implies that $D_{\text{CJ}}\left( v_0,p_0\right)$ is both the value associated with the TZD solution and the lowest observable in a series of overdriven experiments, each performed with $u_{p}$ greater than, but progressively closer to, $u_{\text{CJ}}$, given the same initial state $\left(v_0,p_0\right)$.

Therefore, considering now a set of initial states $\left(v_0,p_0\right)$, overdriven detonations can have the same velocity $D$, or the same value of one of the final state variables $\eta$, only by choosing $u_{\text{p}}$ appropriately for each pair $(v_0,p_0)$. But then, no other final state variable can be invariant. For example, the final states can be on the same isentrope ($\eta\equiv s$), but $D$, $u_{\text{p}}$ and the other final-state variables will change along this isentrope, depending on $\left(v_0,p_0\right)$. For self-sustained CJ detonations, it turns out that the same initial states ensure the invariance of both $D_{\text{CJ}}$ and $s_{\text{CJ}}$. The entropy appears only in the differentials $dh\left( s,p\right)$ (\ref{H_SP}) and $ds\left(p,v\right)$ (\ref{S_PV}), which entails differentiating the Rankine-Hugoniot relations \hyperref[subsec:RHdiff]{(Subsect. \ref{subsec:RHdiff})} to prove the theorem in the form of the equivalence of the constraints $ds_\text{CJ}=0$ and $dD_\text{CJ}=0$ \hyperref[subsec:Proof]{(Subsect. \ref{subsec:Proof})}.

\subsection{\label{subsec:RHdiff}Rankine-Hugoniot differentials}

Introducing the dimensionless hydrodynamic variable
\begin{equation}
z=\;1-\frac{v}{v_0},
\label{NONDIM_VAR_z}
\end{equation}
the differentials of the Rayleigh-Michelson line (\ref{RM}), the Hugoniot
relation (\ref{HUGO}) and the $h\left( p,v\right) $ equation of state (\ref{H_PV}) form the $3\times 3$ linear system for $dv$, $dp$ and $dh$
\begin{align}
\frac{v_0dp}{D^{2}}+\frac{dv}{v_0}=\frac{v_0dp_0}{D^{2}}&+\left(1-2z\right)\frac{dv_0}{v_0}+2z\frac{dD}{D},  \label{R_DIFF} \\
2\frac{dh}{D^{2}}-\left(2-z\right)\frac{v_{0}dp}{D^{2}}&-z\frac{dv}{v_{0}}=\;...  \notag \\
...\;2\frac{dh_{0}}{D^{2}}&-\left(2-z\right)\frac{%
v_{0}dp_{0}}{D^{2}}+z\frac{dv_{0}}{v_{0}}, \label{H_DIFF} \\
\frac{G}{1-z}\frac{dh}{D^{2}}-&\left(G+1\right)\frac{v_0dp}{D^{2}}-M^{-2}\frac{dv}{v_0}=0,  \label{HPV_DIFF}
\end{align}
which thus are linear combinations of $dD$, $dv_0$, $dp_0$ and $dh_0\left( p_0,v_0\right)$. For example, using the notation $F\left(G,z\right)$~(\ref{F}),
\begin{align}
&\left( M^{-2}-1\right) \frac{dv}{v_0}=-zF\frac{dD}{D}+\frac{2-F}{z}\frac{dh_0}{D^{2}}\;... \nonumber \\
&... +\left( 1-F\left( 1-z\right)\right)\frac{dv_0}{v_0}-\frac{2\!-\!z\!-\!F\left(1\!-\!z\right)}{z}\frac{v_0dp_0}{D^{2}}.  \label{dv}
\end{align}

The differential $ds$ of the specific entropy
\begin{equation}
\frac{Tds}{D^{2}}=z^{2}\frac{dD}{D}+\frac{dh_0}{D^{2}}+\left( 1-z\right) z\frac{dv_0}{v_0}-\left(
1-z\right) \frac{v_0dp_0}{D^{2}}  \label{ds}    
\end{equation}
is obtained with $dh\left(s,p\right)$ (\ref{H_SP}) instead of $dh\left(p,v\right)$ (\ref{HPV_DIFF}). The state functions $c$ and $G$ do not appear in $dh\left( s,p\right) $, so neither do $M$ and $F$ in $ds$ (\ref{ds}).

The determinant of system (\ref{R_DIFF})-(\ref{HPV_DIFF}) is $M^{2}-1$, and the right-hand side of (\ref{dv}) has to be set to zero for CJ discontinuities ($M=1$) so that $dv$, $dp$ and $dh$ are finite, hence, from (\ref{dv}), the differentials of $s_{\text{CJ}}(h_0,v_0,p_0)$ and $D_{\text{CJ}}(h_0,v_0,p_0)$,
\begin{align}
\frac{T_{\text{CJ}}ds_{\text{CJ}}}{D_{\text{CJ}}^{2}}& =\frac{2}{F_{\text{CJ}%
}}\frac{dh_0}{D_{\text{CJ}}^{2}}+\frac{z_{\text{CJ}}}{F_{\text{CJ}}}\frac{%
dv_0}{v_0}-\frac{2-z_{\text{CJ}}}{F_{\text{CJ}}}\frac{v_0dp_0}{D_{%
\text{CJ}}^{2}},  \label{IM_ds_Gen} \\
\frac{dD_{\text{CJ}}}{D_{\text{CJ}}}& =\frac{2-F_{\text{CJ}}}{z_{\text{CJ}}^{2}F_{\text{CJ}}}\frac{dh_0}{D_{\text{CJ}}^{2}}+\frac{1-F_{\text{CJ}}\left( 1-z_{\text{CJ}}\right) }{z_{\text{CJ}}F_{\text{CJ}}}\frac{dv_0}{v_0}\;...  \nonumber \\
& ...-\;\frac{2-z_{\text{CJ}}-F_{\text{CJ}}\left( 1-z_{\text{CJ}}\right)}{z_{\text{CJ}}^{2}F_{\text{CJ}}}\frac{v_0dp_0}{D_{\text{CJ}}^{2}}. \label{IM_dD_Gen}
\end{align}%
Replacing $dh_0$ by $dh_0\left( p_0,v_0\right)$ (\ref{H_PV}) written as
\begin{equation}
\frac{dh_0}{D^{2}}=\frac{G_0+1}{G_0}\frac{v_0dp_0}{D^{2}}+\frac{M_0^{-2}}{G_0}\frac{dv_0}{v_0},  \label{EOS_HPV_0}
\end{equation}
and introducing
\begin{align}
a& =z\left( 1-z\right) +\frac{M_0^{-2}}{G_0},\!\!\! & b& =z+\frac{1}{G_0},  \label{Coef_dv_D_Cond} \\ 
A& =\frac{z+\frac{2M_0^{-2}}{G_0}}{F}-a,\!\!\! & B& =\frac{z+\frac{2}{G_0}}{F}-b,
\label{Coef_dv_s_Cond}
\end{align}
the differentials of $s(D,h_0,v_0,p_0)$ (\ref{ds}) and $v(D,h_0,v_0,p_0)$ (\ref{dv}) reduce to differentials of $s(D,v_0,p_0)$ and $v(D,v_0,p_0)$, and so of $v(s,v_0,p_0)$,
\begin{align}
\frac{Tds}{D^{2}}&=\ \ z^2\frac{dD}{D}\,+a\,\frac{dv_0}{v_0}\,+b\,\frac{v_0dp_0}{D^{2}}, \label{ds_D_Cond} \\
\frac{M^{-2}-1}{F/z}\frac{dv}{v_0}&=-z^2\frac{dD}{D}+A\frac{dv_0}{v_0}+B\frac{v_0dp_0}{D^{2}}, \label{dv_D_Cond}
\end{align}
hence, for CJ discontinuities, those of $s_{\text{CJ}}(h_0,v_0,p_0)$ (\ref{IM_ds_Gen}) and $D_{\text{CJ}}(h_0,v_0,p_0)$ (\ref{IM_ds_Gen}) to differentials of $s_{\text{CJ}}(v_0,p_0)$ and $D_{\text{CJ}}(v_0,p_0)$,
\begin{align}
\!\!\!\!\!T_{\text{CJ}}\frac{ds_{\text{CJ}}}{D_{\text{CJ}}^{2}}
&=\left(A_{\text{CJ}}+a_{\text{CJ}}\right)\frac{dv_0}{v_0}+\left(B_{\text{CJ}}+b_{\text{CJ}}\right)\frac{v_0dp_0}{D_{\text{CJ}}^{2}},  \label{ds_CJ_Cond} \\
\!\!\!\!\!z_{\text{CJ}}^{2}\frac{dD_{\text{CJ}}}{D_{\text{CJ}}}&=A_{\text{CJ}} \frac{dv_0}{v_0}+B_{\text{CJ}}\frac{v_0dp_0}{D^{2}}.  \label{dD_CJ_Cond}
\end{align}
The CJ differentials (\ref{ds_CJ_Cond}) and (\ref{ds_CJ_Cond}) can be obtained directly from (\ref{R_DIFF}) and (\ref{H_DIFF}) by using (\ref{EOS_HPV_0}), $dh\left(s,p\right)$ (\ref{H_SP}) and $ds\left( p,v\right)$ (\ref{S_PV}), and the CJ condition $M=1$ in the form $c/v=D/v_0$ (\ref{RH_Mass}). 
The coefficients of their initial variations show that $F_{\text{CJ}}\neq 0$ (\ref{F}) is also the continuity condition for small initial variations to produce small variations of $D_{\text{CJ}}$ and $s_{\text{CJ}}$ \hyperref[sec:Remind]{(Sect. \ref{sec:Remind}}, \hyperref[subsec:Proof]{Subsect. \ref{subsec:Proof}}, \hyperref[sec:CJadm]{App. \ref{sec:CJadm})}.
In the acoustic sonic limit, i.e. $D\rightarrow c_0$, $v/v_0\rightarrow 1$, $z\rightarrow 0$, $F\rightarrow 2$, (\ref{ds}) and (\ref{IM_ds_Gen}) reduce coherently to $dh_0\left(s_0,p_0\right)$ (\ref{H_SP}). This limit is irrelevant to this analysis.
\newline

The differentials (\ref{ds_D_Cond}) and (\ref{dv_D_Cond}) of $s$ and $v$ are essential for the proof \hyperref[subsec:Proof]{(Subsect. \ref{subsec:Proof})}. At constant initial state $(p_0,v_0)$, they give the partial derivatives along the same Hugoniot, i.e. the relative variations induced by a change in the piston speed from one experiment to another, each performed with the same initial state, e.g. (\ref{DERIV_SH_V})-(\ref{DERIV_DH_V}). Subject to the constraint of a constant final-state variable, e.g. $s$ or $D$, they give the relative variations induced by a change in the initial state from one experiment to another. The proof regards the self-sustained constraint $M=1$ as expressing the equivalence of 
the self-sustained detonation and the lower limit of the overdriven detonation \hyperref[subsec:InVarPb]{(Subsect. \ref{subsec:InVarPb})}.
\subsection{\label{subsec:Proof}Proof}

The physical premise is that the variations in the final-state variables induced by a variation in the initial state, from one experiment to another, are finite, regardless of the constraint on that final state. 
Examples are the variations in the CJ state from one initial state to another or those of $p$ and $v$ along an isentrope (\ref{GAMMA}). 
In contrast, the variations at constant initial state, i.e. along the same Hugoniot, 
become infinite when approaching the CJ state ($M\!=\!1$), e.g. $\left.\partial v/\partial D\right)_{v_0,p_0}$ (\ref{DERIV_DH_V}), (\ref{dv_D_Cond}). That of entropy, $\left.\partial s/\partial D\right)_{v_0,p_0}$, 
is unconditionally finite (\ref{DERIV_SH_D}), (\ref{ds_D_Cond}).

It is convenient to distribute the initial states ($v_0,p_0$) on an arbitrary polar curve $p_0^{\ast }\left(v_0\right)$ through a reference point O$^{\ast}(v_{0\ast},p_{0\ast}=p_0^{\ast}\left( v_{0\ast}\right))$ (Fig. \ref{fig:Fig1}). Thus, final states subject to the same constraint lie on a $\left(p\text{-}v\right)$ arc between a point U on a Hugoniot (H) with pole O$\left(v_0,p_0=p_0^{\ast}\left( v_0\right)\right)$ and a point U' on another Hugoniot (H)' with pole O'$\left(v_0+dv_0,p_0+dp_0^{\ast}\right)$, i.e. both on $p_0^{\ast}\left(v_0\right)$.

The differentials of $s(D,v_0,p_0)$ (\ref{ds_D_Cond}) and $v(D,v_0,p_0)$ (\ref{dv_D_Cond}) reduce to differentials of $s^{\ast}(D,v_0)=s(D,v_0,p_0=p_0^{\ast}\left(v_0\right))$ and $v^{\ast}(D,v_0)=v(D,v_0,p_0=p_0^{\ast}\left(v_0\right))$. That gives the partial derivatives of $v^{\ast}$ along final-state $\left(p\text{-}v\right)$ arcs with $s^{\ast}$ or $D$ fixed,
\begin{align}
\frac{M^{-2}-1}{F/z}\left.\frac{\partial v^{\ast}}{\partial v_0}\right)_{s^{\ast}}&\!\!=\left(A+a\right)+\left(B+b\right)\left(\frac{v_0}{D}\right)^{2}\frac{dp_0^{\ast}}{dv_0}, \label{Math_Diff_vs} \\
\frac{M^{-2}-1}{F/z}\left. \frac{\partial v^{\ast}}{\partial v_0}\right)_{D}&\!\!=A+B\left(\frac{v_0}{D}\right)^{2}\frac{dp_0^{\ast }}{dv_0}, \label{Math_Diff_vD} 
\end{align}
and those of $s^{\ast}$ and $D$,
\begin{align}
\frac{v_0T}{D^{2}}\left.\frac{\partial s^{\ast}}{\partial v_0}\right)_{D}&\!\!=-z^{2}\frac{v_0}{D}\left.\frac{\partial D}{\partial v_0}\right)_{s^{\ast}}=a+b\left(\frac{v_0}{D}\right)^{2}\frac{dp_0^{\ast }}{dv_0}, \label{Math_Diff_sD} \\
\frac{v_0}{D}\left.\frac{\partial D}{\partial v_0}\right)_{s^{\ast}}&\!\!=\frac{M^{-2}-1}{zF}\left(\left.\frac{\partial v^{\ast}}{\partial v_0}\right)_{D}\!\!-\left. \frac{\partial v^{\ast}}{\partial v_0}\right)_{s^{\ast}}\right). \label{Math_Diff_Ds}
\end{align}
The first equality in (\ref{Math_Diff_sD}) can be directly obtained from (\ref{DERIV_SH_D}) using the mathematical identity known as the triple product rule or Euler's chain rule,
\begin{align}
\left.\frac{\partial s^{\ast}}{\partial v_0}\right)_{D}\left.\frac{\partial v_0}{\partial D}\right)_{s^{\ast}}\left.\frac{\partial D}{\partial s^{\ast}}\right)_{v_0}=-1. \label{Triple_Rule}
\end{align}
It shows that the partial derivatives, with respect to the initial state, of $s^{\ast}$ at $D$ fixed and of $D$ at $s^{\ast}$ fixed are proportional to each other. If one is zero, so is the other, since $\left.\partial s^{\ast}/\partial D\right)_{v_0} \propto~z^2$ is unconditionally finite (\ref{DERIV_SH_D}), (\ref{ds_D_Cond}).
The identity is valid for any $M\leqslant1$, i.e. any piston speed $u_\text{p}\geqslant u_\text{CJ}$, and any initial state \hyperref[subsec:InVarPb]{(Subsect. \ref{subsec:InVarPb})}. But it does not imply the equivalence of $ds=0$ and $dD=0$. In fact, relation (\ref{Math_Diff_Ds}) shows that it is only possible for final states continuously sonic ($M=1$) -- with $\left.\partial v^{\ast}/\partial v_0\right)_{D}$ and $\left.\partial v^{\ast}/\partial v_0\right)_{s^{\ast}}$ finite -- which proves the theorem
\begin{align}
\left.\frac{\partial s^{\ast}}{\partial v_0}\right)_{D}^{\text{CJ}}=0 \quad &\Leftrightarrow \quad
\left.\frac{\partial D}{\partial v_0}\right)_{s^{\ast}}^{\text{CJ}}=0, \label{DSI} \\
\text{i.e.} \quad ds_\text{CJ}=0 \quad &\Leftrightarrow \quad dD_\text{CJ}=0, \label{DSI_bis}
\end{align}

Therefore, the same initial-state polar $p_0^{\ast}\left(v_0\right)$ produces final states with $D_\text{CJ}$ and $s_\text{CJ}$ constant. The relations (\ref{Math_Diff_vs}) and (\ref{Math_Diff_vD}), or (\ref{ds_CJ_Cond}) and (\ref{dD_CJ_Cond}), then give
\begin{equation} 
-\left(\frac{v_0}{D_{\text{CJ}}}\right)^{2}\frac{dp_0^{\ast }}{dv_0}=\frac{A_{\text{CJ}}+a_{\text{CJ}}}{B_{\text{CJ}}+b_{\text{CJ}}} =\frac{A_{\text{CJ}}}{B_{\text{CJ}}}, \label{VAR_INIT_CJ}
\end{equation}
hence the additional constraint on the CJ state
\begin{equation}
A_{\text{CJ}}b_{\text{CJ}}-a_{\text{CJ}}B_{\text{CJ}}=0, \label{SUPPL_CJ}
\end{equation}
i.e., recalling the notations (\ref{NONDIM_VAR_z}), (\ref{Coef_dv_D_Cond}) and (\ref{Coef_dv_s_Cond}),
\begin{equation}
G_0z_{\text{CJ}}^{2}+2z_{\text{CJ}}-\left( 1-M_{0\text{CJ}}^{-2}\right) =0.
\label{SUPPL_CJ_expl}
\end{equation}

This proof by deduction shows the necessity and the uniqueness of the CJ condition for the theorem to hold. Mathematics distinguishes between deduction and induction. So if $dD_{\text{CJ}}=0$ and $ds_{\text{CJ}}=0$ for the same $dp_0\neq 0$ and the same $dv_0\neq 0$, then the determinant of the system (\ref{ds_CJ_Cond})-(\ref{dD_CJ_Cond}) should be zero, which is indeed what the constraint (\ref{SUPPL_CJ}) expresses. But nothing shows that this can happen, or that the theorem holds only for CJ states. That is illustrated below using the notion of envelope to express the tangency of sets of Hugoniot curves and Rayleigh lines with the same poles (Fig. \ref{fig:Fig1}).
\newline

The Hugoniot curves (H) $:p_{\text{H}}\left( v;p_0,v_0\right)$ (\ref{HUGO}) forms the family $y_{\text{H}}^{\ast }\left( p,v;v_0\right)=0$ with one parameter, $v_0$, if their poles $\left( p_0,v_0\right) $ are distributed on $p_0^{\ast }\left( v_0\right)$,
\begin{align}
& y_{\text{H}}^{\ast }\left( p,v;v_0\right) =\;...  \nonumber \\
& ...\;-h\left(p,v\right) +h_0\left(p_0^{\ast },v_0\right) +\frac{1}{2}\left(p-p_0^{\ast }\right) \left( v_0+v\right).  \label{HUGO_FAMILY}
\end{align}
This family has an envelope if
\begin{equation}
\!\!\left.\frac{\partial y_{\text{H}}^{\ast }}{\partial v_0}\right)_{\!\!p,v}\!\!=0\Leftrightarrow-\left(\frac{v_0}{D}\right)^{2}\!\frac{dp_0^{\ast}}{dv_0}=\frac{z+\frac{2M_0^{-2}}{G_0}}{z+\frac{2}{G_0}}\!\equiv \frac{A+a}{B+b}.  \label{P0_HUGO_ENV}
\end{equation}
The differential of the CJ entropy (\ref{ds_CJ_Cond}) shows that this envelope is an isentrope made up of sonic points. 
Substituting (\ref{P0_HUGO_ENV}) for the derivative of $p_0^{\ast }\left( v_0\right)$ in (\ref{dD_CJ_Cond}) then gives the derivative of $D_{\text{CJ}}$
\begin{equation}
\frac{v_0}{D^{\ast}_{\text{CJ}}}\left.\frac{\partial D^{\ast}_{\text{CJ}}}{\partial v_0}\right)_{s^{\ast}_{\text{CJ}}}=z^{-2}\frac{A_{\text{CJ}}b_{\text{CJ}}-a_{\text{CJ}}B_{\text{CJ}}}{B_{\text{CJ}}+b_{\text{CJ}}}. \label{CJ_Deriv_DCJ}
\end{equation}

Similarly, the Rayleigh-Michelson lines~(R)~$:p_{\text{R}}\left(v,D;p_0,v_0\right)$ (\ref{RM}) 
of which the poles $\left(p_0,v_0\right) $ are distributed on $p_0^{\ast}\left( v_0\right)$ form the family $y_{\text{R}}^{\ast}\left(p,v;v_0\right)=0$ with one parameter, $v_0$, if $D$ is set to $D_{\text{CJ}}^{\ast}(v_0)$ $=D_{\text{CJ}}(v_0,p_0=p_0^{\ast}(v_0))$,
\begin{equation}
y_{\text{R}}^{\ast}\left(p,v;v_0\right) =-p+p_0^{\ast }+\left(\frac{D_{\text{CJ}}^{\ast}}{v_0}\right)^{2}\left(v_0-v\right).  \label{RM_FAMILY}
\end{equation}
This family has an envelope if
\begin{equation}
\!\!\left.\frac{\partial y_{\text{R}}^{\ast}}{\partial v_0}\right)_{\!\!p,v}\!\!=0\Leftrightarrow-\!\left(\frac{v_0}{D_{\text{CJ}}^{\ast}}\right)^{2}\!\frac{dp_0^{\ast}}{dv_0}=1\!-\!2z\!+\!2z\frac{v_0}{D_{\text{CJ}}^{\ast}}\frac{dD_{\text{CJ}}^{\ast}}{dv_0}. \label{P0_RM_ENV_GEN}
\end{equation}
The (R) differential (\ref{R_DIFF}) shows that this condition is satisfied regardless of the variations of $D_{\text{CJ}}^{\ast}$ if this envelope is an isentrope because $-\left.\partial p/\partial v\right)_{s}=(c/v)^2=(D_\text{CJ}/v_0)^2$ (\ref{GAMMA}), (\ref{RH_Mass}). Substituting (\ref{P0_HUGO_ENV}) for $dp_0^{\ast}/dv_0$, or (\ref{CJ_Deriv_DCJ}) for $dD_{\text{CJ}}^{\ast}/dv_0$ in (\ref{P0_RM_ENV_GEN}), results in the same identity, so the same isentrope is an envelope of the (H) and (R) families (\ref{HUGO_FAMILY}) and (\ref{RM_FAMILY}).

Information about the variation of $D_\text{CJ}$ seems not to be deducible from the CJ differentials (\ref{ds_CJ_Cond}) and (\ref{dD_CJ_Cond}) alone but only by taking the CJ limit ($M=1$) of the differential of $v(D,p_0,v_0)$ (\ref{dv_D_Cond}).

\subsection{\label{subsec:CJsuppl}Additional Chapman-Jouguet properties}
\hypertarget{addCJstate}{\noindent{\small{\textbf{\textit {1. CJ state}}}}.}
Recalling the definition (\ref{NONDIM_VAR_z}), the physical solution $z_{\text{CJ}}>0$ to the additional CJ constraint (\ref{SUPPL_CJ_expl}),
\begin{equation}
z_{\text{CJ}}=\frac{\sqrt{1+G_0\left( 1-M_{0\text{CJ}}^{-2}\right) }-1}{G_0}, \label{zCJ_1_PARAM}    
\end{equation} gives the one-variable $(D_{\text{CJ}})$ representation (\ref{CJ_STATE_1_PARAM}) of the CJ state \hyperref[sec:Remind]{(Sect. \ref{sec:Remind})}. For example, from (\ref{VCJ_PARAM}) and (\ref{PCJ_PARAM}),
\begin{align}
&v_{\text{CJ}}=v_0\frac{G_0+1-\sqrt{1+G_0\left( 1-M_{0\text{CJ}}^{-2}\right) }}{G_0}, \label{VCJ_1_PARAM} \\
&p_{\text{CJ}}=p_0+\frac{D_{\text{CJ}}^{2}}{v_0}\frac{\sqrt{1+G_0\left( 1-M_{0\text{CJ}}^{-2}\right) }-1}{G_0}. \label{PCJ_1_PARAM} \\
&\gamma _{\text{CJ}}=\frac{G_0+1-\sqrt{1+G_0\left(1-M_{0\text{CJ}}^{-2}\right) }}{\frac{p_0v_0}{c_{0}^2}G_{0}M_{0\text{CJ}}^{-2}-1+\sqrt{1+G_0\left(1-M_{0\text{CJ}}^{-2}\right)}},   \label{GAMCJ_1_PARAM}
\end{align}
Reciprocally, $D_{\text{CJ}}$ has a one-variable representation. For example, using the dimensionless pressure jump $\pi_{\text{CJ}}=v_0\left( p_{\text{CJ}}-p_0\right)/c_0^{2}$ or the adiabatic exponent $\gamma _{\text{CJ}}$,
\begin{align}
\left(\frac{D_{\text{CJ}}}{c_0}\right)^{2}\!\!\!&=\pi_{\text{CJ}}\left(1+\frac{1}{2\pi_{\text{CJ}}}\right)\left(1+\sqrt{1+\frac{G_0}{\left(1+\frac{1}{2\pi_{\text{CJ}}}\right)^{2}}}\right),  \label{DCJ_1_PARAM_PCJ}
\end{align}
\begin{align}
\left(\frac{D_{\text{CJ}}}{c_0}\right) ^{2}\!\!\!&=\frac{1}{2}\frac{\left(\gamma _{\text{CJ}}+1\right) ^{2}}{\gamma_{\text{CJ}}^{2}-1-G_0}\times
\left\{ 1-2\frac{1+\frac{G_0}{\gamma _{\text{CJ}}+1}}{\gamma _{\text{CJ}}+1}\frac{\gamma _{\text{CJ}}}{\widetilde{\gamma }_0}\right. \;...  \nonumber \\
...&\;+\left. \sqrt{1-4\frac{1+\frac{G_0-\left(
1+G_0\right) \frac{\gamma _{\text{CJ}}}{\widetilde{\gamma }_0}}{\gamma _{\text{CJ}}+1}}{\gamma _{\text{CJ}}+1}\frac{\gamma _{\text{CJ}}}{\widetilde{\gamma }_0}}\right\} ,  \label{DCJ_1_PARAM_GAMCJ}
\end{align}
where $\widetilde{\gamma}_0$ denotes the ratio $c_0^{2}/p_0v_0$, which differs from $\gamma _0$, except for gases \hyperref[sec:Remind]{(Sect. \ref{sec:Remind})}.
The identity
\begin{equation}
G_0=\frac{\alpha _0c_0^{2}}{C_{p0}},\;\alpha _0=\frac{1}{v_0}\left. \frac{\partial v_0}{\partial T_0}\right) _{p_0}, \label{ID_THERM_P_T}
\end{equation}
indicates that the necessary initial data are $c_0$, $C_{p0}$, and $v_0$ measured as a function of $T_0$ at constant $p_0$ so the coefficient of thermal expansion $\alpha _0$ can be determined.
The relation (\ref{DCJ_1_PARAM_GAMCJ}) shows a large sensitivity of $D_{\text{CJ}}$ to $\gamma _{\text{CJ}}$, as is more evident in the gas example below (\ref{DCJ_1_PARAM_IDEAL}).

For an ideal gas, with $W$ the molecular weight, $c$, $C_{p}$, $\alpha $ and $\gamma $ are functions of $T=pv\left(W/R\right)$ only, $G=$ $\gamma-1$ and $\alpha =1/T$. So, for an initially ideal gas,
\begin{align}
\gamma_{\text{CJ}}&=\sqrt{\frac{\gamma_0}{1-\frac{\gamma_0-1}{\gamma _0}M_{0\text{CJ}}^{-2}}},\ %
\left(\frac{D_{\text{CJ}}}{c_0}\right)^{2}=\frac{1-\gamma _0^{-1}}{1-\frac{\gamma_0}{\gamma_{\text{CJ}}^{2}}},  \label{DCJ_1_PARAM_IDEAL} \\
\frac{D_{\text{CJ}}^{2}}{v_0p_{\text{CJ}}}&=\left(1-\left(1-\frac{\gamma_0}{2}\right) \frac{p_0}{p_{\text{CJ}}}\right)\times \; ...  \nonumber \\
&...\;\left\{1+\sqrt{1+\frac{\left( \gamma_0-1\right)
\left( 1-\frac{p_0}{p_{\text{CJ}}}\right) ^{2}}{\left( 1-\left( 1-\frac{\gamma _0}{2}\right) \frac{p_0}{p_{\text{CJ}}}\right)^{2}}}\right\}.
\label{DCJ_1_PARAM_PCJ_IDEAL}
\end{align}

The strong-shock limits ($M_{0\text{CJ}}^{-2}\ll 1$ or $p_0/p_{\text{CJ}}\ll 1$) of $\gamma _{\text{CJ}}$ and $D_{\text{CJ}}^{2}$ are $\sqrt{\gamma_0}$ and $\left( 1+\sqrt{\gamma _0}\right) v_0p_{\text{CJ}}$, respectively (their acoustic limits are $\gamma _0$ and $c_0^{2}$). The typical values $\gamma _0=1.3$, $c_0=330$ m/s and $D_{\text{CJ}}=2000$
m/s give $\gamma _{\text{CJ}}=1.144$, $\sqrt{\gamma _0}=1.140$, with the relative error $100\times \left( \gamma _{\text{CJ}}/\sqrt{\gamma _0}-1\right) =0.316$ \%. The relations (\ref{DCJ_1_PARAM_IDEAL}) and (\ref{DCJ_1_PARAM_PCJ_IDEAL}) apply only to initially ideal gases, but products can be non-ideal if $p_0$ is large enough.
\newline

\hypertarget{addCJisent}{\noindent{\small{\textbf{\textit {2. CJ isentrope}}}}.}
The initial-state polar $p_0^{\ast}\left( v_0\right)$ producing constant $D_{\text{CJ}}$ and $s_{\text{CJ}}$ is solution to the ordinary differential equation formed by substituting $z_{\text{CJ}}$ (\ref{zCJ_1_PARAM}) for $z$ in any of the relations (\ref{VAR_INIT_CJ}), (\ref{P0_RM_ENV_GEN}) or (\ref{P0_HUGO_ENV}). The initial condition is the reference
initial state $\left( p_{0\ast },v_{0\ast }\right) $ with known CJ velocity $D_{\text{CJ}}^{\ast }$. Then, substituting $p_0^{\ast}\left( v_0\right)$ for $p_0$ in (\ref{VCJ_1_PARAM}) and (\ref{RM}) gives
\begin{align}
v_{\text{CJ}}^{\ast }\left( v_0\right) &=v_{\text{CJ}}\left(
v_0,p_0^{\ast }\left( v_0\right) ,D_{\text{CJ}}^{\ast }\right) ,
\label{VCJ_IS} \\
p_{\text{CJ}}^{\ast }\left( v_0\right) &=p_0^{\ast }\left( v_0\right)+\frac{D_{\text{CJ}}^{\ast 2}}{v_0}\left(1-\frac{v_{\text{CJ}}^{\ast
}\left( v_0\right) }{v_0}\right).  \label{PCJ_IS}
\end{align}
The isentrope $p_{\text{S}}^{\ast }\left( v\right) $ is then generated by
eliminating $v_0$ between $v_{\text{CJ}}^{\ast }\left( v_0\right) $ and $%
p_{\text{CJ}}^{\ast }\left( v_0\right) $, which amounts to varying $v_0$ and
representing $p_{\text{CJ}}^{\ast }\left( v_0\right) $ as a function of $%
v_{\text{CJ}}^{\ast }\left( v_0\right)$.

\newpage

\hypertarget{addCJdet}{\noindent{\small{\textbf{\textit {3. CJ determinacy}}}}.}
The theorem holds if the isentropes have finite slopes so the derivatives $\left. \partial v^{\ast}/\partial v_0\right)_{s^{\ast}}$ are finite and non-zero at sonic points \hyperref[subsec:Proof]{(Subsect. \ref{subsec:Proof})}. The condition is obtained by differentiating $c\left( s,v\right)$ (\ref{SOUND}) and the mass balance (\ref{RH_Mass}-a) written as $v=v_0M\left( c/D\right)$,%
\begin{align}
\frac{dv}{v} &=\frac{dv_0}{v_0}+\frac{dc}{c}+\frac{dM}{M}-\frac{dD}{D},
\label{DIFF_MASS_MACH} \\
dc &=\left. \frac{\partial c}{\partial s}\right) _{v}ds+\left.\frac{%
\partial c}{\partial v}\right) _{s}dv,  \label{DIFF_SOUND_SV}
\end{align}%
hence, along an isentrope, and with initial states on an arbitrary polar $p_0^{\ast}\left(v_0\right)$ \hyperref[subsec:Proof]{(Subsect. \ref{subsec:Proof})},
\begin{equation}
\Gamma \frac{v_0}{v}\left.\frac{\partial v^{\ast}}{\partial v_0}\right)_{s^{\ast}}=1-%
\frac{v_0}{D}\left.\frac{\partial D}{\partial v_0}\right)_{s^{\ast}}+\frac{%
v_0}{M}\left.\frac{\partial M^{\ast}}{\partial v_0}\right)_{s^{\ast}},
\end{equation}%
with $\Gamma$ the fundamental derivative of hydrodynamics (\ref{FUND_HYDRO}). If $M=$ const. $=1$, since $\left.\partial D/\partial v_0\right)_{s^{\ast}}^{\text{CJ}}=0$ (\ref{DSI}), the definitions (\ref{H_SP}) and (\ref{GAMMA}) give
\begin{align}
&\!\!\left.\frac{\partial v^{\ast}}{\partial v_0}\right)_{s^{\ast}}^{\text{CJ}}\!\!\!\!\!=-\left(\frac{v_0}{D_{\text{CJ}}}\right)^{2}\!\!\!\left.\frac{\partial p^{\ast}}{\partial v_0}\right)_{s^{\ast}}^{\text{CJ}}\!\!\!\!\!=\Gamma _{\text{CJ}}^{-1}\frac{v_{\text{CJ}}}{v_0},  \label{deriv_vp_son} \\
\frac{v_0}{D_{\text{CJ}}^{2}}&\!\left.\frac{\partial h^{\ast}}{\partial v_0}\right)_{s^{\ast}}^{\text{CJ} }\!\!\!\!\!=-\Gamma _{\text{CJ}}^{-1}\left(\frac{v_{\text{CJ}}}{v_0}\right)^{2}, \label{deriv_h_son} \\
\frac{v_0}{D_{\text{CJ}}}&\!\left.\frac{\partial u^{\ast}}{\partial v_0}\right)_{s^{\ast}}^{{\text{CJ}}}\!\!\!\!\!=\left( 1-\Gamma _{\text{CJ}}^{-1}\right)\frac{v_{\text{CJ}}}{v_0}.  \label{deriv_u_son}
\end{align}
The derivatives of $v$, $p$ and $h$ at $s$ fixed are thus finite and non-zero at CJ points except if $\Gamma _{\text{CJ}}=0$ or $\Gamma _{\text{CJ}}\rightarrow \infty$, respectively. The derivative of $u$ is zero for $\Gamma _{\text{CJ}}=1$.
For the perfect gas \hyperref[sec:CJperf]{(App. \ref{sec:CJperf})}, the partial derivative of $v\left( D;v_0,p_0\right)$ (\ref{PERF_PV_OVRDRV}) with respect to
$v_0$ at $D$ fixed, with $p_0=p_0^{\ast}\left(v_0\right)$, shows that lim$_{D\rightarrow D_{\text{CJ}}}\left.\partial v^{\ast}/\partial v_0\right)_{D}$ is bounded if $\left.\partial D/\partial v_0\right)_{s^{\ast}}=0$. An expression for $\Gamma _{\text{CJ}}=\left.\partial \ln v^{\ast}/\partial v_0\right)_{s^{\ast}}^{\text{CJ}}$ (\ref{deriv_vp_son}) that involves the partial derivatives of $\gamma_0$ and $G_0$ can be obtained by differentiating the constraint (\ref{SUPPL_CJ_expl}) with respect to $v_0$ at $D_{\text{CJ}}$ fixed.
\newline

\hypertarget{addConstr}{\noindent{\small{\textbf{\textit {4. Constraints on the derivatives of} \textit {\!D}$_{\textbf{CJ}}$}}}.}
The partial derivatives of $D_{\text{CJ}}\left(v_0,p_0\right)$ or $D_{\text{CJ}}\left(T_0,p_0\right)$ are not independent, since there are initial-state variations for which $D_{\text{CJ}}$  (or $s_{\text{CJ}}$) is constant, i.e. (\ref{VAR_INIT_CJ}) or (\ref{P0_RM_ENV_GEN}). With $y_0$ denoting either $v_0$ or $T_0$, the triple product rule\begin{equation}
\left. \frac{\partial D_{\text{CJ}}}{\partial y_0}\right) _{p_0}=-\left.\frac{\partial p_0}{\partial y_0}\right)_{D_{\text{CJ}}}\left.\frac{\partial D_{\text{CJ}}}{\partial p_0}\right)_{y_0}
\label{TWO_VAR_ID}
\end{equation}
gives
\begin{align}
\frac{v_0}{D_{\text{CJ}}}&\left.\frac{\partial D_{\text{CJ}}}{\partial
v_0}\right)_{p_0}=\frac{D_{\text{CJ}}}{v_0}\left. \frac{\partial D_{%
\text{CJ}}}{\partial p_0}\right) _{v_0}\times \left( 1-2z_{\text{CJ}%
}\right),  \label{PROP_DERIVS_DCJ_v0} \\
\frac{D_{\text{CJ}}}{v_0}&\left.\frac{\partial D_{\text{CJ}}}{\partial
p_0}\right)_{T_0}=\frac{T_0}{D_{\text{CJ}}}\left. \frac{\partial D_{%
\text{CJ}}}{\partial T_0}\right) _{p_0}\times \;...  \nonumber \\
&...\;\frac{1-\left( 1+\alpha _0T_0G_0\right) \left( 1-2z_{%
\text{CJ}}\right) M_{0\text{CJ}}^{2}}{\left( 1-2z_{\text{CJ}}\right) \alpha
_0T_0}.  \label{PROP_DERIVS_DCJ_p0}
\end{align}%
The latter is obtained from the former and the identities%
\begin{align}
& \frac{T_0}{D_{\text{CJ}}}\left. \frac{\partial D_{\text{CJ}}}{\partial
T_0}\right) _{p_0}=\alpha _0T_0\frac{v_0}{D_{\text{CJ}}}\left. 
\frac{\partial D_{\text{CJ}}}{\partial v_0}\right) _{p_0},
\label{Transf_v0_p0_T0p0} \\
& \frac{D_{\text{CJ}}}{v_0}\left. \frac{\partial D_{\text{CJ}}}{\partial
p_0}\right) _{T_0}=\frac{D_{\text{CJ}}}{v_0}\left. \frac{\partial D_{%
\text{CJ}}}{\partial p_0}\right) _{v_0}\;...  \nonumber \\
& \;\;\;\;\;...\;-M_0^{2}\left( 1+\alpha _0T_0G_0\right) \frac{v_0}{%
D_{\text{CJ}}}\left. \frac{\partial D_{\text{CJ}}}{\partial v_0}\right)
_{p_0},  \label{Transf_v0_p0_p0T0}
\end{align}
so a partial derivative of $D_{\text{CJ}}\left(v_0,p_0\right)$ or $D_{\text{CJ}}\left(T_0,p_0\right)$ determines the other.
Similar constraints exist for $s_{\text{CJ}}$ since $\left. \partial p_0/\partial y_0\right) _{D_{\text{CJ}}}=\left. \partial p_0/\partial y_0\right)_{s_{\text{CJ}}}$ \hyperref[subsec:Proof]{(Subsect. \ref{subsec:Proof})}.
\newline

\hypertarget{addIM}{\noindent{\small{\textbf{\textit {5. The Inverse Method (IM)}}}}.}
This method \cite{{Jones1949},{Stanyuk1955},{Manson1958a}} gives the CJ hydrodynamic variables from experimental values of $D_{\text{CJ}}$ and its partial derivatives with respect to a pair of independent initial-state variables, as discussed by Manson \cite{Manson1958a} and Wood and Fickett \cite{WoodFickett1963}. The IM is not an additional CJ property, but it is useful here to analyze the application of the theorem to liquid explosives \hyperref[sec:Applic-Liq]{(Sect. \ref{sec:Applic-Liq})} because it shares the same physical assumptions as the theorem \hyperref[sec:Remind]{(Sect. \ref{sec:Remind})}. 

The first pair used in this work is $\left(T_0,p_0\right)$. Eliminating $F_{\text{CJ}}$ (\ref{F}) between the partial derivatives $A_{\text{CJ}}$ (\ref{Coef_dv_s_Cond}) and $B_{\text{CJ}}$ (\ref{dD_CJ_Cond}) of $D_{\text{CJ}}\left(v_0,p_0\right)$ gives the CJ state (\ref{NONDIM_VAR_z}) as the solution $z_{\text{CJ}}<1$ of
\begin{equation}
G_0L_{1}z_{\text{CJ}}^{2}+2K_{1}z_{\text{CJ}}-\left( 1-M_{0\text{CJ}}^{-2}\right)
=0,  \label{IVM_VCJ}
\end{equation}
where $L_{1}$ and $K_{1}$ for $D_{\text{CJ}}\left(v_0,p_0\right)$ and $D_{\text{CJ}}\left( T_0,p_0\right)$ are 
\begin{align}
L_{1}&=1+\frac{D_{\text{CJ}}}{v_0}\left. \frac{\partial D_{\text{CJ}}}{\partial p_0}\right) _{v_0}-\frac{v_0}{D_{\text{CJ}}}\left. \frac{\partial D_{\text{CJ}}}{\partial v_0}\right)_{p_0}  \nonumber \\
&=\ 1+\frac{D_{\text{CJ}}}{v_0}\left. \frac{\partial D_{\text{CJ}}}{\partial p_0}\right) _{T_0}\;...  \nonumber \\
&...\;+\frac{1-M_{0\text{CJ}}^{-2}+\alpha _0T_0G_0}{\alpha
_0T_0M_{0\text{CJ}}^{-2}}\frac{T_0}{D_{\text{CJ}}}\left. \frac{\partial D_{\text{CJ}}}{\partial T_0}\right) _{p_0},  \label{IVM_L}
\\
K_{1}&=1+M_{0\text{CJ}}^{-2}\frac{D_{\text{CJ}}}{v_0}\left. \frac{\partial
D_{\text{CJ}}}{\partial p_0}\right) _{v_0}\!\!\!\!-\frac{v_0}{D_{\text{CJ}}}\left. \frac{\partial D_{\text{CJ}}}{\partial v_0}\right) _{p_0}  \nonumber
\\
&=1+M_{0\text{CJ}}^{-2}\frac{D_{\text{CJ}}}{v_0}\left. \frac{\partial
D_{\text{CJ}}}{\partial p_0}\right) _{T_0}\!\!\!\!+\frac{G_0T_0}{D_{\text{CJ}}}\left. \frac{\partial D_{\text{CJ}}}{\partial T_0}\right) _{p_0},
\label{IVM_K}
\end{align}
using (\ref{H_PV}), (\ref{H_TP}), and (\ref{ID_THERM_P_T}). The IM relation (\ref{IVM_VCJ}) reduces to the additional CJ relation (\ref{SUPPL_CJ_expl}) by requiring that the partial derivatives of $D_{\text{CJ}}$ satisfy their compatibility relation (\ref{PROP_DERIVS_DCJ_v0}). Any assumption on the derivatives of $D_{\text{CJ}}$ such that $L_{1}=K_{1}=1$ also reduces (\ref{IVM_VCJ}) to (\ref{SUPPL_CJ_expl}). Manson \cite{Manson1958b} had noted the strong-shock limit $\sqrt{\gamma _0}$ of $\gamma_{\text{CJ}}$ (\ref{DCJ_1_PARAM_IDEAL}) for the ideal gas by neglecting $\left.\partial \ln D_{\text{CJ} }/\partial \ln p_0\right)_{T_0}$ and $\left.\partial \ln D_{\text{CJ}}/\partial \ln T_0\right)_{p_0}$. Actually, such assumptions represent the acoustic solution $M_{0\text{CJ}}=1$ (i.e. $z_{\text{CJ}}=0$) of (\ref{SUPPL_CJ_expl}), which contradicts the distinguished limit required by the large physical values of $M_{0\text{CJ}}^{2}$: the determinants of the $2\times 2$ linear systems (\ref{IVM_L})-(\ref{IVM_K}) for the partial derivatives of $D_{\text{CJ}}\left( v_0,p_0\right)$ and $D_{\text{CJ}}\left(T_0,p_0\right)$, with $L_{1}-1$ and $K_{1}-1$ as non-homogeneous terms, are proportional to $M_{0\text{CJ}}^{-2}-1$, which should be zero if $L_{1}=K_{1}=1$, regardless of the values of the derivatives.

The second pair is $\left(T_0,w_0\right)$, with $p_0$ constant, for a group of isometric mixtures \cite{Wecken1959}. These have the same atomic composition, so the same chemical-equilibrium composition, for any value of a molecular composition parameter, $w_0$, defining all the mixtures in the group \cite{WoodFickett1963}.
Typically, $w_0$ is the fraction of all compounds added to a reference explosive, whose properties are therefore defined by $w_0=0$. The initial equations of state extend to $v_0\left(T_0,p_0,w_0\right)$ and $h_0\left(T_0,p_0,w_0\right)$ whose derivatives at $p_0$ fixed are obtained from calculated or measured functions $v_0\left(T_0,w_0\right)$ and $h_0\left( T_0,w_0\right)$. Setting $dp_0=0$ in (\ref{IM_dD_Gen}) first gives the differential of $D_{\text{CJ}}\left( v_0,h_0\right)$, and eliminating $F_{\text{CJ}}$ (\ref{F}) between its partial derivatives then gives the CJ state as the solution $z_{\text{CJ}}<1$ of
\begin{equation}
L_{2}z_{\text{CJ}}^{2}+2K_{2}z_{\text{CJ}}-1=0,  \label{IVM_VCJ_w0}
\end{equation}%
where $L_{2}$ and $K_{2}$ for $D_{\text{CJ}}\left(v_0,h_0\right)$ and $D_{\text{CJ}}\left( T_0,w_0\right) $ are
\begin{align}
L_{2}&=\ D_{\text{CJ}}\left. \frac{\partial D_{\text{CJ}}}{\partial h_0}\right) _{v_0,p_0}  \nonumber \\
&=\;\frac{\frac{\omega_0}{v_0}\frac{T_0}{D_{\text{CJ}}}\!\left.\frac{\partial D_{\text{CJ}}}{\partial T_0}\right) _{w_0,p_0}\!\!\!-\frac{\alpha_0T_0}{D_{\text{CJ}}}\!\left. \frac{\partial D_{\text{CJ}}}{\partial w_0}\right)_{T_0,p_0}}{\frac{\omega_0}{v_0}\frac{C_{p0}T_0}{D_{\text{CJ}}^{2}}-\frac{\mu_0}{D_{\text{CJ}}^{2}}\alpha _0T_0},  \label{L_w0_T0}
\\
K_{2}&=\ 1-\frac{v_0}{D_{\text{CJ}}}\left. \frac{\partial D_{\text{CJ}}}{\partial v_0}\right) _{h_0,p_0} \nonumber \\
&=1+\frac{\frac{\mu_0}{D_{\text{CJ}}^{2}}\frac{T_0}{D_{\text{CJ}}}\!\left. \frac{\partial D_{\text{CJ}}}{\partial T_0}\right)_{w_0,p_0}\!\!\!-\frac{%
C_{p0}T_0}{D_{\text{CJ}}^{3}}\!\left. \frac{\partial D_{\text{CJ}}}{\partial w_0}\right)_{T_0,p_0}}{\frac{\omega_0}{v_0}\frac{C_{p0}T_0}{D_{\text{CJ}}^{2}}-\frac{\mu_0}{D_{\text{CJ}}^{2}}\alpha_0T_0},  \label{K_w0_T0}
\end{align}
using the identities,
\begin{align}
dv_0&=v_0\alpha_0dT_0+\omega_0dw_0, &\omega_0=\left. \frac{\partial v_0}{\partial w_0}\right)_{T_0,p_0},  \label{dv0_w0_T0} \\
dh_0&=C_{p0}dT_0+\mu_0dw_0, &\mu_0=\left. \frac{\partial h_0}{\partial w_0}\right)_{T_0,p_0}.  \label{dh0_w0_T0}
\end{align}
The pair ($T_0$, $w_0$) is more suitable for liquid explosives because generating sufficiently large and accurate variations of $p_0$ is difficult and because it does not require $c_0$. The pair ($p_0$, $w_0$) can be convenient for gases at high initial pressures \cite{Bauer1988}.

Davis \cite{Davis1981} implemented the IM for condensed explosives ($p_0/p\simeq 10^{-5}$) with the specific internal energy $e_0$ and the density $\rho_0=1/v_0$ as the pair of independent initial-state variables, and $p_0$ constant and negligible. He built $D_{\text{CJ}}\left(e_0,\rho_0\right)$ from Kamlet's method
, and calculated the initial-state polar $e_0\left(\rho_0\right)$ of Hugoniots with the same isentropic envelope, this isentrope and the CJ state. His relations (14) and (31) are equivalent to (\ref{H0_RM_ENV}) and (\ref{H0_HUGO_ENV}) since $e_0=h_0$ if $p_0$ is neglected. Nagayama and Kubota \cite{NagayamaKubota2004} derived an envelope constraint for the R lines from linear laws $D_{\text{CJ}}\left(\rho_0\right)$ with negligible dependence on $e_0$. Their relations (13) and (14) are equivalent to (\ref{H0_RM_ENV}), i.e. $z_{\text{CJ}}=1/2K$ from (\ref{IVM_VCJ_w0}) and (\ref{L_w0_T0}).
\newpage

\hypertarget{addRem}{\noindent{\small{\textbf{\textit {6. Remarks}}}}.}
The envelope conditions \hyperref[subsec:Proof]{(Subsect. \ref{subsec:Proof})} for families of Hugoniot curves (\ref{HUGO}) and Rayleigh-Michelson lines (\ref{RM}), with $p_0$ constant and $v_0$ and $h_0$ the independent variables, are constraints on the variations of $D$ and $h_0$ which, from (\ref{HUGO_FAMILY}), (\ref{P0_RM_ENV_GEN}), and (\ref{NONDIM_VAR_z}), write
\begin{align}
\frac{v_0}{D^{2}}\frac{dh_0}{dv_0}&=-\frac{z}{2}=-\frac{1}{2}\frac{v_0\left(p-p_0\right)}{D^{2}}, \label{H0_HUGO_ENV} \\
\frac{v_0}{D}\frac{dD}{dv_0}&=\;1-\frac{1}{2z}=1-\left( 2\frac{v_0\left(p-p_0\right) }{D^{2}}\right)^{-1}.  \label{H0_RM_ENV}
\end{align}
Substitution in (\ref{ds}) shows the envelope is an isentrope, so $ds_{\text{CJ}}=0$. However, (\ref{H0_RM_ENV}) and (\ref{VCJ_PARAM}) then show that $dD_{\text{CJ}}\ne0$ is necessary to ensure that $v_{\text{CJ}}/v_0>1/2$ and $\gamma _{\text{CJ}}>1$. Therefore, the invariance theorem (\ref{DSI_bis}) is valid only if $p_0$ is varied, even if $p_0/p$ or $p_0v_0/D^{2}$ are negligible.
\section{\label{sec:Applic-Gas}Application to gaseous explosives}
For gaseous explosives, the invariance theorem and the additional CJ properties (Subsects. \hyperref[subsec:Proof]{\ref{subsec:Proof}} and \hyperref[subsec:CJsuppl]{\ref{subsec:CJsuppl}}, resp.) were investigated using the NASA's CEA computer program for ideal detonation products \cite{GordonMcBride-I}. That avoids uncertain interpretations that might arise with more complex equations of state. 

The theorem was discussed based on $I=5$ initial pressures $p_0$ determined by dichotomy for given initial temperatures $T_0$ to produce the same CJ entropy $s_{\text{CJ}}$. The joint invariance of the CJ velocity $D_{\text{CJ}}$ was analyzed using the mean values $\bar{D}_{\text{CJ}}$, the relative deviations $\Delta D_{\text{CJ}}/\bar{D}_{\text{CJ}}$ in percent and their absolute means $m_{D_{\text{CJ}}}$, and the corrected standard deviations $\sigma _{D_{\text{CJ}}}$
\begin{equation}
\bar{D}_{\text{CJ}} =\frac{1}{I}\sum_{i=1}^{I=5}D_{\text{CJ}i}\;,
\ \left( \frac{\Delta D_{\text{CJ}}}{\bar{D}_{\text{CJ}}}\right) _{i}
=100\times \frac{D_{\text{CJ}i}-\bar{D}_{\text{CJ}}}{\bar{D}_{\text{CJ}}},
\end{equation}
\begin{equation}
\\m_{D_{\text{CJ}}} =\frac{1}{I}\sum_{i=1}^{I=5}\left\vert \frac{\Delta D_{%
\text{CJ}}}{\bar{D}_{\text{CJ}}}\right\vert _{i}\;,
\ \sigma _{D_{\text{CJ}}} =\sqrt{\sum_{i=1}^{I=5}\frac{\left( D_{\text{CJ}i}-%
\bar{D}_{\text{CJ}}\right) ^{2}}{I-1}}.
\end{equation}
\hyperref[tab:Tab1]{Tables \ref{tab:Tab1}} show the numerical values of $s_{\text{CJ}}$ and $D_{\text{CJ}}$ for the stoichiometric mixtures \ce{CH4 + 2 O2}, \ce{C3H8 + 5 O2}, \ce{CH4 + 2} Air and \ce{H2 + 0.5} Air, and the 5 ($T_0,p_0 $) pairs. The third pair ($T_0=298.15$ K, $p_0=1$ bar) was chosen as the reference initial state ($v_{0\ast},p_{0\ast}$) (\hyperref[subsec:Proof]{Subsect. \ref{subsec:Proof}}, superscript $\ast$). For each mixture, $m_{D_{\text{CJ}}}$ and $\sigma_{D_{\text{CJ}}}$ are very small, and the values of $D_{\text{CJ}}$ are close to the mean $\bar{D}_{\text{CJ}}$ to $\mathcal{O}\left(0.1\right)$ \% at most. The slightly decreasing trend of $D_{\text{CJ}}$ with increasing $T_0$ could be due to uncertainties in the thermodynamic coefficients, as discussed below.
\begin{table}[t!]
\caption{Joint invariances of the CJ entropy $s_{\text{CJ}}$ and velocity $D_{\text{CJ}}$: mean value $\bar{D}_{\text{CJ}}$, relative deviation $\Delta D_{\text{CJ}} /\bar{D}_{\text{CJ}}$, mean relative deviation $m_{D_{\text{CJ}}}$, corrected standard deviation $\sigma_{D_{\text{CJ}}}$.} 
\label{tab:Tab1}
\begin{ruledtabular}
\begin{tabular}{ccccc}
\multicolumn{2}{l}{\ce{CH4 + 2 O2}}&\multicolumn{2}{l}{$m_{D_{\text{CJ}}}=\,0.08\text{ \%}$}\\
\multicolumn{2}{l}{$\bar{D}_{\text{CJ}}=\,2389.7\text{ m/s}$}&\multicolumn{2}{l}{$\sigma _{D_{\text{CJ}}}\;=\,2.47%
\text{ m/s}$}\\\hline
$T_0$&$p_0$&$s_{\text{CJ}}$&$D_{\text{CJ}}$&$\frac{\Delta D_{\text{CJ}}}{\bar{D}_{\text{CJ}}}$ \\
(K)&(bar)&(kJ/kg/K)&(m/s)&(\%) \\ \hline
\multicolumn{1}{l}{$200.00$} & \multicolumn{1}{l}{\quad $0.6284$} & $id.$ & $2392.9$ & 
\multicolumn{1}{r}{$0.13$} \\ 
\multicolumn{1}{l}{$250.00$} & \multicolumn{1}{l}{\quad $0.8118$} & $id.$ & $2391.2$ & \multicolumn{1}{r}{$0.06$} \\ 
\multicolumn{1}{l}{$298.15^{\ast }$} & \multicolumn{1}{l}{\quad $1.0000^{\ast }$} & $12.6653^{\mathbf{\ast }}$ & $2389.6$ & \multicolumn{1}{r}{$\sim 0.00$}\\ 
\multicolumn{1}{l}{$350.00$} & \multicolumn{1}{l}{\quad $1.2165$} & 
$id.$ & $2388.0$ & 
\multicolumn{1}{r}{$-0.07\hspace{0.0cm}$} \\ 
\multicolumn{1}{l}{$400.00$} & \multicolumn{1}{l}{\quad $1.4410$} & 
$id.$ & $2386.7$ & 
\multicolumn{1}{r}{$-0.12$}%
\end{tabular}
\end{ruledtabular}
\end{table}
\begin{table}[t!]
\vspace{-2.5mm}
\begin{ruledtabular}
\begin{tabular}{ccccc}
\multicolumn{2}{l}{\ce{C3H8 + 5 O2}}&\multicolumn{2}{l}{$m_{D_{\text{CJ}}}=\,0.012\text{ \%}$} \\ 
\multicolumn{2}{l}{$\bar{D}_{\text{CJ}}=\,2356.7\text{ m/s}$}&\multicolumn{2}{l}{$\sigma _{D_{\text{CJ}}}\;=\,0.41%
\text{ m/s}$}\\\hline
$T_0$&$p_0$&$s_{\text{CJ}}$&$D_{\text{CJ}}$&$\frac{\Delta D_{\text{CJ}}}{\bar{D}_{\text{CJ}}}$ \\ 
(K)&(bar)&(kJ/kg/K)&(m/s)&(\%) \\ \hline
\multicolumn{1}{l}{$200.00$} & \multicolumn{1}{l}{\quad $0.6304$} & 
$id.$ & $2357.3$ & 
\multicolumn{1}{r}{$0.03$} \\ 
\multicolumn{1}{l}{$250.00$} & \multicolumn{1}{l}{\quad $0.8127$} & 
$id.$ & $2356.7$ & 
\multicolumn{1}{r}{$\sim 0.00$} \\ 
\multicolumn{1}{l}{$298.15^{\mathbf{\ast }}$} & \multicolumn{1}{l}{\quad $1.0000^{\mathbf{\ast }}$} & $11.9293^{\mathbf{\ast }}$
& $2356.3$ & \multicolumn{1}{r}{$-0.01_{5}$}
\\ 
\multicolumn{1}{l}{$350.00$} & \multicolumn{1}{l}{\quad $1.2165$} & 
$id.$ & $2356.3$ & 
\multicolumn{1}{r}{$-0.01_{5}$} \\ 
\multicolumn{1}{l}{$400.00$} & \multicolumn{1}{l}{\quad $1.4419$} & 
$id.$ & $2356.7$ & 
\multicolumn{1}{r}{$\sim 0.00$}%
\end{tabular}
\end{ruledtabular}
\end{table}
\begin{table}[t!]
\vspace{-2.5mm}
\begin{ruledtabular}
\begin{tabular}{ccccc}
\multicolumn{2}{l}{\ce{CH4 + 2} Air}&\multicolumn{2}{l}{$m_{D_{\text{CJ}}}=\,0.05\text{ \%}$} \\ 
\multicolumn{2}{l}{$\bar{D}_{\text{CJ}}=\,1799.9\text{ m/s}$}&\multicolumn{2}{l}{$\sigma _{D_{\text{CJ}}}\;=\,1.23%
\text{ m/s}$} \\ \hline
$T_0$&$p_0$&$s_{\text{CJ}}$&$D_{\text{CJ}}$&$\frac{\Delta D_{\text{CJ}}}{\bar{D}_{\text{CJ}}}$ \\ 
(K)&(bar)&(kJ/kg/K)&(m/s)&(\%) \\ \hline
\multicolumn{1}{l}{$200.00$} & \multicolumn{1}{l}{\quad $ 0.6044$} & 
$id.$ & $1801.4$ & 
\multicolumn{1}{r}{$0.08$} \\ 
\multicolumn{1}{l}{$250.00$} & \multicolumn{1}{l}{\quad $0.7968$} & 
$id.$ & $1800.7$ & 
\multicolumn{1}{r}{$0.05$} \\ 
\multicolumn{1}{l}{$298.15^{\mathbf{\ast }}$} & \multicolumn{1}{l}{\quad $1.0000^{\ast }$} & $9.4218^{\ast }$ & $%
1799.9$ & \multicolumn{1}{r}{$\sim 0.00$} \\ 
\multicolumn{1}{l}{$350.00$} & \multicolumn{1}{l}{\quad $1.2401$} & 
$id.$ & $1799.1$ & 
\multicolumn{1}{r}{$-0.04$} \\ 
\multicolumn{1}{l}{$400.00$} & \multicolumn{1}{l}{\quad $1.4949$} & 
$id.$ & $1798.3$ & 
\multicolumn{1}{r}{$-0.09$}%
\end{tabular}
\end{ruledtabular}
\end{table}
\begin{table}[t!]
\vspace{-2mm}
\begin{ruledtabular}
\begin{tabular}{ccccc}
\multicolumn{2}{l}{\ce{H2 + 0.5} Air}&\multicolumn{2}{l}{$m_{D_{\text{CJ}}}=\,0.1\text{ \%}$} \\ 
\multicolumn{2}{l}{$\bar{D}_{\text{CJ}}=\,1964.7\text{ m/s}$}&\multicolumn{2}{l}{$\sigma _{D_{\text{CJ}}}\;=\,2.55%
\text{ m/s}$} \\ \hline
$T_0$&$p_0$&$s_{\text{CJ}}$&$D_{\text{CJ}}$&$\frac{\Delta D_{\text{CJ}}}{\bar{D}_{\text{CJ}}}$ \\ 
(K)&(bar)&(kJ/kg/K)&(m/s)&(\%) \\ \hline
\multicolumn{1}{l}{$200.00$} & \multicolumn{1}{l}{\quad $0.6004$} & 
$id.$ & $1967.9$ & 
\multicolumn{1}{r}{$0.16$} \\ 
\multicolumn{1}{l}{$250.00$} & \multicolumn{1}{l}{\quad $0.7941$} & 
$id.$ & $1966.4$ & 
\multicolumn{1}{r}{$0.08$} \\ 
\multicolumn{1}{l}{$298.15^{\mathbf{\ast }}$} & \multicolumn{1}{l}{\quad $1.0000^{\mathbf{\ast }}$} & $10.5927^{\mathbf{\ast }}$
& $1964.8$ & \multicolumn{1}{r}{$\sim 0.00%
$} \\ 
\multicolumn{1}{l}{$350.00$} & \multicolumn{1}{l}{\quad $1.2444$} & 
$id.$ & $1963.1$ & 
\multicolumn{1}{r}{$-0.08$} \\ 
\multicolumn{1}{l}{$400.00$} & \multicolumn{1}{l}{\quad $1.5042$} & 
$id.$ & $1961.5$ & 
\multicolumn{1}{r}{$-0.16$}%
\end{tabular}
\end{ruledtabular}
\end{table}
\begin{table*}[t!]
\caption{Joint invariances of the CJ entropy $s_{\text{CJ}}$ and velocity $D_{\text{CJ}}$ for the \ce{C3H8 + 5 O2} mixture: sensitivity to small changes of initial state.}
\label{tab:Tab2}
\begin{ruledtabular}
\begin{tabular}{cccccccc}
$T_0$ & $p_0$ & $s_{\text{CJ}}$ & $\frac{\Delta s_{\text{CJ}}}{s_{\text{%
CJ}}^{\ast }}$ & $D_{\text{CJ}}$ & $\frac{\Delta D_{\text{CJ}}}{D_{\text{CJ}%
}^{\ast }}$ & $T_{\text{CJ}}$ & $\frac{\Delta T_{\text{CJ}}}{T_{\text{CJ}%
}^{\ast }}$ \\ 
{\small (K)} & {\small (bar)} & {\small (kJ/kg/K)} & {\small (\%)} & {\small %
(m/s)} & {\small (\%)} & {\small (K)} & {\small (\%)} \\ \hline
\multicolumn{1}{l}{$200.00^{\ast}$} & \multicolumn{1}{l}{$%
0.6304^{\ast }$} & \multicolumn{1}{l}{$11.9293^{\ast }$} & $/$ & \multicolumn{1}{l}{$2357.3^{%
\ast }$} & $/$ & \multicolumn{1}{l}{$3799.46^{%
\ast }$} & $/$ \\ 
\multicolumn{1}{l}{$\emph{210.00}$} & \multicolumn{1}{l}{$\emph{0.6660}$} & 
\multicolumn{1}{l}{$\emph{11.9293}^{\ast }$} & $\mathbf{/}$ & 
\multicolumn{1}{l}{$\emph{2357.1}$} & \multicolumn{1}{r}{$\emph{-0.01}$} & 
\multicolumn{1}{l}{$\emph{3801.57}$} & \multicolumn{1}{r}{$\emph{0.06}$} \\ 
\multicolumn{1}{l}{$200.00$} & \multicolumn{1}{l}{$0.6660$} & 
\multicolumn{1}{l}{$11.9093$} & \multicolumn{1}{r}{$-0.17$} & 
\multicolumn{1}{l}{$2359.7$} & \multicolumn{1}{r}{$0.10$} & 
\multicolumn{1}{l}{$3810.15$} & \multicolumn{1}{r}{$0.28$} \\ 
\multicolumn{1}{l}{$210.00$} & \multicolumn{1}{l}{$0.6304$} & 
\multicolumn{1}{l}{$11.9493$} & \multicolumn{1}{r}{$0.17$} & 
\multicolumn{1}{l}{$2354.7$} & \multicolumn{1}{r}{$-0.14$} & 
\multicolumn{1}{l}{$3790.91$} & \multicolumn{1}{r}{$-0.22$} \\ \hline
\multicolumn{1}{l}{$298.15^{\ast }$} & \multicolumn{1}{l}{$%
1.0000^{\ast }$} & \multicolumn{1}{l}{$11.9293^{%
\ast }$} & $/$ & \multicolumn{1}{l}{$2356.3^{%
\ast }$} & $/$ & \multicolumn{1}{l}{$3821.11^{%
\ast }$} & $/$ \\ 
\multicolumn{1}{l}{$\emph{313.06}$} & \multicolumn{1}{l}{$\emph{1.0606}$} & 
\multicolumn{1}{l}{$\emph{11.9293}^{\ast }$} & $\mathbf{/}$ & 
\multicolumn{1}{l}{$\emph{2356.3}$} & \multicolumn{1}{r}{$\emph{0.00}$} & 
\multicolumn{1}{l}{$\emph{3824.64}$} & \multicolumn{1}{r}{$\emph{0.09}$} \\ 
\multicolumn{1}{l}{$298.15$} & \multicolumn{1}{l}{$1.0606$} & 
\multicolumn{1}{l}{$11.9078$} & \multicolumn{1}{r}{$-0.18$} & 
\multicolumn{1}{l}{$2358.9$} & \multicolumn{1}{r}{$0.11$} & 
\multicolumn{1}{l}{$3832.68$} & \multicolumn{1}{r}{$0.30$} \\ 
\multicolumn{1}{l}{$313.06$} & \multicolumn{1}{l}{$1.0000$} & 
\multicolumn{1}{l}{$11.9508$} & \multicolumn{1}{r}{$0.18$} & 
\multicolumn{1}{l}{$2353.6$} & \multicolumn{1}{r}{$-0.11$} & 
\multicolumn{1}{l}{$3813.09$} & \multicolumn{1}{r}{$-0.21$} \\ \hline
\multicolumn{1}{l}{$400.00^{\ast }$} & \multicolumn{1}{l}{$%
1.4419^{\ast }$} & \multicolumn{1}{l}{$11.9293^{%
\ast }$} & $/$ & \multicolumn{1}{l}{$2356.7^{%
\ast }$} & $/$ & \multicolumn{1}{l}{$3846.74^{%
\ast }$} & $/$ \\ 
\multicolumn{1}{l}{$\emph{420.00}$} & \multicolumn{1}{l}{$\emph{1.5371}$} & 
\multicolumn{1}{l}{$\emph{11.9293}^{\ast }$} & $\mathbf{/}$ & 
\multicolumn{1}{l}{$\emph{2356.9}$} & \multicolumn{1}{r}{$\emph{0.01}$} & 
\multicolumn{1}{l}{$\emph{3852.19}$} & \multicolumn{1}{r}{$\emph{0.14}$} \\ 
\multicolumn{1}{l}{$400.00$} & \multicolumn{1}{l}{$1.5371$} & 
\multicolumn{1}{l}{$11.9059$} & \multicolumn{1}{r}{$-0.20$} & 
\multicolumn{1}{l}{$2359.6$} & \multicolumn{1}{r}{$0.12$} & 
\multicolumn{1}{l}{$3859.48$} & \multicolumn{1}{r}{$0.33$} \\ 
\multicolumn{1}{l}{$420.00$} & \multicolumn{1}{l}{$1.4419$} & 
\multicolumn{1}{l}{$11.9527$} & \multicolumn{1}{r}{$0.20$} & 
\multicolumn{1}{l}{$2354.0$} & \multicolumn{1}{r}{$-0.11$} & 
\multicolumn{1}{l}{$3839.46$} & \multicolumn{1}{r}{$-0.19$} \\
\end{tabular}
\end{ruledtabular}
\end{table*}

These small values were validated with a sensitivity analysis using initial states very close to a reference state and the numerical accuracy of CEA as a criterion. \hyperref[tab:Tab2]{Table \ref{tab:Tab2}} shows the results for the mixture \ce{C3H8 + 5 O2} and 3 groups of 4 $\left(T_0,p_0\right)$ pairs. The first pairs of each group are the first, third and fifth in \hyperref[tab:Tab1]{Table \ref{tab:Tab1}-2}. They are the reference state of their group (superscript $*$) since they produce the reference entropy $s_{\text{CJ}}^{\ast}$. 
The second pairs (italics) have $T_0$ values chosen only $5$\% higher than in the first pairs and $p_0$ values determined by dichotomy to produce the reference entropy $s_{\text{CJ}}^{\ast}$. Thus, the relative variations $\Delta D_{\text{CJ}}/D_{\text{CJ}}^{\ast}$ are not greater than $m_{D_{\text{CJ}}}$ \hyperref[tab:Tab1]{(Tab. \ref{tab:Tab1}-2)}, and smaller variations of $T_0$ would have no meaning. The third and fourth pairs are variations at constant $T_0$ and constant $p_0$, respectively, relative to the firsts. The second pairs in each group, which produce the reference entropy $s_{\text{CJ}}^{\ast}$, give the smaller relative variations of $T_{\text{CJ}}$, all at least one order of magnitude greater than the $\mathcal{O}\left(10^{-3}\right) \%$ accuracy $\tilde{d}T/T=0.005$\% of CEA (\cite{GordonMcBride-I}, p.35, eq. 7.24, and p.40). The third and fourth pairs in each group, which do not produce the reference entropy $s_{\text{CJ}}^{\ast }$, give relative deviations of $\mathcal{O}\left(10^{-1}\right) \%$ for $D_{\text{CJ}}$, $s_{\text{CJ}}$ and $T_{\text{CJ}}$,  i.e. 10 times greater than $m_{D_{\text{CJ}}}$. Therefore, the small $\mathcal{O}\left(10^{-2}\right) \%$ relative variations of $D_{\text{CJ}}$ at constant $s_{\text{CJ}}$, and the larger ones $\mathcal{O}\left(10^{-1}\right) \%$ of $s_{\text{CJ}}$ and $D_{\text{CJ}}$ at constant $T_0$ or $p_0$, have both physical and numerical significance and are not due to initial states chosen too close to each other. The relative deviations of $s_{\text{CJ}}$ are slightly smaller than those of $T_{\text{CJ}}$: the combination of $dh\left(s,p\right)$ (\ref{H_SP}), $dh\left( T\right) =C_{p}dT$ (\ref{H_TP}), $pv=RT/W$ and $\gamma =C_{p}/C_{v}$, subject to $\tilde{d}T/T=\tilde{d}p/p$, gives
\begin{equation}
\frac{\tilde{d}s}{s}=\left(2-\gamma ^{-1}\right)\frac{C_{p}}{s}\times \frac{\tilde{d}T}{T}=\mathcal{O}\left(10^{-1}\text{-}1\right)\times \frac{\tilde{d}T}{T}
\end{equation}
since typical $\gamma $, $s$ and $C_{p}$ are $\mathcal{O}\left( 1\right)$, $\mathcal{O}\left( 10\right) $\ kJ/K/kg and $\mathcal{O}\left( 1\text{-}10\right) $\ kJ/K/kg, respectively. At $p_0=1$ bar and $T_0=298.15$ K, CEA gives $\tilde{d}s_{\text{CJ}}/s_{\text{CJ}}=0.33\times \tilde{d}T_{\text{CJ}}/T_{\text{CJ}}$ for \ce{CH4 + 2} Air and $0.89\times \tilde{d}T_{\text{CJ}}/T_{\text{CJ}}$ for \ce{CH4 + 2 O2}.
\newline

\hyperref[tab:Tab3]{Table \ref{tab:Tab3}} shows the initial data for calculating the theoretical CJ state of \ce{C3H8}:\ce{O2} mixtures from their numerical CJ velocity $D_{\text{CJ}}$ using (\ref{DCJ_1_PARAM_IDEAL}), (\ref{VCJ_PARAM}) and (\ref{PCJ_PARAM}).
\hyperref[tab:Tab4]{Table \ref{tab:Tab4}} shows the comparison of the theoretical (\textit{\footnotesize{THEO}}) and numerical (\textit{\footnotesize{NUM}}) CJ properties $r_{\text{CJ}}$=($\gamma_{\text{CJ}}$, $\rho_{\text{CJ}}$, $p_{\text{CJ}}$). All the relative differences  $\epsilon_{r}=100\times\,(1-r_{\text{CJ}}^{\textit{\tiny{THEO}}}/{r_{\text{CJ}}^{\textit{\tiny{NUM}}}})$ are small, ranging from $\mathcal{O}\left(10^{-1}\right)$ to $\mathcal{O}\left(1\right)$ \%, but greater than the $\mathcal{O}\left(10^{-2}\right)$ - $\mathcal{O}\left(10^{-1}\right)$ \% values of $m_{D_{\text{CJ}}}$ (the $r_{\text{CJ}}^{\textit{\tiny{THEO}}}$ and ${r_{\text{CJ}}^{\textit{\tiny{NUM}}}}$ values were rounded after calculating $\epsilon _{r}$). The same is true for other mixtures, e.g. $\epsilon_{\gamma }=-3.4$ \% and $m_{D_{\text{CJ}}}=0.08$~\% for \ce{CH4 + 2 O2} at $T_0=298.15$ K and $p_0=1$ bar.
\newline

These small differences might be due to the sensitivity to the initial thermodynamic coefficients and, here, only to $C_{p0}$. The uncertainties in $s_{\text{CJ}}$, $\gamma _{\text{CJ}}$, $\rho _{\text{CJ}}$ and $p_{\text{CJ}}$ are obtained from $ds\left( p,v\right)$ (\ref{H_SP}), (\ref{VCJ_PARAM}), (\ref{PCJ_PARAM}), $pv=RT/W$, $\gamma _{\text{CJ}}^{2}\simeq \gamma _0=C_{p0}/C_{v0}$ (\ref{DCJ_1_PARAM_IDEAL}) and $C_{p0}-C_{v0}=R/W_0$. The approximations $\gamma _{\text{CJ}}\simeq 1^{+}$ and $M_{0\text{CJ}}^{-2}\ll 1$, and the typical values $ \gamma _{\text{CJ}}^{2}\simeq \gamma _0\simeq G_{\text{CJ}}+1\simeq 1.2$, $s_{\text{CJ}}\simeq 10^{4}$~J/kg, $R\simeq 8$~J/kg/mole and $W_{\text{CJ}}\simeq 2\times 10^{-2}$~kg/mole, then give the estimates
\begin{align}
\frac{\delta s_{\text{CJ}}}{s_{\text{CJ}}}&=\frac{2R/G_{\text{CJ}}}{Ws_{\text{CJ}}}\frac{1-M_{0\text{CJ}}^{-2}}{1+M_{0\text{CJ}}^{-2}/\gamma _0}%
\frac{\delta D_{\text{CJ}}}{D_{\text{CJ}}}\simeq \frac{1}{10}\frac{\delta
D_{\text{CJ}}}{D_{\text{CJ}}},  \label{Delt_sCJ_Init}
\\
\frac{\delta \rho _{\text{CJ}}}{\rho _{\text{CJ}}}&\simeq \frac{\delta p_{\text{CJ}}}{p_{\text{CJ}}}\simeq \frac{-1}{4}\frac{\delta \gamma _0}{\gamma _0}=\frac{-1}{2}%
\frac{\delta C_{p0}}{C_{p0}} .  \label{Delt_RHOCJ_Init}
\end{align}
They show that $D_{\text{CJ}}$ is $10$ times more sensitive than $s_{\text{CJ}}$, validating the above choice to analyze the theorem with initial states that produce the same $s_{\text{CJ}}$ rather than the same $D_{\text{CJ}}$ \hyperref[tab:Tab1]{(Tables \ref{tab:Tab1})}. Also, $\gamma_{\text{CJ}}\simeq\sqrt{\gamma_0}$ is twice as sensitive as $\rho _{\text{CJ}}$ and $p_{\text{CJ}}$, 
and its uncertainty is twice as small as that of $\gamma _0$ (\ref{DCJ_1_PARAM_IDEAL}) and thus the same as that of $C_{p_0}$. The value of $\delta C_{p_0}/C_{p_0}$ depends on $T_0$, $p_0$ and the mixture components and proportions.
\newline

Therefore, the numerical calculation supports the invariance theorem for a wide range of initial conditions. The larger deviations $\Delta D_{\text{CJ}}/\bar{D}_{\text{CJ}}$ at constant $s_{\text{CJ}}$ are very small numbers, numerically and physically valid, and smaller than at constant $p_0$ or $T_0$. The same is true for the differences of the additional CJ properties with the numerical values, which are also no greater than those resulting from the physical uncertainty in the thermochemical coefficients. Similar trends were obtained for \ce{CH4}, \ce{C2H2}, \ce{C2H4}, \ce{C2H6}, and \ce{H2}.

\begin{table*}[t]
\caption{Initial data for calculating the theoretical CJ state from the CJ velocity $D_{\text{CJ}}$ for \ce{C3H8}:\ce{O2} mixtures \hyperref[tab:Tab4]{(Table \ref{tab:Tab4}, \textit{\footnotesize{THEO}})}.}
\label{tab:Tab3}
\begin{ruledtabular}    
\begin{tabular}{ll c cccc c cccc c cccc}
 & & & \multicolumn{4}{c}{ER $=0.8$} & & \multicolumn{4}{c}{ER $=1$} & & \multicolumn{4}{c}{ER $=1.2$} \\
 & & & \multicolumn{4}{c}{$W_0=33.667$ {\small (g/mol) }} & & \multicolumn{4}{c}{$W_0=34.015$ {\small (g/mol) }} & & \multicolumn{4}{c}{$W_0=34.340$ {\small (g/mol) }} \\
\cline{4-7} \cline{9-12} \cline{14-17}
\multicolumn{1}{l}{$T_0$} & \multicolumn{1}{c}{$p_0$} & &
\multicolumn{1}{c}{$\gamma_0$} & \multicolumn{1}{c}{$c_0$} & \multicolumn{1}{c}{$v_0$} & \multicolumn{1}{c}{$D_{\text{CJ}}$} & &
\multicolumn{1}{c}{$\gamma_0$} & \multicolumn{1}{c}{$c_0$} & \multicolumn{1}{c}{$v_0$} & \multicolumn{1}{c}{$D_{\text{CJ}}$} & & 
\multicolumn{1}{c}{$\gamma_0$} & \multicolumn{1}{c}{$c_0$} & \multicolumn{1}{c}{$v_0$} & \multicolumn{1}{c}{$D_{\text{CJ}}$}\\

\multicolumn{1}{l}{\small (K)}   & \multicolumn{1}{c}{\small (bar)} & & &
\multicolumn{1}{c}{\small (m/s)} & \multicolumn{1}{c}{\small (m}$^{3}${\hspace{-0.08cm}\small /kg)} & \multicolumn{1}{c}{\small (m/s)} &  &  & 
\multicolumn{1}{c}{\small (m/s)} & \multicolumn{1}{c}{\small (m}$^{3}${\hspace{-0.08cm}\small /kg)} & \multicolumn{1}{c}{\small (m/s)} &  &  &
\multicolumn{1}{c}{\small (m/s)} & \multicolumn{1}{c}{\small (m}$^{3}${\hspace{-0.08cm}\small /kg)} & \multicolumn{1}{c}{\small (m/s)}\\ \hline

\multicolumn{1}{l}{$200.$} & \multicolumn{1}{l}{
\begin{tabular}{l}
$0.2$ \\ $1$   \\ $5$   
\end{tabular}%
} & & \multicolumn{1}{c}{ %
\begin{tabular}{c}
$id.$ \\ $1.3390$ \\ $id.$
\end{tabular}%
} &\multicolumn{1}{c}{%
\begin{tabular}{c}
$id.$ \\ $257.2$ \\$id.$ 
\end{tabular}%
} &\multicolumn{1}{c}{%
\begin{tabular}{c}
$2.4696$ \\ $0.4939$ \\ $0.0988$
\end{tabular}%
} &\multicolumn{1}{c}{%
\begin{tabular}{l}
$2203.9$ \\ $2269.8$ \\ $2334.7$
\end{tabular}%
}&\,\,\,\,\,\,&
\multicolumn{1}{c}{ 
\begin{tabular}{c}
$id.$ \\ $1.3286$ \\ $id.$
\end{tabular}%
} &\multicolumn{1}{c}{%
\begin{tabular}{c}
$id.$ \\ $254.9$ \\ $id.$ 
\end{tabular}%
} &\multicolumn{1}{c}{%
\begin{tabular}{c}
$2.4444$ \\ $0.4889$ \\ $0.0978$
\end{tabular}%
} &\multicolumn{1}{c}{%
\begin{tabular}{l}
$2306.7$ \\ $2377.6$ \\ $2447.5$
\end{tabular}%
}&\,\,\,\,\,\,&
\multicolumn{1}{c}{ 
\begin{tabular}{c}
$id.$ \\ $1.3194$ \\ $id.$
\end{tabular}%
} &\multicolumn{1}{c}{%
\begin{tabular}{c}
$id.$ \\ $252.8$ \\ $id.$
\end{tabular}%
} &\multicolumn{1}{c}{%
\begin{tabular}{c}
$2.4212$ \\ $0.4842$ \\ $0.0968$
\end{tabular}%
} &\multicolumn{1}{c}{%
\begin{tabular}{l}
$2392.0$ \\ $2466.1$ \\ $2538.8$
\end{tabular}%
}

\\ \hline

\multicolumn{1}{l}{$298.15$} &\multicolumn{1}{l}{ %
\begin{tabular}{l}
$0.2$ \\ $1$   \\ $5$   
\end{tabular}%
} & & \multicolumn{1}{c}{ %
\begin{tabular}{c}
$id.$ \\ $1.3061$ \\ $id.$
\end{tabular}%
} &\multicolumn{1}{c}{%
\begin{tabular}{c}
$id.$ \\ $310.1$ \\ $id.$ 
\end{tabular}%
} &\multicolumn{1}{c}{%
\begin{tabular}{c}
$3.6816$ \\ $0.7363$ \\ $0.1473$
\end{tabular}%
} &\multicolumn{1}{c}{%
\begin{tabular}{l}
$2182.5$ \\ $2249.2$ \\ $2315.4$
\end{tabular}%
}&\,\,\,\,\,\,&
\multicolumn{1}{c}{ 
\begin{tabular}{c}
$id.$ \\ $1.2924$ \\ $id.$
\end{tabular}%
} &\multicolumn{1}{c}{%
\begin{tabular}{c}
$id.$ \\ $306.9$ \\ $id.$ 
\end{tabular}%
} &\multicolumn{1}{c}{%
\begin{tabular}{c}
$3.6439$ \\ $0.7288$ \\ $0.1458$
\end{tabular}%
} &\multicolumn{1}{c}{%
\begin{tabular}{l}
$2284.6$ \\ $2356.3$ \\ $2427.6$
\end{tabular}%
}&\,\,\,\,\,\,&
\multicolumn{1}{c}{ 
\begin{tabular}{c}
$id.$ \\ $1.2807$ \\ $id.$
\end{tabular}%
} &\multicolumn{1}{c}{%
\begin{tabular}{c}
$id.$ \\ $304.1$ \\ $000.0$
\end{tabular}%
} &\multicolumn{1}{c}{%
\begin{tabular}{c}
$3.6094$ \\ $0.7219$ \\ $0.1444$
\end{tabular}%
} &\multicolumn{1}{c}{%
\begin{tabular}{l}
$2369.8$ \\ $2444.7$ \\ $2518.9$
\end{tabular}%
}

\\ \hline

\multicolumn{1}{l}{$400.$} &\multicolumn{1}{l}{
\begin{tabular}{l}
$0.2$ \\ $1$   \\ $5$   
\end{tabular}%
} & & \multicolumn{1}{c}{ %
\begin{tabular}{c}
$id.$ \\ $1.2716$ \\ $id.$
\end{tabular}%
} &\multicolumn{1}{c}{%
\begin{tabular}{c}
$id.$ \\ $354.4$ \\$id.$ 
\end{tabular}%
} &\multicolumn{1}{c}{%
\begin{tabular}{c}
$4.9393$ \\ $0.9878$ \\ $0.1976$
\end{tabular}%
} &\multicolumn{1}{c}{%
\begin{tabular}{l}
$2165.5$ \\ $2233.2$ \\ $2300.6$
\end{tabular}%
}&\,\,\,\,\,\,&
\multicolumn{1}{c}{ 
\begin{tabular}{c}
$id.$ \\ $1.2563$ \\ $id.$
\end{tabular}%
} &\multicolumn{1}{c}{%
\begin{tabular}{c}
$id.$ \\ $350.5$ \\$id.$ 
\end{tabular}%
} &\multicolumn{1}{c}{%
\begin{tabular}{c}
$4.8887$ \\ $0.9777$ \\ $0.1956$
\end{tabular}%
} &\multicolumn{1}{c}{%
\begin{tabular}{l}
$2267.6$ \\ $2340.1$ \\ $2412.6$
\end{tabular}%
}&\,\,\,\,\,\,&
\multicolumn{1}{c}{ 
\begin{tabular}{c}
$id.$ \\ $1.2434$ \\ $id.$
\end{tabular}%
} &\multicolumn{1}{c}{%
\begin{tabular}{c}
$id.$ \\ $347.0$ \\$id.$ 
\end{tabular}%
} &\multicolumn{1}{c}{%
\begin{tabular}{c}
$4.8425$ \\ $0.9685$ \\ $0.1937$
\end{tabular}%
} &\multicolumn{1}{c}{%
\begin{tabular}{l}
$2352.9$ \\ $2428.6$ \\ $2504.2$
\end{tabular}%
}

\end{tabular}
\end{ruledtabular}
\end{table*}
\begin{table*}[t]
\caption{Comparison of numerical (\textit{\footnotesize{NUM}}) and theoretical (\textit{\footnotesize{THEO}}) CJ properties ($r_{\text{CJ}}$) of \ce{C3H8}:\ce{O2} mixtures for 3 equivalence ratios ER and 3 initial temperatures $T_0$ and pressures $p_0$.}
\label{tab:Tab4}
\begin{ruledtabular}
\begin{tabular}{lllccc}
$T_0$ &$p_0$ &\hspace{0.3cm}$r_{\text{CJ}}$ & ER $=0.8$ & ER $=1$ & ER $%
=1.2$ \\ 
{\small (K)} & {\hspace{-0.1cm}\small (bar)} &  & 
\hspace{0.2cm}%
\begin{tabular}{lcc}
\hline
\hspace{0.4cm}\textit{\scriptsize{NUM}} & \hspace{0.1cm}\textit{\scriptsize{THEO}} & \hspace{0.1cm}$\epsilon _{r}$ (\%)%
\end{tabular}
& 
\hspace{0.2cm}%
\begin{tabular}{lcc}
\hline
\hspace{0.4cm}\textit{\scriptsize{NUM}} & \hspace{0.1cm}\textit{\scriptsize{THEO}} & \hspace{0.1cm}$\epsilon _{r}$ (\%)%
\end{tabular}
& 
\hspace{0.2cm}%
\begin{tabular}{lcc}
\hline
\hspace{0.4cm}\textit{\scriptsize{NUM}} & \hspace{0.1cm}\textit{\scriptsize{THEO}} & \hspace{0.1cm}$\epsilon _{r}$ (\%)%
\end{tabular}
\\ \hline
& $0.2$ & \multicolumn{1}{c}{%
\begin{tabular}{l}
$\gamma _{\text{CJ}}$\\
$\rho _{\text{CJ}}/\rho _0$ \\ 
$p_{\text{CJ}}/p_0$
\end{tabular}%
} & 
\begin{tabular}{rrr}
$1.125$ & $1.159$ & $\emph{-3.03}$\\
$1.870$ & $1.844$ & $\emph{1.38}$ \\ 
$46.746$ & $46.010$ & $\emph{1.58}$
\end{tabular}
& 
\begin{tabular}{rrr}
$1.127$ & $1.154$ & $\emph{-2.41}$\\
$1.870$ & $1.849$ & $\emph{1.14}$ \\ 
$51.635$ & $50.966$ & $\emph{1.29}$
\end{tabular}
& 
\begin{tabular}{rrr}
$1.130$ & $1.150$ & $\emph{-1.81}$\\
$1.870$ & $1.854$ & $\emph{0.86}$ \\ 
$55.950$ & $55.402$ & $\emph{0.98}$
\end{tabular}
\\ \cline{2-6}
$200.$ & $1$ & \multicolumn{1}{c}{%
\begin{tabular}{l}
$\gamma _{\text{CJ}}$\\
$\rho _{\text{CJ}}/\rho _0$ \\ 
$p_{\text{CJ}}/p_0$
\end{tabular}%
} & 
\begin{tabular}{rrr}
$1.134$ & $1.159$ & $\emph{-2.23}$\\
$1.864$ & $1.845$ & $\emph{1.02}$ \\ 
$49.354$ & $48.775$ & $\emph{1.17}$ 
\end{tabular}
& 
\begin{tabular}{rrr}
$1.136$ & $1.154$ & $\emph{-1.60}$\\
$1.865$ & $1.850$ & $\emph{0.77}$ \\ 
$54.602$ & $54.121$ & $\emph{0.88}$ 
\end{tabular}
& 
\begin{tabular}{rrr}
$1.139$ & $1.150$ & $\emph{-0.96}$\\
$1.863$ & $1.855$ & $\emph{0.47}$ \\ 
$59.180$ & $58.861$ & $\emph{0.54}$ 
\end{tabular}
\\ \cline{2-6}
& $5$ & \multicolumn{1}{c}{%
\begin{tabular}{l}
$\gamma _{\text{CJ}}$\\
$\rho _{\text{CJ}}/\rho _0$ \\ 
$p_{\text{CJ}}/p_0$
\end{tabular}%
} & 
\begin{tabular}{rrr}
$1.142$ & $1.159$ & $\emph{-1.50}$\\
$1.859$ & $1.846$ & $\emph{0.69}$ \\ 
$51.990$ & $51.580$ & $\emph{0.79}$ 
\end{tabular}
& 
\begin{tabular}{rrr}
$1.144$ & $1.154$ & $\emph{-0.86}$\\
$1.859$ & $1.851$ & $\emph{0.43}$ \\ 
$57.612$ & $57.325$ & $\emph{0.50}$ 
\end{tabular}
& 
\begin{tabular}{rrr}
$1.148$ & $1.150$ & $\emph{-0.19}$\\
$1.858$ & $1.856$ & $\emph{0.11}$ \\ 
$62.436$ & $62.357$ & $\emph{0.13}$ 
\end{tabular}
\\ \hline
& $0.2$ & \multicolumn{1}{c}{%
\begin{tabular}{l}
$\gamma _{\text{CJ}}$\\
$\rho _{\text{CJ}}/\rho _0$ \\ 
$p_{\text{CJ}}/p_0$
\end{tabular}%
} & 
\begin{tabular}{rrr}
$1.123$ & $1.146$ & $\emph{-1.98}$\\
$1.861$ & $1.844$ & $\emph{0.92}$ \\ 
$30.939$ & $30.617$ & $\emph{1.04}$ 
\end{tabular}
& 
\begin{tabular}{rrr}
$1.125$ & $1.139$ & $\emph{-1.23}$\\
$1.863$ & $1.852$ & $\emph{0.58}$ \\ 
$34.170$ & $33.947$ & $\emph{0.65}$ 
\end{tabular}
& 
\begin{tabular}{rrr}
$1.128$ & $1.134$ & $\emph{-0.53}$\\
$1.863$ & $1.858$ & $\emph{0.27}$ \\ 
$37.031$ & $36.919$ & $\emph{0.30}$ 
\end{tabular}

\\ \cline{2-6}

$298.15$\hspace{0.2cm} & $1$ & \multicolumn{1}{c}{%
\begin{tabular}{l}
$\gamma _{\text{CJ}}$\\
$\rho _{\text{CJ}}/\rho _0$ \\ 
$p_{\text{CJ}}/p_0$
\end{tabular}%
} & 
\begin{tabular}{rrr}
$1.132$ & $1.145$ & $\emph{-1.18}$\\
$1.856$ & $1.846$ & $\emph{0.55}$ \\ 
$32.696$ & $32.491$ & $\emph{0.63}$ 
\end{tabular}
& 
\begin{tabular}{rrr}
$1.134$ & $1.139$ & $\emph{-0.43}$\\
$1.857$ & $1.854$ & $\emph{0.20}$ \\ 
$36.165$ & $36.084$ & $\emph{0.23}$ 
\end{tabular}
& 
\begin{tabular}{rrr}
$1.137$ & $1.134$ & $\emph{0.32}$\\
$1.857$ & $1.860$ & $\emph{-0.12}$ \\ 
$39.206$ & $39.262$ & $\emph{-0.14}$ 
\end{tabular}
\\ \cline{2-6}
& $5$ & \multicolumn{1}{c}{%
\begin{tabular}{l}
$\gamma _{\text{CJ}}$\\
$\rho _{\text{CJ}}/\rho _0$ \\ 
$p_{\text{CJ}}/p_0$
\end{tabular}%
} & 
\begin{tabular}{rrr}
$1.140$ & $1.145$ & $\emph{-0.43}$\\
$1.852$ & $1.848$ & $\emph{0.20}$ \\ 
$34.486$ & $34.406$ & $\emph{0.23}$ 
\end{tabular}
& 
\begin{tabular}{rrr}
$1.143$ & $1.139$ & $\emph{0.34}$\\
$1.852$ & $1.855$ & $\emph{-0.16}$ \\ 
$38.204$ & $38.273$ & $\emph{-0.18}$ 
\end{tabular}
& 
\begin{tabular}{rrr}
$1.146$ & $1.133$ & $\emph{1.11}$\\
$1.852$ & $1.861$ & $\emph{-0.50}$ \\ 
$41.418$ & $41.654$ & $\emph{-0.57}$ 
\end{tabular}
\\ \hline
& $0.2$ & \multicolumn{1}{c}{%
\begin{tabular}{l}
$\gamma _{\text{CJ}}$\\
$\rho _{\text{CJ}}/\rho _0$ \\ 
$p_{\text{CJ}}/p_0$
\end{tabular}%
} & 
\begin{tabular}{rrr}
$1.122$ & $1.131$ & $\emph{-0.79}$\\
$1.852$ & $1.845$ & $\emph{0.38}$ \\ 
$22.843$ & $22.747$ & $\emph{0.42}$
\end{tabular}
& 
\begin{tabular}{rrr}
$1.124$ & $1.124^{-}$ & $\emph{-0.04}$\\
$1.855$ & $1.855$ & $\emph{-0.00}$ \\ 
$25.232$ & $25.233$ & $\emph{-0.00}$
\end{tabular}
& 
\begin{tabular}{rrr}
$1.126$ & $1.117$ & $\emph{0.79}$\\
$1.855$ & $1.862$ & $\emph{-0.39}$ \\ 
$27.352$ & $27.471$ & $\emph{-0.43}$
\end{tabular}
\\ \cline{2-6}
$400.$ & $1$ & \multicolumn{1}{c}{%
\begin{tabular}{l}
$\gamma _{\text{CJ}}$\\
$\rho _{\text{CJ}}/\rho _0$ \\ 
$p_{\text{CJ}}/p_0$
\end{tabular}%
} & 
\begin{tabular}{rrr}
$1.131$ & $1.131$ & $\emph{-0.01}$\\
$1.848$ & $1.848$ & $\emph{-0.01}$ \\ 
$24.162$ & $24.164$ & $\emph{-0.01}$
\end{tabular}
& 
\begin{tabular}{rrr}
$1.133$ & $1.123$ & $\emph{0.85}$\\
$1.850$ & $1.857$ & $\emph{-0.39}$ \\ 
$26.726$ & $26.843$ & $\emph{-0.44}$

\end{tabular}
& 
\begin{tabular}{rrr}
$1.136$ & $1.117$ & $\emph{1.63}$\\
$1.850$ & $1.864$ & $\emph{-0.78}$ \\ 
$28.982$ & $29.238$ & $\emph{-0.88}$

\end{tabular}
\\ \cline{2-6}
& $5$ & \multicolumn{1}{c}{%
\begin{tabular}{l}
$\gamma _{\text{CJ}}$\\
$\rho _{\text{CJ}}/\rho _0$ \\ 
$p_{\text{CJ}}/p_0$
\end{tabular}%
} & 
\begin{tabular}{rrr}
$1.139$ & $1.131$ & $\emph{0.77}$\\
$1.843$ & $1.850$ & $\emph{-0.36}$ \\ 
$25.512$ & $25.618$ & $\emph{-0.41}$

\end{tabular}
& 
\begin{tabular}{rrr}
$1.142$ & $1.123$ & $\emph{1.62}$\\
$1.845$ & $1.859$ & $\emph{-0.76}$ \\ 
$28.262$ & $28.505$ & $\emph{-0.86}$

\end{tabular}
& 
\begin{tabular}{rrr}
$1.145$ & $1.117$ & $\emph{2.43}$\\
$1.845$ & $1.866$ & $\emph{-1.17}$ \\ 
$30.652$ & $31.059$ & $\emph{-1.33}$

\end{tabular}

\end{tabular}
\end{ruledtabular}
\end{table*}
\clearpage
\section{\label{sec:Applic-Liq}Application to liquid explosives}
Four carbon liquids were studied, namely nitromethane (NM: \ce{CH3NO2}), isopropyl nitrate (IPN: \ce{C3H7NO3}), hot trinitrotoluene (LTNT: \ce{C7H5N3O6}), and the stoichiometric mixture NPNA3 (\ce{C3H10N4O11}) consisting of 1 volume of 2-nitropropane (\ce{C3H7NO2}) and 3 volumes of nitric acid (\ce{HNO3}).
The analysis is based on the detonation pressures ($p$) given by the experiments (\textit{\footnotesize{EXP}}), the theorem (\textit{\footnotesize{THEO}}) \hyperlink{addCJstate}{(Subsect. \ref{subsec:CJsuppl}, §1)}, and the Inverse Method (\textit{\footnotesize{IM}}) \hyperlink{addIM}{(Subsect. \ref{subsec:CJsuppl}, §5)}. The subscript CJ is used here for the experimental values, although they may not represent those of CJ detonations.

\hyperref[tab:Tab5]{Table \ref{tab:Tab5}} compares values of $p_{\text{CJ}}^{\textit{\tiny{EXP}}}$, $p_{\text{CJ}}^{\textit{\tiny{THEO}}}$ and $p_{\text{CJ}}^{\textit{\tiny{IM}}}$. The values of $p_{\text{CJ}}^{\textit{\tiny{THEO}}}$ were calculated from (\ref{PCJ_1_PARAM}) with the experimental velocities $D_{\text{CJ}}^{\textit{\tiny{EXP}}}\left(T_0,p_0\right)$, and those of $p_{\text{CJ}}^{\textit{\tiny{IM}}}$ from (\ref{IVM_VCJ}), (\ref {IVM_L}), (\ref{IVM_K}) with these velocities and their confident derivatives shown in the second rows and columns of Tables \hyperref[tab:Tab6]{\ref{tab:Tab6}} and \hyperref[tab:Tab7]{\ref{tab:Tab7}-left}. The values of $p_{\text{CJ}}^{\textit{\tiny{THEO}}}$ are greater than those of $p_{\text{CJ}}^{\textit{\tiny{EXP}}}$. 
The same is true for those of $p_{\text{CJ}}^{\textit{\tiny{IM}}}$ of NM, albeit with smaller differences. The adiabatic exponents $\gamma_{\text{CJ}}^{\textit{\tiny{THEO}}}$ (\ref{GAMCJ_1_PARAM}) are consistently smaller. Although the $p_{\text{CJ}}^{\textit{\tiny{THEO}}}$ and $p_{\text{CJ}}^{\textit{\tiny{IM}}}$ values have the correct magnitude, their differences from the $p_{\text{CJ}}^{\textit{\tiny{EXP}}}$ values are significant. The analysis eliminates the effects of data and measurement uncertainties and discusses the hydrodynamic framework of the modelling. In the main, a single inviscid phase, as represented by (\ref{H_SP})-(\ref{S_PV}) and (\ref{RH_Mass})-(\ref{RH_Energ}), may not be relevant for the detonation products of these explosives.

Although old, the initial properties in \hyperref[tab:Tab5]{Table \ref{tab:Tab5}} are accurate and still referred to, e.g. \cite{{SheffieldEtAl2001},{ZhangEtAl2002}} for IPN. For liquids, they can vary slowly over time and from one batch to another, which may explain the small differences between authors for NM. Here, no references ensure that the detonation measurements with the same batch of explosive were made over short enough periods. Four data sets for NM -- $I$, $II$, $III$, $IV$ -- were used to statistically assess the sensitivity of the calculations to small differences in these properties.
For NM-$II$, they were taken from Brochet and Fisson \cite{BrochetFisson1969}, and for NM-$I$ from Davis et al. \cite{DavisEtAl1965} except for $c_0$, taken from \cite{BrochetFisson1969}.
For NM-$III$, they are those from Lysne and Hardesty \cite{LysneHardesty1973} except for $C_{p_0}$, calculated with the fit $C_{p_0}${\small (J/kg/K)}$=1720.9+0.54724\times T_0${\small (C)} of Jones and Giauque's measurements \cite{JonesGiauque1947} between the melting ($245$ K) and ambient ($298$ K) temperatures.
For NM-$IV$, $\rho_0$ and $\alpha_0$ were calculated with the fit $\rho_0$ {\small (kg/m$^{-3}$)}$=1152.0-1.1395\times T_0${\small (C)}$-1.665\times 10^{-3}\times T_0^{2}${\small (C)} from Berman and West \cite{BermanWest1967}.
For IPN, the data were taken from \cite{BrochetFisson1969}, for NPNA3 from Bernard et al. \cite{BernardEtAl1966}, and for LTNT from \cite{DavisEtAl1965} and \cite{Garn1960} except for $c_0$, taken as the constant $a$ of the asymptote $D=a+bu$ to Garn's shock Hugoniot measurements \cite{Garn1959}.

The experimental velocities and pressures for NM-$I$ and LTNT were taken from \cite{DavisEtAl1965}, and for NM-$II$, $III$, $IV$ and IPN from \cite{BrochetFisson1969}. For NPNA3, $D_{\text{CJ}}^{\textit{\tiny{EXP}}}$ was taken from \cite{BernardEtAl1966}, but $p_{\text{CJ}}^{\textit{\tiny{EXP}}}$ was not found. The pairs of independent partial derivatives of $D_{\text{CJ}}^{\textit{\tiny{EXP}}}$ required to implement the Inverse Method were found only for NM and IPN. Tables \hyperref[tab:Tab6]{\ref{tab:Tab6}} and \hyperref[tab:Tab7]{\ref{tab:Tab7}-left} give those of $D_{\text{CJ}}^{\textit{\tiny{EXP}}}\left(T_0,p_0\right)$ for NM and IPN \cite{{BrochetFisson1969},{DavisEtAl1965}}. \hyperref[tab:Tab7]{Table \ref{tab:Tab7}-right} gives those of $D_{\text{CJ}}^{\textit{\tiny{EXP}}}\left(T_0,w_0\right)$ at constant $p_0$ for NM obtained from isometric mixtures of NM and acenina at mass fractions $w_0$. Acenina is the equimolar mixture of methyl cyanide (\ce{CH3CN}), nitric acid (\ce{HNO3}) and water (\ce{H2O}), so its atomic composition is identical to that of NM (\ce{CH3NO2}) \cite{DavisEtAl1965}. Only the derivative $\partial D_{\text{CJ}}^{\textit{\tiny{EXP}}}/\partial T_0)_{p_0}$ was found for LTNT and NPNA3 \hyperref[tab:Tab5]{(Tab. \ref{tab:Tab8})} \cite{{DavisEtAl1965},{BernardEtAl1966}}.

For NM, \hyperref[tab:Tab5]{Table \ref{tab:Tab5}} shows a low sensitivity of the theoretical pressure $p_{\text{CJ}}^{\textit{\tiny{THEO}}}$ to the initial properties, in contrast to its IM pressure $p_{\text{CJ}}^{\textit{\tiny{IM}}}$ calculated from $D_{\text{CJ}}^{\textit{\tiny{EXP}}}\left(T_0,p_0\right)$ and its derivatives. \hyperref[tab:Tab6]{Table \ref{tab:Tab6}} also shows a high sensitivity of $p_{\text{CJ}}^{\textit{\tiny{IM}}}$ to these derivatives. \hyperref[tab:Tab7]{Table \ref{tab:Tab7}-left} shows the same for IPN, that to $\partial D_{\text{CJ}}^{\textit{\tiny{EXP}}}/\partial p_0)_{T_0}$ appearing higher because its uncertainty interval is 10 times that of NM. \hyperref[tab:Tab7]{Table \ref{tab:Tab7}-right} shows that the $p_{\text{CJ}}^{\textit{\tiny{IM}}}$ value for NM-$I$ calculated from $D_{\text{CJ}}^{\textit{\tiny{EXP}}}(T_0,w_0)$ agrees well with the measured value $12.7$ GPa \hyperref[tab:Tab5]{(Tab. \ref{tab:Tab5})}, with a weak sensitivity to any of the derivatives of $D_{\text{CJ}}^{\textit{\tiny{EXP}}}(T_0,w_0)$ and -- not shown for brevity~-- to those of $\omega_0$ (\ref{dv0_w0_T0}), $\mu_0$ (\ref{dh0_w0_T0}), and the initial properties.

Both IM options give the right magnitude, but the uncertainty in the derivatives of $D_{\text{CJ}}^{\textit{\tiny{EXP}}}$ makes it difficult to consider one as more reliable. For example, the $\simeq 10 \%$ range of those of $D_{\text{CJ}}^{\textit{\tiny{EXP}}}(T_0,p_0)$ for NM \hyperref[tab:Tab6]{(Tab. \ref{tab:Tab6})} results in a $\simeq 30 \%$ range of $p_{\text{CJ}}^{\textit{\tiny{IM}}}$ which almost includes both $p_{\text{CJ}}^{\textit{\tiny{EXP}}}$ and $p_{\text{CJ}}^{\textit{\tiny{THEO}}}$. For IPN \hyperref[tab:Tab7]{(Tab. \ref{tab:Tab7}-left)}, the uncertainty in $\partial D_{\text{CJ}}^{\textit{\tiny{EXP}}}/\partial p_0)_{T_0}$ is too large.
Similarly, the common derivative $\partial D_{\text{CJ}}^{\textit{\tiny{EXP}}}/\partial T_0)_{p_0}$ to  $D_{\text{CJ}}^{\textit{\tiny{EXP}}}\left(T_0,p_0\right)$ and $D_{\text{CJ}}^{\textit{\tiny{EXP}}}\left(T_0,w_0\right)$ can be used to calculate $\partial D_{\text{CJ}}/\partial p_0)_{T_0}$. \hyperref[tab:Tab8]{Table  \ref{tab:Tab8}} compares its values iteratively adjusted so that $p_{\text{CJ}}^{\textit{\tiny{IM}}}=p_{\text{CJ}}^{\textit{\tiny{EXP}}}$ using the IM relation (\ref{IVM_VCJ}) with those calculated so that $p_{\text{CJ}}^{\textit{\tiny{IM}}}=p_{\text{CJ}}^{\textit{\tiny{THEO}}}$ using the additional constraint (\ref{PROP_DERIVS_DCJ_p0}). For NM-$I$, the value adjusted with $p_{\text{CJ}}^{\textit{\tiny{EXP}}}=12.7$ GPa is about equal to the mean value $2.0$ m/s/MPa, and the value calculated with $p_{\text{CJ}}^{\textit{\tiny{THEO}}}$ is only slightly less than its lower limit \hyperref[tab:Tab6]{(Tab. \ref{tab:Tab6})}. They differ by $0.25$ m/s/MPa ($\simeq 12\%$) which only slightly exceeds the uncertainty range of $0.2$ m/s/MPa ($\simeq 10\%$). For IPN, the two values are well within the uncertainty range of $2$ m/s/MPa and differ by $0.44$ m/s/MPa ($\simeq 13\%$). For LTNT, $\partial D_{\text{CJ}}^{\textit{\tiny{EXP}}}/\partial p_0)_{T_0}$ is unknown, but the two values differ by only $0.048$ m/s/MPa ($\simeq 5 \%$) which is smaller than the uncertainty range of $\partial D_{\text{CJ}}^{\textit{\tiny{EXP}}}/\partial p_0)_{T_0}$  for NM-$I$. For NPNA3, the experimental values of $\partial D_{\text{CJ}}/\partial p_0)_{T_0}$ and $p_{\text{CJ}}$ are unknown, so the calculated values are indicative.

\begin{table*}[t!]
\caption{Comparison of detonation pressures and adiabatic exponents at $p_0=1$ bar for nitromethane (NM), isopropyl nitrate (IPN), liquid trinitrotoluene (LTNT), and NPNA3. Symbol $\emptyset$: not enough or no data. \textit{\footnotesize{EXP}}:
experiments. \textit{\footnotesize{THEO}}: additional CJ properties. \textit{\footnotesize{IM}}: Inverse Method with $D_{\text{CJ}}^{\textit{\tiny{EXP}}}(T_0,p_0)$ and its medium uncertainty range derivatives, \hyperref[tab:Tab6]{cf. Tab. \ref{tab:Tab6}}.}
\label{tab:Tab5}
\begin{ruledtabular}
\begin{tabular}{lccccccccc}
&$T_0$ & $\rho _0$ & $\alpha _0$ & $C_{p_0}$ & $c_0$ & $G_0$ & $D_{\text{CJ}}^{\textit{\tiny{EXP}}}$
&\begin{tabular}{c}
$p_{\text{CJ}}$ {\small (GPa)}
\end{tabular}
&\begin{tabular}{c}
$\gamma_{\text{CJ}}$
\end{tabular}
\\
& {\small (C)} & {\small (kg/m$^{3}$)} & {\small (1/K)} & {\small (J/kg/K)} & {\small (m/s)} & & {\small (m/s)}
&\begin{tabular}{rrr}\hline
\ \ \textit{\scriptsize{EXP}} & \textit{\scriptsize{IM}} & \textit{\scriptsize{THEO}}
\end{tabular}
&\begin{tabular}{rrr}\hline
\ \ \textit{\scriptsize{EXP}} & \textit{\scriptsize{IM}} & \textit{\scriptsize{THEO}}
\end{tabular}
\\ \hline
NM $I$ & $4$ & $1159$ & $1.16\times10^{-3}$ & $1733$ & $1423$ & $1.36$ & $6334$
&\begin{tabular}{ccc}
$14.8$ & $13.2$ & $17.6$
\end{tabular}
&\begin{tabular}{ccc}
$2.14$ & $2.53$ & $1.65$
\end{tabular}
\\
\quad \quad $II$ & $4$ & $1156$ & $1.19\times10^{-3}$ & $1747$ & $1423$ & $1.38$ & $6330$
&\begin{tabular}{ccc}
$12.7$ & $12.9$ & $17.5$
\end{tabular}
&\begin{tabular}{ccc}
$2.65$ & $2.58$ & $1.65$
\end{tabular}
\\
\quad \quad $III$ & $4$ & $1151$ & $1.22\times10^{-3}$ & $1723$ & $1400$ & $1.39$ & $6330$
&\begin{tabular}{ccc}
$12.7$ & $13.6$ & $17.4$
\end{tabular}
&\begin{tabular}{ccc}
$2.63$ & $2.39$ & $1.65$
\end{tabular}
\\
\quad \quad $IV$ & $4$ & $1147$ & $1.00\times10^{-3}$ & $1723$ & $1400$ & $1.14$ & $6330$
&\begin{tabular}{ccc}
$12.7$ & $15.8$ & $17.9$
\end{tabular}
&\begin{tabular}{ccc}
$2.62$ & $1.91$ & $1.57$
\end{tabular}
\\ \hline
IPN & $40$ & $1017$ & $1.23\times10^{-3}$ & $1867$ & $1049$ & $0.72$ & $5330$
&\begin{tabular}{ccc}
$08.7$ & $13.1$ & $12.1$
\end{tabular}
&\begin{tabular}{ccc}
$2.32$ & $1.21$ & $1.40$
\end{tabular}
\\ \hline
LTNT & $93$ & $1450$ & $0.70\times10^{-3}$ & $1573$ & $2140$ & $2.04$ & $6590$
&\begin{tabular}{ccccc}
$18.2$&\,& $\emptyset $&\,  & $21.0$
\end{tabular}
&\begin{tabular}{ccccc}
$2.46$&\, & $\emptyset $&\,  & $2.00$
\end{tabular}
\\ \hline
NPNA3 & $25$ & $1275$ & $1.11\times10^{-3}$ & $1512$ & $1184$ & $1.03$ & $6670$
&\begin{tabular}{ccccc}
&$\emptyset $\ \ \ \ &\ & $\emptyset $\ \ \ & 
$22.8$
\end{tabular}
&\begin{tabular}{ccccc}
& $\emptyset $ &\ \ \ \ & $\emptyset $\ \ \ & 
$1.49$
\end{tabular}
\end{tabular}
\end{ruledtabular}

\end{table*}
\begin{table*}[t!]
\caption{Sensitivity of the Inverse Method pressures $p_{\text{CJ}}^{\textit{\tiny{IM}}}$\ (GPa) and adiabatic exponents $\gamma_{\text{CJ}}^{\textit{\tiny{IM}}}$ to the uncertainties of the initial data and the derivatives of measured detonation velocities $D_{\text{CJ}}^{\textit{\tiny{EXP}}}(T_0,p_0)$ for nitromethane (NM, \hyperref[tab:Tab5]{Tab. \ref{tab:Tab5}}) at $p_0=1$ bar. Symbol $\emptyset$: no solution to (\ref{IVM_VCJ}).}
\label{tab:Tab6}
\begin{ruledtabular}
\begin{tabular}{l cc cc}

\multicolumn{2}{l}{$\left. \partial D_{\text{CJ}}^{\textit{\tiny{EXP}}}/\partial
p_0\right)_{T_0}$}
&\multicolumn{3}{c}{$\left. \partial D_{\text{CJ}}^{\textit{\tiny{EXP}}}/\partial T_0\right)_{p_0}%
\pm 0.18\;\text{ (m/s/K)}$} \vspace{0.1cm} \\ \cline{3-5}
\multicolumn{2}{l}{$\pm 0.1$ {\small (m/s/MPa)}} &  \multicolumn{1}{c}{$\,\,-4.14\,\,$} & \multicolumn{1}{c}{$\,\,-3.96\,\,$} & \multicolumn{1}{c}{$\,\,-3.78\,\,$} \\ \hline

\begin{tabular}{c}
$ $\\
\quad$1.9$
\end{tabular}&
\begin{tabular}{c}
$ $\\
$p_{\text{CJ}}^{\textit{\tiny{IM}}}$ \\ 
$\gamma_{\text{CJ}}^{\textit{\tiny{IM}}}$
\end{tabular}&

\begin{tabular}{cccc}
 $I$ & $II$ & $III$ & $IV$ \\ \hline
 $\,\,16.5\,\,$ & $\,\,16.1\,\,$ & $\,\,17.8\,\,$ & $\,\,\,\,\,\,\emptyset\,\,\,\,\,\,$ \\
 $\,\,1.81\,\,$ & $\,\,1.87\,\,$ & $\,\,1.59\,\,$ & $\,\,\,\,\,\,\emptyset\,\,\,\,\,\,$ \\ \hline
\end{tabular}&

\begin{tabular}{cccc}
 $I$ & $II$ & $III$ & $IV$ \\ \hline
  $\,\,14.7\,\,$ & $\,\,14.4\,\,$ & $\,\,15.4\,\,$ & $\,\,19.0\,\,$ \\
  $\,\,2.17\,\,$ & $\,\,2.22\,\,$ & $\,\,2.00\,\,$ & $\,\,1.41\,\,$ \\ \hline
\end{tabular}&

\begin{tabular}{cccc}
 $I$ & $II$ & $III$ & $IV$ \\ \hline
  $\,\,13.4\,\,$ & $\,\,13.2\,\,$ & $\,\,13.9\,\,$ & $\,\,16.1\,\,$ \\
  $\,\,2.46\,\,$ & $\,\,2.50\,\,$ & $\,\,2.32\,\,$ & $\,\,1.85\,\,$ \\ \hline
\end{tabular}\\  \cline{1-2} 

\quad$\,2.0$&
\begin{tabular}{c}
$p_{\text{CJ}}^{\textit{\tiny{IM}}}$ \\ 
$\gamma _{\text{CJ}}^{\textit{\tiny{IM}}}$
\end{tabular}&

\begin{tabular}{cccc}
  $\,\,14.3\,\,$ & $\,\,14.0\,\,$ & $\,\,15.0\,\,$ & $\,\,18.5\,\,$ \\
  $\,\,2.25\,\,$ & $\,\,2.30\,\,$ & $\,\,2.08\,\,$ & $\,\,1.49\,\,$ \\ \hline
\end{tabular}&

\begin{tabular}{cccc}
  $\,\,13.2\,\,$ & $\,\,12.9\,\,$ & $\,\,13.6\,\,$ & $\,\,15.8\,\,$ \\
  $\,\,2.53\,\,$ & $\,\,2.58\,\,$ & $\,\,2.39\,\,$ & $\,\,1.91\,\,$ \\ \hline
\end{tabular}&

\begin{tabular}{cccc}
  $\,\,12.3\,\,$ & $\,\,12.1\,\,$ & $\,\,12.6\,\,$ & $\,\,14.3\,\,$ \\
  $\,\,2.79\,\,$ & $\,\,2.83\,\,$ & $\,\,2.66\,\,$ & $\,\,2.22\,\,$ \\ \hline
\end{tabular}\\  \cline{1-2} 

\quad$\,2.1$&
\begin{tabular}{c}
$p_{\text{CJ}}^{\textit{\tiny{IM}}}$ \\ 
$\gamma _{\text{CJ}}^{\textit{\tiny{IM}}}$
\end{tabular}&

\begin{tabular}{cccc}
  $\,\,12.9\,\,$ & $\,\,12.7\,\,$ & $\,\,13.3\,\,$ & $\,\,15.5\,\,$ \\
  $\,\,2.61\,\,$ & $\,\,2.65\,\,$ & $\,\,2.46\,\,$ & $\,\,1.96\,\,$
\end{tabular}&

\begin{tabular}{cccc}
  $\,\,12.1\,\,$ & $\,\,11.9\,\,$ & $\,\,12.4\,\,$ & $\,\,14.0\,\,$ \\ 
  $\,\,2.85\,\,$ & $\,\,2.90\,\,$ & $\,\,2.72\,\,$ & $\,\,2.28\,\,$
\end{tabular}&

\begin{tabular}{cccc}
 $\,\,11.4\,\,$ & $\,\,11.3\,\,$ & $\,\,11.6\,\,$ & $\,\,13.0\,\,$ \\
 $\,\,3.08\,\,$ & $\,\,3.12\,\,$ & $\,\,2.96\,\,$ & $\,\,2.54\,\,$
\end{tabular}
 
\end{tabular}
\end{ruledtabular}

\end{table*}
\begin{table*}[t!]
\caption{Sensitivity of the Inverse Method pressures $p_{\text{CJ}}^{\textit{\tiny{IM}}}$\ (GPa) and adiabatic exponents $\gamma_{\text{CJ}}^{\textit{\tiny{IM}}}$\ to the uncertainties of derivatives of experimental detonation velocities. Left: isopropyl nitrate (IPN, \hyperref[tab:Tab5]{Tab. \ref{tab:Tab5}}) with $D_{\text{CJ}}^{\textit{\tiny{EXP}}}(T_0,p_0)$. Right: nitromethane (NM-$I$, \hyperref[tab:Tab5]{Tab. \ref{tab:Tab5}}) with $D_{\text{CJ}}^{\textit{\tiny{EXP}}}(T_0,w_0)$ for acenina mass fraction $w_0=0$, $\mu_0\equiv\left.\partial h_0/\partial w_0\right) _{T_0,p_0}=$ $-2.021\pm 0.17$ MJ/kg, $\omega_0\equiv\left. \partial v_0/\partial w_0\right)_{T_0,p_0}=1.5\pm 0.2 \times 10^{-5}$ m$^{3}$/kg. Symbol $\emptyset$: no solution to (\ref{IVM_VCJ}).}
\label{tab:Tab7}
\begin{ruledtabular}
\begin{tabular}{cccc cccc}
\multicolumn{4}{c}{
\begin{tabular}{lll lll}

\multicolumn{1}{l}{$\left.\partial D_{\text{CJ}}^{\textit{\tiny{EXP}}}/\partial
p_0\right)_{T_0}$}&
\multicolumn{5}{r}{\quad \quad$\left. \partial D_{\text{CJ}}^{\textit{\tiny{EXP}}}/\partial T_0\right) _{p_0}%
\pm 0.10 \text{ (m/s/K)}$} \vspace{0.1cm} \\ \cline{3-6}
$\pm 1.0 \text{ (m/s/MPa)}$ &  &  &\quad \quad $-4.13$ & $-4.03$ & $-3.93$ \\ \hline

\quad\quad$2$&
\begin{tabular}{l}
$p_{\text{CJ}}^{\textit{\tiny{IM}}}$ \\ 
$\gamma_{\text{CJ}}^{\textit{\tiny{IM}}}$%
\end{tabular}& &\quad \quad 
\begin{tabular}{l}
\,\,\,\,\,\,\,$\emptyset$\\
\,\,\,\,\,\,\,$\emptyset$
\end{tabular}&
\begin{tabular}{l}
\,\,\,\,\,\,\,$\emptyset$\\
\,\,\,\,\,\,\,$\emptyset$
\end{tabular}&
\begin{tabular}{l}
\,\,\,\,\,\,\,$\emptyset$\\
\,\,\,\,\,\,\,$\emptyset$
\end{tabular}\\  \cline{1-3} \cline{4-6}

\quad\quad$3$&
\begin{tabular}{l}
$p_{\text{CJ}}^{\textit{\tiny{IM}}}$ \\ 
$\gamma_{\text{CJ}}^{\textit{\tiny{IM}}}$%
\end{tabular}& &\quad \quad
\begin{tabular}{l}
\,\,\,$\emph{15.9}$\\
\,\,\,$<\emph{1}$
\end{tabular}&
\begin{tabular}{l}
\,\,\,$13.1$\\
\,\,\,$1.21$
\end{tabular}&
\begin{tabular}{l}
\,\,\,$11.8$\\
\,\,\,$1.45$
\end{tabular}\\  \cline{1-3} \cline{4-6}

\quad\quad$4$&
\begin{tabular}{l}
$p_{\text{CJ}}^{\textit{\tiny{IM}}}$ \\ 
$\gamma_{\text{CJ}}^{\textit{\tiny{IM}}}$%
\end{tabular}& &\quad \quad 
\begin{tabular}{l}
\,\,\,$7.3$\\
\,\,\,$2.94$
\end{tabular}&
\begin{tabular}{l}
\,\,\,$7.2$\\
\,\,\,$3.03$
\end{tabular}&
\begin{tabular}{l}
\,\,\,$7.0$\\
\,\,\,$3.12$
\end{tabular}\\
 
\end{tabular}
}
&
\multicolumn{3}{c}{
\quad\quad\begin{tabular}{lll lll} 

\multicolumn{1}{l}{$\left. \partial D^{2\textit{\tiny{\,EXP}}}_{\text{CJ}}/\partial
w_0\right)_{T_0,p_0}$}&
\multicolumn{5}{c}{\quad $\left. \partial D_{\text{CJ}}^{\textit{\tiny{EXP}}}/\partial T_0\right)_{w_0,p_0}\pm 0.18 \text{ (m/s/K)}$} \vspace{0.1cm} \\ \cline{3-6}
$\pm 0.18$ {\small (mm/$\mu$s)}$^2$ &  &  &\quad $-4.14$ & $-3.96$ & $-3.78$ \\ \hline

\quad\quad$-8.16$&
\begin{tabular}{l}
$p_{\text{CJ}}^{\textit{\tiny{IM}}}$ \\ 
$\gamma_{\text{CJ}}^{\textit{\tiny{IM}}}$%
\end{tabular} & &\quad
\begin{tabular}{l} 
\,\,\,$12.4$\\
\,\,\,$2.74$
\end{tabular}&
\begin{tabular}{l}
\,\,\,$12.6$\\
\,\,\,$2.70$
\end{tabular}&
\begin{tabular}{l}
\,\,\,$12.7$\\
\,\,\,$2.66$
\end{tabular}\\  \cline{1-3} \cline{4-6}

\quad\quad$-7.98$&
\begin{tabular}{l}
$p_{\text{CJ}}^{\textit{\tiny{IM}}}$ \\ 
$\gamma _{\text{CJ}}^{\textit{\tiny{IM}}}$%
\end{tabular} & &\quad
\begin{tabular}{l}
\,\,\,$12.5$\\
\,\,\,$2.73$
\end{tabular}&
\begin{tabular}{l}
\,\,\,$12.6$\\
\,\,\,$2.69$
\end{tabular}&
\begin{tabular}{l}
\,\,\,$12.7$\\
\,\,\,$2.65$
\end{tabular}\\  \cline{1-3} \cline{4-6}

\quad\quad$-7.80$&
\begin{tabular}{l}
$p_{\text{CJ}}^{\textit{\tiny{IM}}}$ \\ 
$\gamma _{\text{CJ}}^{\textit{\tiny{IM}}}$%
\end{tabular} & &\quad
\begin{tabular}{c}
\,\,\,$12.5$\\
\,\,\,$2.72$
\end{tabular}&
\begin{tabular}{l}
\,\,\,$12.6$\\
\,\,\,$2.67$
\end{tabular}&
\begin{tabular}{l}
\,\,\,$12.8$\\
\,\,\,$2.63$
\end{tabular}\\
 
\end{tabular}
}
\\
\end{tabular}
\end{ruledtabular}

\end{table*}
\begin{table}[t!]
\caption{Values of $\left.\partial D_{\text{CJ}}/\partial p_0\right)_{T_0}$ obtained from $\left.\partial D_{\text{CJ}}^{\textit{\tiny{EXP}}}/\partial T_0\right)_{p_0}$ so that $p_{\text{CJ}}^{\textit{\tiny{IM}}}$ matches $ p_{\text{CJ}}^{\textit{\tiny{THEO}}}$ or $ p_{\text{CJ}}^{\textit{\tiny{EXP}}}$.}
\label{tab:Tab8}
\begin{ruledtabular}
\begin{tabular}{lccccc}
&$\left.\partial D_{\text{CJ}}^{\textit{\tiny{EXP}}}/\partial T_0\right)_{p_0}$ & & \!\!\!\!$\left.\partial D_{\text{CJ}}/\partial p_0\right)_{T_0}$& $p_{\text{CJ}}$ & $\gamma_{\text{CJ}}$
\\
& {\small (m/s/K)} & & {\small (m/s/MPa)} & {\small (GPa)}
\\ \hline
NM $I$ & $-3.96$
&
\begin{tabular}{r}
\textit{\scriptsize{THEO}} \\
\textit{\scriptsize{EXP}}
\end{tabular}
&
\begin{tabular}{l}
$1.788$\\
$2.039$
\end{tabular}
&
\begin{tabular}{l}
$17.550$\\
$12.696$
\end{tabular}
&
\begin{tabular}{l}
$1.649$\\
$2.663$
\end{tabular}
\\
\hline
IPN & $-4.03$
&
\begin{tabular}{r}
\textit{\scriptsize{THEO}} \\
\textit{\scriptsize{EXP}}
\end{tabular}
&
\begin{tabular}{l}
$3.058$\\
$3.502$
\end{tabular}
&
\begin{tabular}{l}
$12.059$\\
\, $8.699$
\end{tabular}
&
\begin{tabular}{l}
$1.396$\\
$2.321$
\end{tabular}
\\ \hline
LTNT & $-3.50$
&
\begin{tabular}{r}
\textit{\scriptsize{THEO}} \\
\textit{\scriptsize{EXP}}
\end{tabular}
&
\begin{tabular}{l}
$0.908$\\
$0.955$
\end{tabular}
&
\begin{tabular}{l}
$20.963$\\
$18.215$
\end{tabular}
&
\begin{tabular}{l}
$2.004$\\
$2.457$
\end{tabular}
\\ \hline
NPNA3 & $-5.60$ & \textit{\scriptsize{THEO}} &$3.333$ & $22.806$ & $1.487$
\end{tabular}
\end{ruledtabular}
\end{table}


The paucity and imprecision of the partial derivatives of $D_{\text{CJ}}^{\textit{\tiny{EXP}}}(T_0,p_0)$ or $D_{\text{CJ}}^{\textit{\tiny{EXP}}}(T_0,w_0)$ make it difficult for the IM to predict accurately CJ pressures of condensed explosives. Davis et al. \cite{{DavisEtAl1965},{FickDav2000}} questioned the CJ hypothesis for liquid explosives because their IM pressure $12.6$ GPa for NM obtained from $D_{\text{CJ}}^{\textit{\tiny{EXP}}}(T_0,w_0)$ measurements was lower than their experimental estimate of $14.8$ GPa (\hyperref[tab:Tab5]{Tab. \ref{tab:Tab5}}), which Petrone \cite{Petrone1968} considered too large. Davis \cite{Davis1981} then carefully analyzed the issue of sensitivity to the velocity derivatives and their coefficients in (\ref{IVM_L})-(\ref{IVM_K}) and (\ref{L_w0_T0})-(\ref{K_w0_T0}). The theoretical pressures $p_{\text{CJ}}^{\textit{\tiny{THEO}}}$ are calculated from the only value of $D_{\text{CJ}}^{\textit{\tiny{EXP}}}$ and should therefore be more accurate than the IM pressures $p_{\text{CJ}}^{\textit{\tiny{IM}}}$. However, they exceed the experimental values $p_{\text{CJ}}^{\textit{\tiny{EXP}}}$ by more than the typical uncertainty of $\pm 2 $~GPa.

Therefore, data and measurement uncertainties cannot support or refute the theorem despite the fair agreement of $p_{\text{CJ}}^{\textit{\tiny{THEO}}}$ and $p_{\text{CJ}}^{\textit{\tiny{EXP}}}$ for LTNT. That suggests instead discussing the physics of the observations and the assumptions of the hydrodynamic framework since these are common to the theorem and the IM \hyperref[sec:Remind]{(Sect. \ref{sec:Remind})}.

Experiments with condensed explosives are usually performed in finite-diameter tubes, which produces sonic-frozen regimes of curved detonation \hyperref[sec:CJPos]{(Sect. \ref{sec:CJPos})}. This diameter effect results from the very high pressures -- $\sim \mathcal{O}\left( 10\right)$ GPa -- which expand the tube, and hence the reaction zone, transversely to the direction of propagation, e.g. \cite{{WoodKirkwood1954},{Bdzil1981},{ChiqueteShort2019}}. The measured velocities and pressures are often lower than the CJ values because of incomplete chemical reactions at the frozen sonic locus. Reaching the planar limit described by the TZD expansion \hyperref[subsec:InVarPb]{(Subsect. \ref{subsec:InVarPb})} and the ZND planar reaction zone \cite{Higgins2012} requires large-diameter tubes. Linear extrapolation of measured velocities to infinite diameters, e.g. \hyperref[tab:Tab5]{Tab. \ref{tab:Tab5}}, may underestimate the equilibrium $D_{\text{CJ}}^{\textit{\tiny{EXP}}}$ because of the possible convexity of the velocity-diameter dependence at large diameters.

Sharpe's numerical simulations of ignition by an overdriven detonation \cite{Sharpe2000} show that, in the long-time limit, a stable reaction zone reaches either the equilibrium or a sonic-frozen CJ state depending on whether the system geometry is planar or not. Rapid pressure and temperature drops beyond the reaction zone, i.e. short run distances, certainly freeze the chemical equilibrium. In the TZD (planar) flow \hyperref[subsec:InVarPb]{(Subsect. \ref{subsec:InVarPb})}, the derivatives tend to zero with increasing detonation run distance, as do the physical ZND derivatives with decreasing distance to the end of the reaction zone. In systems of hyperbolic differential equations, e.g. the Euler equations, the derivatives are discontinuous through sonic loci. Thus, in an inviscid reactive flow, a sonic-frozen interface separates an expansion and an incomplete reaction zone. Therefore, large tubes should also be long enough for the reaction processes to reach chemical equilibrium and for that equilibrium to shift in the expanding products. However, the longer they are, the less detectable the derivative jumps become at the sonic locus. A slope discontinuity on a measured plot of pressure or material speed may not be the CJ-equilibrium locus separating the ZND and TZD flows and may be difficult to extract from the signal noise.

Incomplete reactions due to a weak divergence of the detonation zone can result here from a two-step decomposition of the \ce{NO2} grouping common to the four explosives of this analysis. In the compact semi-developed form, NM writes: \ce{CH3(NO2)}, IPN: \ce{(CH3)2(H)CO(NO2)}, LTNT: \ce{C6H2(CH3)(NO2)3}, NP: \ce{C(CH3)2(H)(NO2)}, and NA: \ce{O(H)(NO2)}, so NPNA3 contains 4 \ce{NO2} groupings per volume of NP. In gases, \ce{NO2} first decomposes into \ce{NO} which then decomposes into \ce{N2} (cf. refs. in \cite{DesbordesPresles2012}). Parker and Wolfhard \cite{ParkerWolfhard1953} and Branch et al. \cite{BranchEtal1991} observed a two-front laminar flame in \ce{CH4 /NO2 /O2} and \ce{CH2O/NO2 /O2} mixtures on a flat burner. Presles et al. \cite{PreslesEtal1996} identified a two-level cellular structure of detonation \hyperref[sec:CJPos]{(Sect. \ref{sec:CJPos})} in gaseous NM, where the transverse waves of the smaller cells propagate on the fronts of the larger ones. The first step gives the lower flame front and the smaller detonation cells. Whether this applies to liquids is uncertain, but the divergence of the detonation zone may slow the reaction sufficiently for the expansion head to enter the reaction zone and set at the position of the intermediate step \hyperref[sec:Remind]{(Sect. \ref{sec:Remind})}. Non-ideal detonation regimes resulting from multi-step heat release, possibly low-velocity with pressures well below the CJ values, are well-known phenomena of detonation dynamics.

Here, no arbitrary increase of the velocities in \hyperref[tab:Tab5]{Table V} can bring $p_{\text{CJ}}^{\textit{\tiny{IM}}}$ and $p_{\text{CJ}}^{\textit{\tiny{THEO}}}$ closer together, and, from $\Delta p/p \simeq 2 \Delta D/D$ (\ref{PCJ_PARAM}), the same with $p_{\text{CJ}}^{\textit{\tiny{EXP}}}$ and $p_{\text{CJ}}^{\textit{\tiny{THEO}}}$. Agreement is unlikely to be obtained for small increases in velocity $D_{\text{CJ}}^{\textit{\tiny{EXP}}}$, e.g. extrapolations from measurements in cylinders wider and longer than usual, or those defined from mean values and the typical measurement uncertainty of $\pm 10$ m/s.

Thus, neither the inaccuracy nor the representativeness of the measurements can explain the disagreement between $p_{\text{CJ}}^{\textit{\tiny{THEO}}}$ and $p_{\text{CJ}}^{\textit{\tiny{EXP}}}$. The theorem uses experimental detonation velocities to calculate detonation pressures under the usual assumptions of the hydrodynamics (Sects. \hyperref[sec:Remind]{\ref{sec:Remind}} and \hyperref[sec:DSITh]{\ref{sec:DSITh}}). If these velocities are accepted as CJ, and if the calculated pressures are too high compared with the measured pressures, then it is a legitimate conclusion that these assumptions are not suitable for the liquid explosives considered here.

For example, NM, LTNT and IPN have negative oxygen balances, which produces large amounts of carbon. Their reaction zones and expanding products should be described as non-homogeneous fluids using multiphase balance laws and constitutive relations suitable to processes such as carbon condensation \cite{{BergerViard1962},{Bastea2017},{BatraevEtal2018}}, bubble formation and thermal and mechanical non-equilibria. 
Measured velocities and pressures in highly carbonated gases have indeed lower values than those calculated with models of detonation products that do not include carbon condensation \cite{{Kistiakovski1952}, {Kistiakovski1955}, {Kistiakovski1956}}. The case of the stoichiometric composition NPNA3 cannot be discussed because no experimental pressure is available.

\section{\label{sec:Disc}Discussion and conclusions}
This work brings out two unnoticed properties of the CJ equilibrium model of detonation in the framework of classical hydrodynamics, i.e. the initial state and the detonation products of the explosive are single-phase fluids at thermal chemical equilibrium, with temperature $T$ and pressure $p$ as the independent variables. The first one is that the CJ velocity and specific entropy are invariant under the same variations of initial temperature $T_0$ with initial pressure $p_0$ \hyperref[subsec:Proof]{(Subsect. \ref{subsec:Proof})}. The second one is a set of additional relations, for example for calculating the CJ state, including the adiabatic exponent and the isentrope, from the only CJ velocity, or the CJ velocity from any one of the CJ variables \hyperref[subsec:CJsuppl]{(Subsect. \ref{subsec:CJsuppl})}, without an equation state for the detonation products.

They are no substitute for detailed thermochemical calculations which give not only the CJ state but also the velocity and the composition based on explicit equilibrium $(T,p)$ equations of state, e.g. that of the ideal gas in the NASA's CEA computer program \cite{GordonMcBride-I}, or BKW and JCZ3 developments or reparametrizations for condensed explosives \cite{{CowperthwaiteZwisler1976},{FriedSouers1996}}. 

That justifies asking what the gain is over the usual method of measuring a pair of variables, such as $p_{\text{CJ}}$ and $D_{\text{CJ}}$, to calibrate equations of state through numerical CJ calculations. If anything, this provides a semi-empirical criterion for discussing whether a given pair can represent the CJ equilibrium state and thus for improving the measurement conditions or modelling assumptions.

For ideal gases, the predicted CJ properties, given the CJ velocity, compare accurately with those from detailed chemical equilibrium calculations for detonation products that are ideal gases \hyperref[sec:Applic-Gas]{(Sect. \ref{sec:Applic-Gas})}. However, for four liquid carbon explosives, the predicted pressures are higher than the measured values \hyperref[sec:Applic-Liq]{(Sect. \ref{sec:Applic-Liq})}. The sensitivity analysis excludes the effect of uncertainties in the initial properties and the detonation velocity and pressure measurements. Since the theorem and the derived properties hinge on no particular form of equilibrium equations of state, the hydrodynamic framework, i.e. detonation products viewed as single-phase fluids at chemical equilibrium and relaxed degrees of freedom of the molecules, should be questioned for these explosives.




The relations (\ref{H_SP})-(\ref{S_PV}) are differentials of equations of state used here to represent initial-state variations at constant initial composition and the final-state variations resulting from these initial-state variations. Regarding the final state, the fact that these equations of state have two independent variables, e.g. $T$ and $p$, implies that the final chemical composition varies with $T$ and $p$ according to Gibbs's laws of chemical equilibrium \cite{GordonMcBride-I}. Since $T$ and $p$ vary with the initial state, these differentials are unsuitable for representing variations of final states whose chemical compositions would be in frozen equilibrium. There is no reason why different initial states should produce different final states but the same final composition. 
Therefore, the theorem applies only to CJ equilibrium detonations.

This could apply to any sufficiently rich carbon explosive, gaseous, liquid or solid, because of carbon condensation \cite{{Kistiakovski1952}, {Kistiakovski1955}, {Kistiakovski1956}, {BatraevEtal2018}} and, if liquid, the presence or formation of bubbles. The analysis would benefit from initial and detonation data for carbonless or stoichiometric carbon liquid explosives. One possibility is ammonium nitrate \ce{NH4NO3} above its melting temperature ($443$ K), another is the stoichiometric composition NPNA3 \hyperref[sec:Applic-Gas]{(Sect. \ref{sec:Applic-Gas})}. However, both raise safety issues, the first because of its metastability at elevated temperatures, the second because of its chemical aggressiveness and probable carcinogenicity.

These properties easily derive from the classical Rankine-Hugoniot relations embedded in the single-phase adiabatic Euler equations. Compressible multiphase fluid dynamics with thermal and mechanical non-equilibrium at elevated pressures and temperatures remains a theoretical and numerical challenge. Averaged balance laws and constitutive relationships derived from heat transfer and mixing rules are workarounds to fit into the single-phase paradigm. 
The additional CJ properties can be used to discuss the physics in these homogenization approaches and better combine experiments and models.


The theorem requires that no variation in the initial state induces infinite variations in the CJ state (Sects. \hyperref[sec:Remind]{\ref{sec:Remind}} and \hyperlink{addCJdet}{\ref{subsec:CJsuppl} §3}, \hyperref[sec:CJadm]{App. \ref{sec:CJadm}}). That is an analogue in the thermodynamic plane to the transonic condition in the physical space, i.e. finite derivatives at a sonic locus in a laminar flow of an inviscid fluid, such as the planar or weakly diverging ZND flows \cite{{WoodKirkwood1954},{Bdzil1981},{Higgins2012}}.
\newline



The Euler equations for inviscid fluids, completed with relevant constitutive relations, form a closed hyperbolic system. A data distribution on a non-characteristic side of a discontinuity defines a well-posed Cauchy problem without using entropy, e.g. \cite{{Vidal1999a},{Vidal1999b}}. The sonic side of the CJ front is an example of characteristic distribution.
The initial state and the velocity of the front give the distribution, or the initial state and one characteristic-state variable give the velocity of the front. Entropy was used in this analysis to obtain additional properties of the CJ sonic state without an equation of state. This may illustrate a general property of horizons in hyperbolic systems. 
The CJ locus is the sound-like separatrix between the time-like Taylor-Zel'dovich-D\"{o}ring expansion, where $u+c \leqslant D_\text{CJ}$ \hyperref[subsec:InVarPb]{(Subsect. \ref{subsec:InVarPb})}, and the space-like Zel'dovich-von Neuman-D\"{o}ring reaction zone, where $u +c \geqslant D_\text{CJ}$, i.e. a Cauchy event horizon in the TZD expansion for an observer in the ZND reaction zone, and vice versa. 
\newline

This work emphasizes that the strict mathematical framework of the inviscid homogeneous fluid has a limited capacity for physical representativeness.

\appendix
\newpage
\section{\label{sec:CJadm}\break Chapman-Jouguet admissibility}
The equilibrium expansion behind a CJ detonation front is homentropic and self-similar \hyperref[subsec:InVarPb]{(Subsect. \ref{subsec:InVarPb})}. The backward-facing Riemann invariant is uniform, that is, $du-\left(v/c\right) dp=0$, and, since $u_{\text{p}}<u_{\text{CJ}}$, the material speed $u$ (as well as $p$ and $v^{-1}$) and the frontward-facing disturbance velocity $u+c=x/t$ have to decrease from the CJ front so expansion can spread out. Differentiating $u+c$ and expressing $p$ and $c$ as functions of $s$ and $v$ give $\Gamma^{-1}d\left( u+c\right) = du= vdp/c=-cdv/v$ \cite{Thomson1971}, hence $\Gamma >0$. Similarly, $T$ decreases if $G>0$ (\ref{GRUN}).
The relations (\ref{R_SLP})-(\ref{H_SLP}), (\ref{DERIV_SR_V})-(\ref{DERIV_DH_V}), and (\ref{DIFF_MASS_MACH})-(\ref{DIFF_SOUND_SV}) give 
\begin{gather}
\left. \frac{\partial ^{2}p_{\text{H}}}{\partial v^{2}}\right) _{\text{CJ}}\!\!\!=%
\frac{2}{F_{\text{CJ}}}\left. \frac{\partial ^{2}p_{\text{S}}}{\partial v^{2}%
}\right) _{\text{CJ}}\!\!\!=\frac{4\Gamma _{\text{CJ}}}{F_{\text{CJ}}} \frac{%
D_{\text{CJ}}^{2}}{v_0^{3}}\frac{v_0}{v_{\text{CJ}}},  \label{CURVAT_H_S}\\
\!\!\!-G_{\text{CJ}}\left. \frac{\partial ^{2}s_{\text{R}}}{\partial v^{2}}\right)
_{\text{CJ}}\!\!\!=\frac{F_{\text{CJ}}}{\frac{v_0}{v_{\text{CJ}}}-1}\left. \frac{%
\partial ^{2}s_{\text{H}}}{\partial v^{2}}\right) _{\text{CJ}}\!\!\!=2\Gamma _{%
\text{CJ}} \frac{D_{\text{CJ}}^{2}}{v_0^{2}T_{\text{CJ}}},
\label{DIFF_SECOND_SRH} \\
\left. \frac{\partial M_{\text{H}}}{\partial v}\right) _{\text{CJ}}\!\!\!=\frac{%
\Gamma _{\text{CJ}}}{v_{\text{CJ}}}.  \label{DIFF_M_H}
\end{gather}
The curvatures of a Hugoniot and an isentrope thus have the same sign if $F_{\text{CJ}}>0$, that is, if $G_{\text{CJ}}<2/\left( v_0/v_{\text{CJ}}-1\right) $, that of the Hugoniot then being the larger if $G_{\text{CJ}}>0$, which is the case for most fluids. Also, $F_{\text{CJ}}\neq 0$ is the condition for finite Hugoniot curvature (\ref{CURVAT_H_S}) and entropy variations at a CJ point \hyperref[subsec:RHdiff]{(Subsect. \ref{subsec:RHdiff})} for physical isentropes ($\Gamma \neq 0$, \hyperlink{addCJdet}{(Subsect. \ref{subsec:CJsuppl}, §3)}
The derivative (\ref{DIFF_M_H}) of $M$ with respect to $v$ along a Hugoniot at a CJ point shows, since $\Gamma _{\text{CJ}}>0$, that $M<1$ above, and $M>1$ below, a CJ point, hence $F_{\text{CJ}}>0$, $\left. \partial ^{2}p_{\text{H}}/\partial v^{2}\right) _{\text{CJ}}>0$ and $\left. \partial ^{2}s_{\text{H}}/\partial v^{2}\right) _{\text{CJ}}>0$ from (\ref{CURVAT_H_S}) and (\ref{DIFF_SECOND_SRH}).
Therefore, a CJ detonation point is admissible only on a convex Hugoniot arc. 
Its physical branch is above the CJ point since $s$ increases and $M$ decreases with decreasing $v$. Other approaches use concavity of entropy $s\left( e,v\right) $ or 
convexity of energy $e\left( s,v\right)$.
\section{\label{sec:CJperf}\break Chapman-Jouguet relations for the perfect gas}
The perfect gas is the ideal gas with constant heat capacities $\bar{C}_{v}=\left( R/W\right) /\left( \bar{\gamma}-1\right)$ and $\bar{C}_{p}=\left( R/W\right) \bar{\gamma}/\left( \bar{\gamma}-1\right) 
$, with $W$ the molecular weight and $R=8.31451$ J/mol.K the gas constant. The adiabatic exponent $\gamma $ reduces to the constant ratio $\bar{\gamma}=\bar{C}_{p}/\bar{C}_{v}$, the Gruneisen coefficient $G$ to $\bar{\gamma}-1$, the
fundamental derivative $\Gamma $ to $\left( \bar{\gamma}+1\right) /2$, and an isentrope to $pv^{\bar{\gamma}}=$ const. Using $T\left( p,v\right) =\left( W/R\right) pv$, $dh$ (\ref{H_TP}) reduces to $dh\left( T\right) =C_{p}\left( T\right) dT$ whose integrals give the difference (\ref{PERF_HPV_EOS}) of enthalpies of the products at $\left( T,p\right) $ and the fresh gas at $\left(T_0,p_0\right)$ (neglecting the differences of their $W$ and $\bar{\gamma}$), and (\ref{HUGO}) then gives the Hugoniot (H) curve (\ref{PERF_HUGO}), i.e.

\begin{equation}
h\left( p,v\right) -h_0\left( p_0,v_0\right)=\frac{\bar{\gamma}\left(
pv-p_0v_0\right) }{\bar{\gamma}-1}-Q_0,  \label{PERF_HPV_EOS}
\end{equation}
\begin{equation}
p_{\text{H}}\left( v;v_0,p_0\right)=p_0\times \frac{1-\frac{\bar{%
\gamma}-1}{\bar{\gamma}+1}\left( \frac{v}{v_0}-\frac{2Q_0}{p_0v_0}%
\right) }{\frac{v}{v_0}-\frac{\bar{\gamma}-1}{\bar{\gamma}+1}}.
\label{PERF_HUGO}
\end{equation}
A CJ state is given by (\ref{VCJ_PARAM}) and (\ref{PCJ_PARAM}) with $\bar{\gamma}$ substituted for $\gamma _{\text{CJ}}$. A CJ velocity $D_{\text{CJ}}$ is a solution to the 2$^{nd}$ degree equation obtained by substituting $v_{\text{CJ}}$ (\ref{VCJ_PARAM}) and $p_{\text{CJ}}$ (\ref{PCJ_PARAM}) for $v$ and $p$ in (\ref{PERF_HUGO}). The supersonic compressive solution (subscript CJc, \hyperref[sec:Remind]{Sect. \ref{sec:Remind}}) is the CJ-detonation velocity $D_{\text{CJc}}\left(v_0,p_0\right)$,
\begin{align}
D_{\text{CJc}}&=\widetilde{D}_{\text{CJ}}\left(\frac{1}{2}+\widetilde{M}_{0\text{CJ}}^{-2}+\frac{1}{2}\sqrt{1+4\widetilde{M}%
_{0\text{CJ}}^{-2}}\right) ^{\frac{1}{2}},  \label{PERF_DCJ}
\\
\widetilde{D}_{\text{CJ}}^{2}&=2\left( \bar{\gamma}^{2}-1\right) Q_0,\quad 
\widetilde{M}_{0\text{CJ}}=\widetilde{D}_{\text{CJ}}/c_0, \label{PERF_DCJ_APPROX}
\end{align}
with dominant value $\widetilde{D}_{\text{CJ}}$ if $\widetilde{M}_{0\text{CJ}}^{-2}<<1$ and acoustic (non-reactive) limit $c_0$ ($Q_0=0$).
The subsonic expansive solution (subscript CJx) is the CJ-deflagration velocity $D_{\text{CJx}}$, deduced from $D_{\text{CJc}}$ by changing the sign before the square root in (\ref{PERF_DCJ}), hence the relation
\begin{equation}
D_{\text{CJc}}D_{\text{CJx}}=c_0^{2}\quad \text{or}\quad M_{0\text{CJc}}M_{0\text{CJx}}=1,  \label{DCJ_DET_DEFL}
\end{equation}
which had not been noticed before and shows that $D_{\text{CJx}}$ has dominant value $\widetilde{D}_{\text{CJ}}/\widetilde{M}_{0\text{CJ}}^{2}\equiv c_0/\widetilde{M}_{0\text{CJ}}$. One CJ state can be expressed with the other, 
\begin{equation}
\hspace{-0.25cm}\frac{p_{\text{CJx}}}{p_{\text{CJc}}}=\frac{M_{0\text{CJc}}^{-2}}{\bar{\gamma%
}}\frac{1+\bar{\gamma}M_{0\text{CJc}}^{-2}}{1+\bar{\gamma}^{-1}M_{0\text{CJc}%
}^{-2}},\frac{v_{\text{CJx}}}{v_{\text{CJc}}}=M_{0\text{CJc}}^{4}\frac{p_{\text{CJx}}}{p_{\text{CJc}}}. \label{CJ_DET_DEFL}
\end{equation}

There are two overdriven detonation solutions ($Q_0>0$, $D\geqslant D_{\text{CJc}}$, Fig. 2). Only the upper, U, is a physical intersect of a (R) line (\ref{RM}) and the (H) curve (\ref{PERF_HUGO}) (subsonic, $M<1$, \hyperref[sec:Remind]{Sect. \ref{sec:Remind}}). It writes

\begin{align}
\frac{v_0\left(p-p_0\right) }{D^{2}}=&1-\frac{v}{v_0}=\frac{1-M_0^{-2}-\sqrt{\Delta _{D}}}{\bar{\gamma}+1},
\label{PERF_PV_OVRDRV} \\
\Delta _{D}=&\left( 1-\left( \frac{D_{\text{CJc}}}{D}\right) ^{2}\right)\left( 1-\left( \frac{D_{\text{CJx}}}{D}\right) ^{2}\right) \notag \\
=&\left( 1-M_0^{-2}\right) ^{2}-\left( \frac{\widetilde{D}_{\text{CJ}}}{D}\right) ^{4}. \label{DeltaD}
\end{align}
The lower, L, is non-physical (supersonic, $M>1$). It is obtained by changing the sign before $\sqrt{\Delta _{D}}$
above. Both reduce to the shock solution N by setting $Q_0=0$, so $\sqrt{\Delta _{D}}=1-M_0^{-2}$. The theoretical CJ deflagration viewed as an adiabatic discontinuity with the same initial state as the CJ detonation is not admissible (subsonic, $M_{0\text{CJx}}<1$): (\ref{MACH}) is not satisfied \hyperref[sec:Remind]{(Sect. \ref{sec:Remind})}. It was useful here for completeness and because  the relations (\ref{PERF_PV_OVRDRV})-(\ref{DeltaD}) more obviously return the CJ relations (\ref{VCJ_PARAM})-(\ref{PCJ_PARAM}) if $\Delta_{D}=0$, i.e. $v_{\text{CJc}}$ and $p_{\text{CJc}}$ if $D=D_{\text{CJc}}$, and $v_{\text{CJx}}$ and $p_{\text{CJx}}$ if $D=D_{\text{CJx}}$. From (\ref{DCJ_DET_DEFL}), $\left( D_{\text{CJx}}/D\right) ^{2}=$ $\left( c_0^{2}/DD_{\text{CJc}}\right) ^{2}\leqslant M_{0\text{CJc}}^{-4}\ll 1$ that negligibly contributes to $\Delta _{D}$ compared to $\left( D_{\text{CJc}}/D\right)^{2}$. The typical values $c_0=300$ m/s and $D_{\text{CJc}}=2000$ m/s give $D_{\text{CJx}}=45$ m/s.
\newpage

\bibliography{DSI_2024_BIB}

\providecommand{\noopsort}[1]{}\providecommand{\singleletter}[1]{#1}
\begin{thebibliography}{75}%
\makeatletter
\providecommand \@ifxundefined [1]{%
 \@ifx{#1\undefined}
}%
\providecommand \@ifnum [1]{%
 \ifnum #1\expandafter \@firstoftwo
 \else \expandafter \@secondoftwo
 \fi
}%
\providecommand \@ifx [1]{%
 \ifx #1\expandafter \@firstoftwo
 \else \expandafter \@secondoftwo
 \fi
}%
\providecommand \natexlab [1]{#1}%
\providecommand \enquote  [1]{``#1''}%
\providecommand \bibnamefont  [1]{#1}%
\providecommand \bibfnamefont [1]{#1}%
\providecommand \citenamefont [1]{#1}%
\providecommand \href@noop [0]{\@secondoftwo}%
\providecommand \href [0]{\begingroup \@sanitize@url \@href}%
\providecommand \@href[1]{\@@startlink{#1}\@@href}%
\providecommand \@@href[1]{\endgroup#1\@@endlink}%
\providecommand \@sanitize@url [0]{\catcode `\\12\catcode `\$12\catcode `\&12\catcode `\#12\catcode `\^12\catcode `\_12\catcode `\%12\relax}%
\providecommand \@@startlink[1]{}%
\providecommand \@@endlink[0]{}%
\providecommand \url  [0]{\begingroup\@sanitize@url \@url }%
\providecommand \@url [1]{\endgroup\@href {#1}{\urlprefix }}%
\providecommand \urlprefix  [0]{URL }%
\providecommand \Eprint [0]{\href }%
\providecommand \doibase [0]{https://doi.org/}%
\providecommand \selectlanguage [0]{\@gobble}%
\providecommand \bibinfo  [0]{\@secondoftwo}%
\providecommand \bibfield  [0]{\@secondoftwo}%
\providecommand \translation [1]{[#1]}%
\providecommand \BibitemOpen [0]{}%
\providecommand \bibitemStop [0]{}%
\providecommand \bibitemNoStop [0]{.\EOS\space}%
\providecommand \EOS [0]{\spacefactor3000\relax}%
\providecommand \BibitemShut  [1]{\csname bibitem#1\endcsname}%
\let\auto@bib@innerbib\@empty
\bibitem [{\citenamefont {Jouguet}(1901)}]{Jouguet1901}%
  \BibitemOpen
  \bibfield  {author} {\bibinfo {author} {\bibfnamefont {E.}~\bibnamefont {Jouguet}},\ }\bibfield  {title} {\bibinfo {title} {Sur la propagation des discontinuit\'{e}s dans les fluides},\ }\href@noop {} {\bibfield  {journal} {\bibinfo  {journal} {C. R. Acad. Sci. Paris}\ }\textbf {\bibinfo {volume} {132}},\ \bibinfo {pages} {673} (\bibinfo {year} {1901})}\BibitemShut {NoStop}%
\bibitem [{\citenamefont {Jones}(1949)}]{Jones1949}%
  \BibitemOpen
  \bibfield  {author} {\bibinfo {author} {\bibfnamefont {H.}~\bibnamefont {Jones}},\ }\bibfield  {title} {\bibinfo {title} {The properties of gases at high pressures that can be deduced from explosion experiments},\ }in\ \href@noop {} {\emph {\bibinfo {booktitle} {3$^{\text{rd}}$ Symp. on Combustion, Flame and Explosion Phenomena}}}\ (\bibinfo  {publisher} {Williams and Wilkins, Baltimore},\ \bibinfo {year} {1949})\ pp.\ \bibinfo {pages} {590--594}\BibitemShut {NoStop}%
\bibitem [{\citenamefont {Stanyukovich}(1960)}]{Stanyuk1955}%
  \BibitemOpen
  \bibfield  {author} {\bibinfo {author} {\bibfnamefont {K.~P.}\ \bibnamefont {Stanyukovich}},\ }\href@noop {} {\emph {\bibinfo {title} {Non-stationary flows in continuous media}}}\ (\bibinfo  {publisher} {Pergamon, London (transl. State Publishers of Technical and Theoretical Literature, Moscow, 1955)},\ \bibinfo {year} {1960})\BibitemShut {NoStop}%
\bibitem [{\citenamefont {Manson}(1958{\natexlab{a}})}]{Manson1958a}%
  \BibitemOpen
  \bibfield  {author} {\bibinfo {author} {\bibfnamefont {N.}~\bibnamefont {Manson}},\ }\bibfield  {title} {\bibinfo {title} {Une nouvelle relation de la th\'{e}orie hydrodynamique des ondes de d\'{e}tonation},\ }\href@noop {} {\bibfield  {journal} {\bibinfo  {journal} {C. R. Acad. Sci. Paris}\ }\textbf {\bibinfo {volume} {246}},\ \bibinfo {pages} {2860} (\bibinfo {year} {1958}{\natexlab{a}})}\BibitemShut {NoStop}%
\bibitem [{\citenamefont {Vieille}(1900)}]{Vieille1900}%
  \BibitemOpen
  \bibfield  {author} {\bibinfo {author} {\bibfnamefont {P.}~\bibnamefont {Vieille}},\ }\bibfield  {title} {\bibinfo {title} {R\^{o}le des discontinuit\'{e}s dans la propagation des ph\'{e}nom\`{e}nes explosifs},\ }\href@noop {} {\bibfield  {journal} {\bibinfo  {journal} {C. R. Acad. Sci. Paris}\ }\textbf {\bibinfo {volume} {130}},\ \bibinfo {pages} {413} (\bibinfo {year} {1900})}\BibitemShut {NoStop}%
\bibitem [{\citenamefont {Fickett}\ and\ \citenamefont {Davis}(2000)}]{FickDav2000}%
  \BibitemOpen
  \bibfield  {author} {\bibinfo {author} {\bibfnamefont {W.}~\bibnamefont {Fickett}}\ and\ \bibinfo {author} {\bibfnamefont {W.~C.}\ \bibnamefont {Davis}},\ }\href@noop {} {\emph {\bibinfo {title} {Detonation: theory and experiment}}}\ (\bibinfo  {publisher} {Dover Publications, Inc.},\ \bibinfo {year} {2000})\BibitemShut {NoStop}%
\bibitem [{\citenamefont {Higgins}(2012)}]{Higgins2012}%
  \BibitemOpen
  \bibfield  {author} {\bibinfo {author} {\bibfnamefont {A.}~\bibnamefont {Higgins}},\ }\bibfield  {title} {\bibinfo {title} {Steady one-dimensional detonation},\ }in\ \href@noop {} {\emph {\bibinfo {booktitle} {Shock Waves Sciences and Technology Reference Library, Vol.6: Detonation dynamics}}}\ (\bibinfo  {publisher} {Springer-Verlag, Berlin, Heidelberg},\ \bibinfo {year} {2012})\ pp.\ \bibinfo {pages} {33--105}\BibitemShut {NoStop}%
\bibitem [{\citenamefont {Zel'dovich}\ and\ \citenamefont {Kompaneets}(1960)}]{ZeldoKompa1960}%
  \BibitemOpen
  \bibfield  {author} {\bibinfo {author} {\bibfnamefont {Y.~B.}\ \bibnamefont {Zel'dovich}}\ and\ \bibinfo {author} {\bibfnamefont {A.~S.}\ \bibnamefont {Kompaneets}},\ }\href@noop {} {\emph {\bibinfo {title} {Theory of detonation}}}\ (\bibinfo  {publisher} {Academic Press, New York (transl. Gostekhizdat, Moscow, 1955)},\ \bibinfo {year} {1960})\BibitemShut {NoStop}%
\bibitem [{\citenamefont {Wood}\ and\ \citenamefont {Kirkwood}(1954)}]{WoodKirkwood1954}%
  \BibitemOpen
  \bibfield  {author} {\bibinfo {author} {\bibfnamefont {W.~W.}\ \bibnamefont {Wood}}\ and\ \bibinfo {author} {\bibfnamefont {J.~G.}\ \bibnamefont {Kirkwood}},\ }\bibfield  {title} {\bibinfo {title} {Diameter effect in condensed explosives. the relation between velocity and radius of curvature of the detonation wave},\ }\href@noop {} {\bibfield  {journal} {\bibinfo  {journal} {J. Chem. Phys.}\ }\textbf {\bibinfo {volume} {2(11)}},\ \bibinfo {pages} {1920} (\bibinfo {year} {1954})}\BibitemShut {NoStop}%
\bibitem [{\citenamefont {He}\ and\ \citenamefont {Clavin}(1994)}]{HeClavin1994}%
  \BibitemOpen
  \bibfield  {author} {\bibinfo {author} {\bibfnamefont {L.}~\bibnamefont {He}}\ and\ \bibinfo {author} {\bibfnamefont {P.}~\bibnamefont {Clavin}},\ }\bibfield  {title} {\bibinfo {title} {On the direct initiation of gaseous detonations by an energy source},\ }\href@noop {} {\bibfield  {journal} {\bibinfo  {journal} {J. Fluid Mech.}\ }\textbf {\bibinfo {volume} {277}},\ \bibinfo {pages} {227} (\bibinfo {year} {1994})}\BibitemShut {NoStop}%
\bibitem [{\citenamefont {Kasimov}\ and\ \citenamefont {Stewart}(2004)}]{KasimovStewart2004}%
  \BibitemOpen
  \bibfield  {author} {\bibinfo {author} {\bibfnamefont {A.~R.}\ \bibnamefont {Kasimov}}\ and\ \bibinfo {author} {\bibfnamefont {D.~S.}\ \bibnamefont {Stewart}},\ }\bibfield  {title} {\bibinfo {title} {On the dynamics of self-sustained one-dimensional detonations: a numerical study in the shock-attached frame},\ }\href@noop {} {\bibfield  {journal} {\bibinfo  {journal} {Phys. Fluids}\ }\textbf {\bibinfo {volume} {16(10)}},\ \bibinfo {pages} {3566} (\bibinfo {year} {2004})}\BibitemShut {NoStop}%
\bibitem [{\citenamefont {Short}\ \emph {et~al.}(2020)\citenamefont {Short}, \citenamefont {Voelkel},\ and\ \citenamefont {Chiquete}}]{ShortEtAl2020}%
  \BibitemOpen
  \bibfield  {author} {\bibinfo {author} {\bibfnamefont {M.}~\bibnamefont {Short}}, \bibinfo {author} {\bibfnamefont {S.~J.}\ \bibnamefont {Voelkel}},\ and\ \bibinfo {author} {\bibfnamefont {C.}~\bibnamefont {Chiquete}},\ }\bibfield  {title} {\bibinfo {title} {Steady detonation propagation in thin channels with strong confinement},\ }\href@noop {} {\bibfield  {journal} {\bibinfo  {journal} {J. Fluid Mech.}\ }\textbf {\bibinfo {volume} {889, \emph{A3}}} (\bibinfo {year} {2020})}\BibitemShut {NoStop}%
\bibitem [{\citenamefont {Sharpe}(2000)}]{Sharpe2000}%
  \BibitemOpen
  \bibfield  {author} {\bibinfo {author} {\bibfnamefont {G.~J.}\ \bibnamefont {Sharpe}},\ }\bibfield  {title} {\bibinfo {title} {The structure of planar and curved detonation waves with reversible reactions},\ }\href@noop {} {\bibfield  {journal} {\bibinfo  {journal} {Phys. Fluids}\ }\textbf {\bibinfo {volume} {12(11)}},\ \bibinfo {pages} {3007} (\bibinfo {year} {2000})}\BibitemShut {NoStop}%
\bibitem [{\citenamefont {Denisov}\ and\ \citenamefont {Troshin}(1959)}]{DenisovTroshin1959}%
  \BibitemOpen
  \bibfield  {author} {\bibinfo {author} {\bibfnamefont {Y.~N.}\ \bibnamefont {Denisov}}\ and\ \bibinfo {author} {\bibfnamefont {Y.~K.}\ \bibnamefont {Troshin}},\ }\bibfield  {title} {\bibinfo {title} {Pulsating and spinning detonation of gaseous detonation in tubes},\ }\href@noop {} {\bibfield  {journal} {\bibinfo  {journal} {Dokl. Akad. Nauk. SSSR}\ }\textbf {\bibinfo {volume} {125}},\ \bibinfo {pages} {110} (\bibinfo {year} {1959})}\BibitemShut {NoStop}%
\bibitem [{\citenamefont {Desbordes}\ and\ \citenamefont {Presles}(2012)}]{DesbordesPresles2012}%
  \BibitemOpen
  \bibfield  {author} {\bibinfo {author} {\bibfnamefont {D.}~\bibnamefont {Desbordes}}\ and\ \bibinfo {author} {\bibfnamefont {H.-N.}\ \bibnamefont {Presles}},\ }\bibfield  {title} {\bibinfo {title} {Multi-scaled cellular detonation},\ }in\ \href@noop {} {\emph {\bibinfo {booktitle} {Shock Waves Sciences and Technology Reference Library, Vol.6: Detonation dynamics}}}\ (\bibinfo  {publisher} {Springer-Verlag, Berlin, Heidelberg},\ \bibinfo {year} {2012})\ pp.\ \bibinfo {pages} {281--338}\BibitemShut {NoStop}%
\bibitem [{\citenamefont {Monnier}\ \emph {et~al.}(2022)\citenamefont {Monnier}, \citenamefont {Rodriguez}, \citenamefont {Vidal},\ and\ \citenamefont {Zitoun}}]{Monnier2022}%
  \BibitemOpen
  \bibfield  {author} {\bibinfo {author} {\bibfnamefont {V.}~\bibnamefont {Monnier}}, \bibinfo {author} {\bibfnamefont {V.}~\bibnamefont {Rodriguez}}, \bibinfo {author} {\bibfnamefont {P.}~\bibnamefont {Vidal}},\ and\ \bibinfo {author} {\bibfnamefont {R.}~\bibnamefont {Zitoun}},\ }\bibfield  {title} {\bibinfo {title} {An analysis of three-dimensional patterns of experimental detonation cells},\ }\href {https://doi.org/10.1016/j.combustflame.2022.112310} {\bibfield  {journal} {\bibinfo  {journal} {Combust. Flame}\ }\textbf {\bibinfo {volume} {245}},\ \bibinfo {pages} {112310} (\bibinfo {year} {2022})}\BibitemShut {NoStop}%
\bibitem [{\citenamefont {Monnier}\ \emph {et~al.}(2023)\citenamefont {Monnier}, \citenamefont {Vidal}, \citenamefont {Rodriguez},\ and\ \citenamefont {Zitoun}}]{Monnier2023}%
  \BibitemOpen
  \bibfield  {author} {\bibinfo {author} {\bibfnamefont {V.}~\bibnamefont {Monnier}}, \bibinfo {author} {\bibfnamefont {P.}~\bibnamefont {Vidal}}, \bibinfo {author} {\bibfnamefont {V.}~\bibnamefont {Rodriguez}},\ and\ \bibinfo {author} {\bibfnamefont {R.}~\bibnamefont {Zitoun}},\ }\bibfield  {title} {\bibinfo {title} {From graph theory and geometric probabilities to a representative width for three-dimensional detonation cells},\ }\href {https://doi.org/10.1016/j.combustflame.2023.112996} {\bibfield  {journal} {\bibinfo  {journal} {Combust. Flame}\ }\textbf {\bibinfo {volume} {256}},\ \bibinfo {pages} {112996} (\bibinfo {year} {2023})}\BibitemShut {NoStop}%
\bibitem [{\citenamefont {Edwards}\ and\ \citenamefont {Short}(2019)}]{EdwardsShort2019}%
  \BibitemOpen
  \bibfield  {author} {\bibinfo {author} {\bibfnamefont {L.}~\bibnamefont {Edwards}}\ and\ \bibinfo {author} {\bibfnamefont {M.}~\bibnamefont {Short}},\ }\bibfield  {title} {\bibinfo {title} {Modeling of the cellular structure of detonation in liquid explosives},\ }in\ \href@noop {} {\emph {\bibinfo {booktitle} {abstract H05.008, APS Division of Fluid Dynamics}}}\ (\bibinfo {year} {2019})\BibitemShut {NoStop}%
\bibitem [{\citenamefont {Urtiew}\ and\ \citenamefont {Kusubov}(1970)}]{UrtiewKusubov1970}%
  \BibitemOpen
  \bibfield  {author} {\bibinfo {author} {\bibfnamefont {P.~A.}\ \bibnamefont {Urtiew}}\ and\ \bibinfo {author} {\bibfnamefont {A.~S.}\ \bibnamefont {Kusubov}},\ }\bibfield  {title} {\bibinfo {title} {Wall traces of detonation in nitromethane-acetone mixtures},\ }in\ \href@noop {} {\emph {\bibinfo {booktitle} {5$^{\,\text{th}}$ Symp. (Int.) Detonation}}}\ (\bibinfo  {publisher} {ONR},\ \bibinfo {year} {1970})\ pp.\ \bibinfo {pages} {105--114}\BibitemShut {NoStop}%
\bibitem [{\citenamefont {Persson}\ and\ \citenamefont {Bjarnholt}(1970)}]{PerssonBjarnholt1970}%
  \BibitemOpen
  \bibfield  {author} {\bibinfo {author} {\bibfnamefont {P.~A.}\ \bibnamefont {Persson}}\ and\ \bibinfo {author} {\bibfnamefont {G.}~\bibnamefont {Bjarnholt}},\ }\bibfield  {title} {\bibinfo {title} {A photographic technique for mapping failure waves and other instability phenomena in liquid explosives detonation},\ }in\ \href@noop {} {\emph {\bibinfo {booktitle} {5$^{\,\text{th}}$ Symp. (Int.) Detonation}}}\ (\bibinfo  {publisher} {ONR},\ \bibinfo {year} {1970})\ pp.\ \bibinfo {pages} {115--118}\BibitemShut {NoStop}%
\bibitem [{\citenamefont {Tarver}\ and\ \citenamefont {Urtiew}(2010)}]{TarverUrtiew2010}%
  \BibitemOpen
  \bibfield  {author} {\bibinfo {author} {\bibfnamefont {C.~M.}\ \bibnamefont {Tarver}}\ and\ \bibinfo {author} {\bibfnamefont {P.~A.}\ \bibnamefont {Urtiew}},\ }\bibfield  {title} {\bibinfo {title} {Theory and modeling of liquid explosive detonation},\ }\href@noop {} {\bibfield  {journal} {\bibinfo  {journal} {Journal of Energetic Materials}\ }\textbf {\bibinfo {volume} {28(4)}},\ \bibinfo {pages} {299} (\bibinfo {year} {2010})}\BibitemShut {NoStop}%
\bibitem [{\citenamefont {Tarver}(1982{\natexlab{a}})}]{Tarver1982a}%
  \BibitemOpen
  \bibfield  {author} {\bibinfo {author} {\bibfnamefont {C.~M.}\ \bibnamefont {Tarver}},\ }\bibfield  {title} {\bibinfo {title} {Chemical energy release in one-dimensional detonation waves in gaseous explosives},\ }\href {https://doi.org/10.1016/0010-2180(82)90011-6} {\bibfield  {journal} {\bibinfo  {journal} {Combust. Flame}\ }\textbf {\bibinfo {volume} {46}},\ \bibinfo {pages} {111} (\bibinfo {year} {1982}{\natexlab{a}})}\BibitemShut {NoStop}%
\bibitem [{\citenamefont {Tarver}(1982{\natexlab{b}})}]{Tarver1982b}%
  \BibitemOpen
  \bibfield  {author} {\bibinfo {author} {\bibfnamefont {C.~M.}\ \bibnamefont {Tarver}},\ }\bibfield  {title} {\bibinfo {title} {Chemical energy release in the cellular structure of gaseous detonation waves},\ }\href {https://doi.org/10.1016/0010-2180(82)90012-8} {\bibfield  {journal} {\bibinfo  {journal} {Combust. Flame}\ }\textbf {\bibinfo {volume} {46}},\ \bibinfo {pages} {135} (\bibinfo {year} {1982}{\natexlab{b}})}\BibitemShut {NoStop}%
\bibitem [{\citenamefont {Vargas}\ \emph {et~al.}(2022)\citenamefont {Vargas}, \citenamefont {M\'{e}vel}, \citenamefont {Lino~da Silva},\ and\ \citenamefont {Lacoste}}]{Vargas2022}%
  \BibitemOpen
  \bibfield  {author} {\bibinfo {author} {\bibfnamefont {J.}~\bibnamefont {Vargas}}, \bibinfo {author} {\bibfnamefont {R.}~\bibnamefont {M\'{e}vel}}, \bibinfo {author} {\bibfnamefont {M.}~\bibnamefont {Lino~da Silva}},\ and\ \bibinfo {author} {\bibfnamefont {D.~A.}\ \bibnamefont {Lacoste}},\ }\bibfield  {title} {\bibinfo {title} {Development of a steady detonation reactor with state-to-state thermochemical modeling},\ }\href@noop {} {\bibfield  {journal} {\bibinfo  {journal} {Shock Waves}\ }\textbf {\bibinfo {volume} {32}},\ \bibinfo {pages} {679} (\bibinfo {year} {2022})}\BibitemShut {NoStop}%
\bibitem [{\citenamefont {Vargas}\ \emph {et~al.}(2023)\citenamefont {Vargas}, \citenamefont {Chatelain}, \citenamefont {Lacoste}, \citenamefont {Huang},\ and\ \citenamefont {M\'{e}vel}}]{Vargas2023}%
  \BibitemOpen
  \bibfield  {author} {\bibinfo {author} {\bibfnamefont {J.}~\bibnamefont {Vargas}}, \bibinfo {author} {\bibfnamefont {K.~P.}\ \bibnamefont {Chatelain}}, \bibinfo {author} {\bibfnamefont {D.~A.}\ \bibnamefont {Lacoste}}, \bibinfo {author} {\bibfnamefont {X.}~\bibnamefont {Huang}},\ and\ \bibinfo {author} {\bibfnamefont {R.}~\bibnamefont {M\'{e}vel}},\ }\bibfield  {title} {\bibinfo {title} {{Non-equilibrium effect in H$_{2}$-O$_{2}$-Diluent mixtures using the ZND reactor model (presentation 78)}},\ }in\ \href@noop {} {\emph {\bibinfo {booktitle} {29$^{\text{th}}$ ICDERS}}}\ (\bibinfo {year} {2023})\BibitemShut {NoStop}%
\bibitem [{\citenamefont {Dremin}(1999)}]{Dremin1999}%
  \BibitemOpen
  \bibfield  {author} {\bibinfo {author} {\bibfnamefont {A.~N.}\ \bibnamefont {Dremin}},\ }\href@noop {} {\emph {\bibinfo {title} {Towards detonation theory}}}\ (\bibinfo  {publisher} {Springer, New York},\ \bibinfo {year} {1999})\BibitemShut {NoStop}%
\bibitem [{\citenamefont {Tarver}(1982{\natexlab{c}})}]{Tarver1982c}%
  \BibitemOpen
  \bibfield  {author} {\bibinfo {author} {\bibfnamefont {C.~M.}\ \bibnamefont {Tarver}},\ }\bibfield  {title} {\bibinfo {title} {Chemical energy release in self-sustaining detonation waves in condensed explosives},\ }\href {https://doi.org/10.1016/0010-2180(82)90013-X} {\bibfield  {journal} {\bibinfo  {journal} {Combust. Flame}\ }\textbf {\bibinfo {volume} {46}},\ \bibinfo {pages} {157} (\bibinfo {year} {1982}{\natexlab{c}})}\BibitemShut {NoStop}%
\bibitem [{\citenamefont {Tarver}(2012)}]{Tarver2012}%
  \BibitemOpen
  \bibfield  {author} {\bibinfo {author} {\bibfnamefont {C.~M.}\ \bibnamefont {Tarver}},\ }\bibfield  {title} {\bibinfo {title} {Condensed matter detonation: theory and practice},\ }in\ \href@noop {} {\emph {\bibinfo {booktitle} {Shock Waves Sciences and Technology Reference Library, Vol.6: Detonation dynamics}}}\ (\bibinfo  {publisher} {Springer-Verlag, Berlin, Heidelberg},\ \bibinfo {year} {2012})\ pp.\ \bibinfo {pages} {339--372}\BibitemShut {NoStop}%
\bibitem [{\citenamefont {Kistiakovski}\ \emph {et~al.}(1952)\citenamefont {Kistiakovski}, \citenamefont {Knight},\ and\ \citenamefont {Malin}}]{Kistiakovski1952}%
  \BibitemOpen
  \bibfield  {author} {\bibinfo {author} {\bibfnamefont {G.~B.}\ \bibnamefont {Kistiakovski}}, \bibinfo {author} {\bibfnamefont {H.~T.}\ \bibnamefont {Knight}},\ and\ \bibinfo {author} {\bibfnamefont {M.~E.}\ \bibnamefont {Malin}},\ }\bibfield  {title} {\bibinfo {title} {{Gaseous detonations. IV. The acetylene-oxygen mixtures}},\ }\href@noop {} {\bibfield  {journal} {\bibinfo  {journal} {J. Chem. Phys.}\ }\textbf {\bibinfo {volume} {20}},\ \bibinfo {pages} {884} (\bibinfo {year} {1952})}\BibitemShut {NoStop}%
\bibitem [{\citenamefont {Kistiakovski}\ and\ \citenamefont {Zinman}(1955)}]{Kistiakovski1955}%
  \BibitemOpen
  \bibfield  {author} {\bibinfo {author} {\bibfnamefont {G.~B.}\ \bibnamefont {Kistiakovski}}\ and\ \bibinfo {author} {\bibfnamefont {W.~G.}\ \bibnamefont {Zinman}},\ }\bibfield  {title} {\bibinfo {title} {{Gaseous detonations. VII. A study of thermodynamic equilibrium in acetylene-oxygen waves}},\ }\href@noop {} {\bibfield  {journal} {\bibinfo  {journal} {J. Chem. Phys.}\ }\textbf {\bibinfo {volume} {23}},\ \bibinfo {pages} {1889} (\bibinfo {year} {1955})}\BibitemShut {NoStop}%
\bibitem [{\citenamefont {Kistiakovski}\ and\ \citenamefont {Mangelsdorf}(1952)}]{Kistiakovski1956}%
  \BibitemOpen
  \bibfield  {author} {\bibinfo {author} {\bibfnamefont {G.~B.}\ \bibnamefont {Kistiakovski}}\ and\ \bibinfo {author} {\bibfnamefont {P.~C.~J.}\ \bibnamefont {Mangelsdorf}},\ }\bibfield  {title} {\bibinfo {title} {{Gaseous detonations. VIII. Two-Stage detonations in acetylene-oxygen mixtures}},\ }\href@noop {} {\bibfield  {journal} {\bibinfo  {journal} {J. Chem. Phys.}\ }\textbf {\bibinfo {volume} {25}},\ \bibinfo {pages} {516} (\bibinfo {year} {1952})}\BibitemShut {NoStop}%
\bibitem [{\citenamefont {Batraev}\ \emph {et~al.}(2018)\citenamefont {Batraev}, \citenamefont {Vasil{'}ev}, \citenamefont {Ul{'}yanitskii}, \citenamefont {Shtertser},\ and\ \citenamefont {Rybin}}]{BatraevEtal2018}%
  \BibitemOpen
  \bibfield  {author} {\bibinfo {author} {\bibfnamefont {I.~S.}\ \bibnamefont {Batraev}}, \bibinfo {author} {\bibfnamefont {A.~A.}\ \bibnamefont {Vasil{'}ev}}, \bibinfo {author} {\bibfnamefont {V.~Y.}\ \bibnamefont {Ul{'}yanitskii}}, \bibinfo {author} {\bibfnamefont {A.~A.}\ \bibnamefont {Shtertser}},\ and\ \bibinfo {author} {\bibfnamefont {D.~K.}\ \bibnamefont {Rybin}},\ }\bibfield  {title} {\bibinfo {title} {Investigation of gas detonation in over-rich mixtures of hydrocarbons with oxygen},\ }\href@noop {} {\bibfield  {journal} {\bibinfo  {journal} {Combustion, Explosion, and Shock Waves}\ }\textbf {\bibinfo {volume} {54}},\ \bibinfo {pages} {207} (\bibinfo {year} {2018})}\BibitemShut {NoStop}%
\bibitem [{\citenamefont {Berger}\ and\ \citenamefont {Viard}(1962)}]{BergerViard1962}%
  \BibitemOpen
  \bibfield  {author} {\bibinfo {author} {\bibfnamefont {J.}~\bibnamefont {Berger}}\ and\ \bibinfo {author} {\bibfnamefont {J.}~\bibnamefont {Viard}},\ }\href@noop {} {\emph {\bibinfo {title} {Physique des explosifs solides (p.186-190)}}}\ (\bibinfo  {publisher} {Dunod, Paris},\ \bibinfo {year} {1962})\BibitemShut {NoStop}%
\bibitem [{\citenamefont {Bastea}(2017)}]{Bastea2017}%
  \BibitemOpen
  \bibfield  {author} {\bibinfo {author} {\bibfnamefont {S.}~\bibnamefont {Bastea}},\ }\bibfield  {title} {\bibinfo {title} {Nanocarbon condensation in detonation},\ }\href@noop {} {\bibfield  {journal} {\bibinfo  {journal} {Nature Scientific Reports}\ }\textbf {\bibinfo {volume} {7}},\ \bibinfo {pages} {42151} (\bibinfo {year} {2017})}\BibitemShut {NoStop}%
\bibitem [{\citenamefont {Gordon}\ and\ \citenamefont {McBride}(1994)}]{GordonMcBride-I}%
  \BibitemOpen
  \bibfield  {author} {\bibinfo {author} {\bibfnamefont {S.}~\bibnamefont {Gordon}}\ and\ \bibinfo {author} {\bibfnamefont {B.}~\bibnamefont {McBride}},\ }\href@noop {} {\emph {\bibinfo {title} {Computer program for calculation of complex chemical equilibrium compositions and applications, I. Analysis (Ref. 1311)}}},\ \bibinfo {type} {Tech. Rep.}\ (\bibinfo  {institution} {NASA},\ \bibinfo {year} {1994})\BibitemShut {NoStop}%
\bibitem [{\citenamefont {Duhem}(1909)}]{Duhem1909}%
  \BibitemOpen
  \bibfield  {author} {\bibinfo {author} {\bibfnamefont {P.}~\bibnamefont {Duhem}},\ }\bibfield  {title} {\bibinfo {title} {Sur la propagation des ondes de choc au sein des fluides},\ }\href@noop {} {\bibfield  {journal} {\bibinfo  {journal} {Z. Phys. Chem.}\ }\textbf {\bibinfo {volume} {69}},\ \bibinfo {pages} {160} (\bibinfo {year} {1909})}\BibitemShut {NoStop}%
\bibitem [{\citenamefont {Bethe}(1942)}]{Bethe1942}%
  \BibitemOpen
  \bibfield  {author} {\bibinfo {author} {\bibfnamefont {H.~A.}\ \bibnamefont {Bethe}},\ }\href@noop {} {\emph {\bibinfo {title} {The theory of shock waves for an arbitrary equation of state}}}\ (\bibinfo  {publisher} {Report 545, OSRD},\ \bibinfo {year} {1942})\BibitemShut {NoStop}%
\bibitem [{\citenamefont {Weyl}(1949)}]{Weyl1949}%
  \BibitemOpen
  \bibfield  {author} {\bibinfo {author} {\bibfnamefont {H.}~\bibnamefont {Weyl}},\ }\bibfield  {title} {\bibinfo {title} {Shock waves in arbitrary fluids},\ }\href@noop {} {\bibfield  {journal} {\bibinfo  {journal} {Comm. Pure Appl. Math.}\ }\textbf {\bibinfo {volume} {2}},\ \bibinfo {pages} {103} (\bibinfo {year} {1949})}\BibitemShut {NoStop}%
\bibitem [{\citenamefont {Thomson}(1971)}]{Thomson1971}%
  \BibitemOpen
  \bibfield  {author} {\bibinfo {author} {\bibfnamefont {P.~A.}\ \bibnamefont {Thomson}},\ }\bibfield  {title} {\bibinfo {title} {A fundamental derivative in gasdynamics},\ }\href@noop {} {\bibfield  {journal} {\bibinfo  {journal} {Phys. Fluids}\ }\textbf {\bibinfo {volume} {14(9)}},\ \bibinfo {pages} {1843} (\bibinfo {year} {1971})}\BibitemShut {NoStop}%
\bibitem [{\citenamefont {D'yakov}(1954)}]{Dyakov1954}%
  \BibitemOpen
  \bibfield  {author} {\bibinfo {author} {\bibfnamefont {S.~P.}\ \bibnamefont {D'yakov}},\ }\bibfield  {title} {\bibinfo {title} {On the stability of shock waves},\ }\href@noop {} {\bibfield  {journal} {\bibinfo  {journal} {Zh. Eksp. Teor. Fiz.}\ }\textbf {\bibinfo {volume} {27}},\ \bibinfo {pages} {288} (\bibinfo {year} {1954})}\BibitemShut {NoStop}%
\bibitem [{\citenamefont {Kontorovich}(1957)}]{Kontorovich1957}%
  \BibitemOpen
  \bibfield  {author} {\bibinfo {author} {\bibfnamefont {V.~M.}\ \bibnamefont {Kontorovich}},\ }\bibfield  {title} {\bibinfo {title} {Concerning the stability of shock waves},\ }\href@noop {} {\bibfield  {journal} {\bibinfo  {journal} {JETP}\ }\textbf {\bibinfo {volume} {6(6)}},\ \bibinfo {pages} {1179} (\bibinfo {year} {1957})}\BibitemShut {NoStop}%
\bibitem [{\citenamefont {Bates}\ and\ \citenamefont {Montgomery}(2000)}]{BatesMontgomery2000}%
  \BibitemOpen
  \bibfield  {author} {\bibinfo {author} {\bibfnamefont {J.~W.}\ \bibnamefont {Bates}}\ and\ \bibinfo {author} {\bibfnamefont {D.~C.}\ \bibnamefont {Montgomery}},\ }\bibfield  {title} {\bibinfo {title} {{The D'yakov-Kontorovich instability of shock waves in real gases}},\ }\href@noop {} {\bibfield  {journal} {\bibinfo  {journal} {Phys. Rev. Letters}\ }\textbf {\bibinfo {volume} {84(6)}},\ \bibinfo {pages} {1180} (\bibinfo {year} {2000})}\BibitemShut {NoStop}%
\bibitem [{\citenamefont {Brun}(2013)}]{Brun2013}%
  \BibitemOpen
  \bibfield  {author} {\bibinfo {author} {\bibfnamefont {L.}~\bibnamefont {Brun}},\ }\href@noop {} {\emph {\bibinfo {title} {The spontaneous acoustic emission of the shock front in a perfect fluid: solving a riddle}}}\ (\bibinfo  {publisher} {Report CEA-R-6337, CEA},\ \bibinfo {year} {2013})\BibitemShut {NoStop}%
\bibitem [{\citenamefont {Clavin}\ and\ \citenamefont {Searby}(2016)}]{Clavin2016}%
  \BibitemOpen
  \bibfield  {author} {\bibinfo {author} {\bibfnamefont {P.}~\bibnamefont {Clavin}}\ and\ \bibinfo {author} {\bibfnamefont {G.}~\bibnamefont {Searby}},\ }\href@noop {} {\emph {\bibinfo {title} {Combustion waves and fronts in flows: flames, shocks, detonations, ablation fronts and explosion of stars}}}\ (\bibinfo  {publisher} {Cambridge University Press},\ \bibinfo {year} {2016})\BibitemShut {NoStop}%
\bibitem [{\citenamefont {Landau}(1958)}]{Landau1944}%
  \BibitemOpen
  \bibfield  {author} {\bibinfo {author} {\bibfnamefont {L.}~\bibnamefont {Landau}},\ }\href@noop {} {\emph {\bibinfo {title} {1944, cited in Landau L. \& Lifchitz E., Fluid Mechanics, Chapt. IX, §88}}}\ (\bibinfo  {publisher} {Pergamon, Oxford},\ \bibinfo {year} {1958})\BibitemShut {NoStop}%
\bibitem [{\citenamefont {Lax}(1957)}]{Lax1957}%
  \BibitemOpen
  \bibfield  {author} {\bibinfo {author} {\bibfnamefont {P.~D.}\ \bibnamefont {Lax}},\ }\bibfield  {title} {\bibinfo {title} {Hyperbolic systems of conservation laws, {II}},\ }\href@noop {} {\bibfield  {journal} {\bibinfo  {journal} {Comm. Pure and Appl. Math.}\ }\textbf {\bibinfo {volume} {10}},\ \bibinfo {pages} {537} (\bibinfo {year} {1957})}\BibitemShut {NoStop}%
\bibitem [{\citenamefont {Fowles}(1975)}]{Fowles1975}%
  \BibitemOpen
  \bibfield  {author} {\bibinfo {author} {\bibfnamefont {G.~R.}\ \bibnamefont {Fowles}},\ }\bibfield  {title} {\bibinfo {title} {Subsonic-supersonic condition for shocks},\ }\href@noop {} {\bibfield  {journal} {\bibinfo  {journal} {Phys. Fluids}\ }\textbf {\bibinfo {volume} {18(7)}},\ \bibinfo {pages} {776} (\bibinfo {year} {1975})}\BibitemShut {NoStop}%
\bibitem [{\citenamefont {Taylor}(1941)}]{Taylor1950}%
  \BibitemOpen
  \bibfield  {author} {\bibinfo {author} {\bibfnamefont {G.~I.}\ \bibnamefont {Taylor}},\ }\bibfield  {title} {\bibinfo {title} {The dynamics of the combustion products behind plane and spherical detonation fronts in explosives},\ }\href@noop {} {\bibfield  {journal} {\bibinfo  {journal} {Proc. Roy. Soc.}\ }\textbf {\bibinfo {volume} {A 200}},\ \bibinfo {pages} {235} (\bibinfo {year} {1950 (1941)})}\BibitemShut {NoStop}%
\bibitem [{\citenamefont {D\"{o}ring}\ and\ \citenamefont {Burkhardt}(1944)}]{DoringBurkhardt1944}%
  \BibitemOpen
  \bibfield  {author} {\bibinfo {author} {\bibfnamefont {W.}~\bibnamefont {D\"{o}ring}}\ and\ \bibinfo {author} {\bibfnamefont {G.}~\bibnamefont {Burkhardt}},\ }\href@noop {} {\emph {\bibinfo {title} {Beitr\"{a}ge zur Theorie der Detonation (Ref. Bericht n${{}^\circ}$1939)}}},\ \bibinfo {type} {Tech. Rep.}\ (\bibinfo  {institution} {Deutsche Luftfahrtforschung},\ \bibinfo {year} {1944})\BibitemShut {NoStop}%
\bibitem [{\citenamefont {Wood}\ and\ \citenamefont {Fickett}(1963)}]{WoodFickett1963}%
  \BibitemOpen
  \bibfield  {author} {\bibinfo {author} {\bibfnamefont {W.~W.}\ \bibnamefont {Wood}}\ and\ \bibinfo {author} {\bibfnamefont {W.}~\bibnamefont {Fickett}},\ }\bibfield  {title} {\bibinfo {title} {{Investigation of the CJ hypothesis by the "Inverse Method"}},\ }\href@noop {} {\bibfield  {journal} {\bibinfo  {journal} {Phys. Fluids}\ }\textbf {\bibinfo {volume} {6(5)}},\ \bibinfo {pages} {648} (\bibinfo {year} {1963})}\BibitemShut {NoStop}%
\bibitem [{\citenamefont {Manson}(1958{\natexlab{b}})}]{Manson1958b}%
  \BibitemOpen
  \bibfield  {author} {\bibinfo {author} {\bibfnamefont {N.}~\bibnamefont {Manson}},\ }\bibfield  {title} {\bibinfo {title} {Semi-empirical determination of gas characteristics in the {C}hapman-{J}ouguet state},\ }\href@noop {} {\bibfield  {journal} {\bibinfo  {journal} {Combust. Flame}\ }\textbf {\bibinfo {volume} {2(2)}},\ \bibinfo {pages} {226} (\bibinfo {year} {1958}{\natexlab{b}})}\BibitemShut {NoStop}%
\bibitem [{\citenamefont {Wecken}(1959)}]{Wecken1959}%
  \BibitemOpen
  \bibfield  {author} {\bibinfo {author} {\bibfnamefont {F.}~\bibnamefont {Wecken}},\ }\href@noop {} {\emph {\bibinfo {title} {Note Technique n${{}^\circ}$459 (avril)}}},\ \bibinfo {type} {Tech. Rep.}\ (\bibinfo  {institution} {Institut Franco-Allemand de Saint-Louis},\ \bibinfo {year} {1959})\BibitemShut {NoStop}%
\bibitem [{\citenamefont {Bauer}\ \emph {et~al.}(1988)\citenamefont {Bauer}, \citenamefont {Vidal}, \citenamefont {Manson},\ and\ \citenamefont {Heuz\'e}}]{Bauer1988}%
  \BibitemOpen
  \bibfield  {author} {\bibinfo {author} {\bibfnamefont {P.}~\bibnamefont {Bauer}}, \bibinfo {author} {\bibfnamefont {P.}~\bibnamefont {Vidal}}, \bibinfo {author} {\bibfnamefont {N.}~\bibnamefont {Manson}},\ and\ \bibinfo {author} {\bibfnamefont {O.}~\bibnamefont {Heuz\'e}},\ }\bibinfo {title} {{Applicability of the Inverse Method to the determination of CJ parameters for gaseous mixtures at elevated pressures (11$^{\text{th}}$ ICDERS, Warsaw, 1987)}},\ in\ \href {https://doi.org/10.2514/5.9781600865886.0064.0076} {\emph {\bibinfo {booktitle} {Dynamics of Explosions}}}\ (\bibinfo  {publisher} {AIAA Inc.},\ \bibinfo {year} {1988})\ pp.\ \bibinfo {pages} {64--76}\BibitemShut {NoStop}%
\bibitem [{\citenamefont {Davis}(1981)}]{Davis1981}%
  \BibitemOpen
  \bibfield  {author} {\bibinfo {author} {\bibfnamefont {W.~C.}\ \bibnamefont {Davis}},\ }\bibfield  {title} {\bibinfo {title} {Equation of state from detonation velocity measurements},\ }\href@noop {} {\bibfield  {journal} {\bibinfo  {journal} {Combust. Flame}\ }\textbf {\bibinfo {volume} {41}},\ \bibinfo {pages} {171} (\bibinfo {year} {1981})}\BibitemShut {NoStop}%
\bibitem [{\citenamefont {Nagayama}\ and\ \citenamefont {Kubota}(2004)}]{NagayamaKubota2004}%
  \BibitemOpen
  \bibfield  {author} {\bibinfo {author} {\bibfnamefont {K.}~\bibnamefont {Nagayama}}\ and\ \bibinfo {author} {\bibfnamefont {S.}~\bibnamefont {Kubota}},\ }\bibfield  {title} {\bibinfo {title} {{Approximate method for predicting the Chapman-Jouguet state of condensed explosives}},\ }\href@noop {} {\bibfield  {journal} {\bibinfo  {journal} {Propellants, Explosives, Pyrotechnics}\ }\textbf {\bibinfo {volume} {29(2)}},\ \bibinfo {pages} {118} (\bibinfo {year} {2004})}\BibitemShut {NoStop}%
\bibitem [{\citenamefont {Sheffield}\ \emph {et~al.}(2001)\citenamefont {Sheffield}, \citenamefont {Davis}, \citenamefont {Engelke}, \citenamefont {Alcon}, \citenamefont {Baer},\ and\ \citenamefont {Renlund}}]{SheffieldEtAl2001}%
  \BibitemOpen
  \bibfield  {author} {\bibinfo {author} {\bibfnamefont {S.~A.}\ \bibnamefont {Sheffield}}, \bibinfo {author} {\bibfnamefont {L.~L.}\ \bibnamefont {Davis}}, \bibinfo {author} {\bibfnamefont {R.}~\bibnamefont {Engelke}}, \bibinfo {author} {\bibfnamefont {R.~R.}\ \bibnamefont {Alcon}}, \bibinfo {author} {\bibfnamefont {M.~R.}\ \bibnamefont {Baer}},\ and\ \bibinfo {author} {\bibfnamefont {A.~M.}\ \bibnamefont {Renlund}},\ }\bibfield  {title} {\bibinfo {title} {Hugoniot and shock initiation studies of isopropyl nitrate},\ }in\ \href@noop {} {\emph {\bibinfo {booktitle} {12$^{\,\text{th}}$ APS Topical Conf. on Shock Compression of Condensed Matter}}}\ (\bibinfo {year} {2001})\BibitemShut {NoStop}%
\bibitem [{\citenamefont {Zhang}\ \emph {et~al.}(2002)\citenamefont {Zhang}, \citenamefont {Murray}, \citenamefont {Yoshinaka},\ and\ \citenamefont {Higgins}}]{ZhangEtAl2002}%
  \BibitemOpen
  \bibfield  {author} {\bibinfo {author} {\bibfnamefont {F.}~\bibnamefont {Zhang}}, \bibinfo {author} {\bibfnamefont {S.~B.}\ \bibnamefont {Murray}}, \bibinfo {author} {\bibfnamefont {A.}~\bibnamefont {Yoshinaka}},\ and\ \bibinfo {author} {\bibfnamefont {A.}~\bibnamefont {Higgins}},\ }\bibfield  {title} {\bibinfo {title} {Shock initiation and detonability of isopropyl nitrate},\ }in\ \href@noop {} {\emph {\bibinfo {booktitle} {12$^{\,\text{th}}$ Symp. (Int.) Detonation, San Diego, CA}}}\ (\bibinfo  {publisher} {ONR},\ \bibinfo {year} {2002})\ pp.\ \bibinfo {pages} {781--790}\BibitemShut {NoStop}%
\bibitem [{\citenamefont {Brochet}\ and\ \citenamefont {Fisson}(1970)}]{BrochetFisson1969}%
  \BibitemOpen
  \bibfield  {author} {\bibinfo {author} {\bibfnamefont {C.}~\bibnamefont {Brochet}}\ and\ \bibinfo {author} {\bibfnamefont {F.}~\bibnamefont {Fisson}},\ }\bibfield  {title} {\bibinfo {title} {{D}\'{e}termination de la pression de d\'{e}tonation dans un explosif condens\'{e} homog\`{e}ne, {E}{xplosifs n$^{\circ}4$}, 113-120 (1969), and {M}onopropellant detonation: isopropyl nitrate},\ }\href@noop {} {\bibfield  {journal} {\bibinfo  {journal} {Astronaut. Acta}\ }\textbf {\bibinfo {volume} {15}},\ \bibinfo {pages} {419} (\bibinfo {year} {1970})}\BibitemShut {NoStop}%
\bibitem [{\citenamefont {Davis}\ \emph {et~al.}(1965)\citenamefont {Davis}, \citenamefont {Craig},\ and\ \citenamefont {Ramsay}}]{DavisEtAl1965}%
  \BibitemOpen
  \bibfield  {author} {\bibinfo {author} {\bibfnamefont {W.~C.}\ \bibnamefont {Davis}}, \bibinfo {author} {\bibfnamefont {B.~G.}\ \bibnamefont {Craig}},\ and\ \bibinfo {author} {\bibfnamefont {J.~B.}\ \bibnamefont {Ramsay}},\ }\bibfield  {title} {\bibinfo {title} {Failure of the {C}hapman-{J}ouguet theory for liquid and solid explosives},\ }\href@noop {} {\bibfield  {journal} {\bibinfo  {journal} {Phys. Fluids}\ }\textbf {\bibinfo {volume} {8(12)}},\ \bibinfo {pages} {2169} (\bibinfo {year} {1965})}\BibitemShut {NoStop}%
\bibitem [{\citenamefont {Lysne}\ and\ \citenamefont {Hardesty}(1973)}]{LysneHardesty1973}%
  \BibitemOpen
  \bibfield  {author} {\bibinfo {author} {\bibfnamefont {P.~C.}\ \bibnamefont {Lysne}}\ and\ \bibinfo {author} {\bibfnamefont {D.~R.}\ \bibnamefont {Hardesty}},\ }\bibfield  {title} {\bibinfo {title} {Fundamental equation of state of liquid nitromethane to 100 kbar},\ }\href@noop {} {\bibfield  {journal} {\bibinfo  {journal} {J. Chem. Phys.}\ }\textbf {\bibinfo {volume} {59(12)}},\ \bibinfo {pages} {6512} (\bibinfo {year} {1973})}\BibitemShut {NoStop}%
\bibitem [{\citenamefont {Jones}\ and\ \citenamefont {Giauque}(1947)}]{JonesGiauque1947}%
  \BibitemOpen
  \bibfield  {author} {\bibinfo {author} {\bibfnamefont {W.~M.}\ \bibnamefont {Jones}}\ and\ \bibinfo {author} {\bibfnamefont {W.~F.}\ \bibnamefont {Giauque}},\ }\bibfield  {title} {\bibinfo {title} {{The entropy of nitromethane. Heat capacity of solid and liquid. Vapor pressure, heats of fusion and vaporization}},\ }\href@noop {} {\bibfield  {journal} {\bibinfo  {journal} {J. Am. Chem. Soc.}\ }\textbf {\bibinfo {volume} {69(5)}},\ \bibinfo {pages} {983} (\bibinfo {year} {1947})}\BibitemShut {NoStop}%
\bibitem [{\citenamefont {Berman}\ and\ \citenamefont {West}(1967)}]{BermanWest1967}%
  \BibitemOpen
  \bibfield  {author} {\bibinfo {author} {\bibfnamefont {H.~A.}\ \bibnamefont {Berman}}\ and\ \bibinfo {author} {\bibfnamefont {E.~D.}\ \bibnamefont {West}},\ }\bibfield  {title} {\bibinfo {title} {Density and vapor pressure of nitromethane 26${{}^\circ}$ to 200${{}^\circ}${C}},\ }\href@noop {} {\bibfield  {journal} {\bibinfo  {journal} {J. Chem. and Eng. Data}\ }\textbf {\bibinfo {volume} {12(2)}},\ \bibinfo {pages} {197} (\bibinfo {year} {1967})}\BibitemShut {NoStop}%
\bibitem [{\citenamefont {Bernard}\ \emph {et~al.}(1966)\citenamefont {Bernard}, \citenamefont {Brossard}, \citenamefont {Claude},\ and\ \citenamefont {Manson}}]{BernardEtAl1966}%
  \BibitemOpen
  \bibfield  {author} {\bibinfo {author} {\bibfnamefont {Y.}~\bibnamefont {Bernard}}, \bibinfo {author} {\bibfnamefont {J.}~\bibnamefont {Brossard}}, \bibinfo {author} {\bibfnamefont {P.}~\bibnamefont {Claude}},\ and\ \bibinfo {author} {\bibfnamefont {N.}~\bibnamefont {Manson}},\ }\bibfield  {title} {\bibinfo {title} {Caract\'{e}ristiques des d\'{e}tonations dans les m\'{e}langes liquides de nitropropane {II} avec l'acide nitrique},\ }\href@noop {} {\bibfield  {journal} {\bibinfo  {journal} {C. R. Acad. Sci. Paris}\ }\textbf {\bibinfo {volume} {263}},\ \bibinfo {pages} {1097} (\bibinfo {year} {1966})}\BibitemShut {NoStop}%
\bibitem [{\citenamefont {Garn}(1960)}]{Garn1960}%
  \BibitemOpen
  \bibfield  {author} {\bibinfo {author} {\bibfnamefont {W.~B.}\ \bibnamefont {Garn}},\ }\bibfield  {title} {\bibinfo {title} {{Detonation pressure of liquid TNT}},\ }\href@noop {} {\bibfield  {journal} {\bibinfo  {journal} {J. Chem. Phys.}\ }\textbf {\bibinfo {volume} {32(3)}},\ \bibinfo {pages} {653} (\bibinfo {year} {1960})}\BibitemShut {NoStop}%
\bibitem [{\citenamefont {Garn}(1959)}]{Garn1959}%
  \BibitemOpen
  \bibfield  {author} {\bibinfo {author} {\bibfnamefont {W.~B.}\ \bibnamefont {Garn}},\ }\bibfield  {title} {\bibinfo {title} {{Determination of the unreacted {H}ugoniot for liquid TNT}},\ }\href@noop {} {\bibfield  {journal} {\bibinfo  {journal} {J. Chem. Phys.}\ }\textbf {\bibinfo {volume} {30(3)}},\ \bibinfo {pages} {819} (\bibinfo {year} {1959})}\BibitemShut {NoStop}%
\bibitem [{\citenamefont {Petrone}(1968)}]{Petrone1968}%
  \BibitemOpen
  \bibfield  {author} {\bibinfo {author} {\bibfnamefont {F.~J.}\ \bibnamefont {Petrone}},\ }\bibfield  {title} {\bibinfo {title} {Validity of the classical detonation wave structure for condensed explosives},\ }\href@noop {} {\bibfield  {journal} {\bibinfo  {journal} {Phys. Fluids}\ }\textbf {\bibinfo {volume} {11(7)}},\ \bibinfo {pages} {1473} (\bibinfo {year} {1968})}\BibitemShut {NoStop}%
\bibitem [{\citenamefont {Bdzil}(1981)}]{Bdzil1981}%
  \BibitemOpen
  \bibfield  {author} {\bibinfo {author} {\bibfnamefont {J.~B.}\ \bibnamefont {Bdzil}},\ }\bibfield  {title} {\bibinfo {title} {Steady-state two-dimensional detonation},\ }\href {https://doi.org/doi:10.1017/S0022112081002085} {\bibfield  {journal} {\bibinfo  {journal} {J. Fluid Mech.}\ }\textbf {\bibinfo {volume} {108}},\ \bibinfo {pages} {195–226} (\bibinfo {year} {1981})}\BibitemShut {NoStop}%
\bibitem [{\citenamefont {Chiquete}\ and\ \citenamefont {Short}(2019)}]{ChiqueteShort2019}%
  \BibitemOpen
  \bibfield  {author} {\bibinfo {author} {\bibfnamefont {M.}~\bibnamefont {Chiquete}}\ and\ \bibinfo {author} {\bibfnamefont {M.}~\bibnamefont {Short}},\ }\bibfield  {title} {\bibinfo {title} {Characteristic path analysis of confinement influence on steady two-dimensional detonation propagation},\ }\href {https://doi.org/https://doi.org/10.1017/jfm.2018.995} {\bibfield  {journal} {\bibinfo  {journal} {J. Fluid Mech.}\ }\textbf {\bibinfo {volume} {863}},\ \bibinfo {pages} {789} (\bibinfo {year} {2019})}\BibitemShut {NoStop}%
\bibitem [{\citenamefont {Parker}\ and\ \citenamefont {Wolfhard}(1953)}]{ParkerWolfhard1953}%
  \BibitemOpen
  \bibfield  {author} {\bibinfo {author} {\bibfnamefont {W.~G.}\ \bibnamefont {Parker}}\ and\ \bibinfo {author} {\bibfnamefont {H.~G.}\ \bibnamefont {Wolfhard}},\ }\bibfield  {title} {\bibinfo {title} {{Some characteristics of flames supported by NO and NO$_{2}$}},\ }in\ \href {https://doi.org/https://doi.org/10.1016/S0082-0784(53)80058-5} {\emph {\bibinfo {booktitle} {4$^{\,\text{th}}$ Symp. (Int.) Combust.}}}\ (\bibinfo  {publisher} {The Combustion Institute},\ \bibinfo {year} {1953})\ pp.\ \bibinfo {pages} {420--428}\BibitemShut {NoStop}%
\bibitem [{\citenamefont {Branch}\ \emph {et~al.}(1991)\citenamefont {Branch}, \citenamefont {Sadequ}, \citenamefont {Alfarayedhi},\ and\ \citenamefont {Van~Tiggelen}}]{BranchEtal1991}%
  \BibitemOpen
  \bibfield  {author} {\bibinfo {author} {\bibfnamefont {M.~C.}\ \bibnamefont {Branch}}, \bibinfo {author} {\bibfnamefont {M.~E.}\ \bibnamefont {Sadequ}}, \bibinfo {author} {\bibfnamefont {A.~A.}\ \bibnamefont {Alfarayedhi}},\ and\ \bibinfo {author} {\bibfnamefont {P.~J.}\ \bibnamefont {Van~Tiggelen}},\ }\bibfield  {title} {\bibinfo {title} {{Measurements of the structure of laminar, premixed flames of CH$_{4}$/NO$_{2}$/O$_{2}$ and CH$_{2}$O/NO$_{2}$/O$_{2}$ mixtures}},\ }\href {https://doi.org/https://doi.org/10.1016/0010-2180(91)90071-I} {\bibfield  {journal} {\bibinfo  {journal} {Combust. Flame}\ }\textbf {\bibinfo {volume} {83}},\ \bibinfo {pages} {228} (\bibinfo {year} {1991})}\BibitemShut {NoStop}%
\bibitem [{\citenamefont {Presles}\ \emph {et~al.}(1996)\citenamefont {Presles}, \citenamefont {Desbordes}, \citenamefont {Guirard},\ and\ \citenamefont {Guerraud}}]{PreslesEtal1996}%
  \BibitemOpen
  \bibfield  {author} {\bibinfo {author} {\bibfnamefont {H.~N.}\ \bibnamefont {Presles}}, \bibinfo {author} {\bibfnamefont {D.}~\bibnamefont {Desbordes}}, \bibinfo {author} {\bibfnamefont {M.}~\bibnamefont {Guirard}},\ and\ \bibinfo {author} {\bibfnamefont {C.}~\bibnamefont {Guerraud}},\ }\bibfield  {title} {\bibinfo {title} {Gaseous nitromethane and nitromethane–oxygen mixtures: a new detonation structure},\ }\href {https://doi.org/https://doi.org/10.1007/BF02515194} {\bibfield  {journal} {\bibinfo  {journal} {Shock Waves}\ }\textbf {\bibinfo {volume} {6}},\ \bibinfo {pages} {111–114} (\bibinfo {year} {1996})}\BibitemShut {NoStop}%
\bibitem [{\citenamefont {Cowperthwaite}\ and\ \citenamefont {Zwisler}(1976)}]{CowperthwaiteZwisler1976}%
  \BibitemOpen
  \bibfield  {author} {\bibinfo {author} {\bibfnamefont {M.}~\bibnamefont {Cowperthwaite}}\ and\ \bibinfo {author} {\bibfnamefont {W.~H.}\ \bibnamefont {Zwisler}},\ }\bibfield  {title} {\bibinfo {title} {{The JCZ equation of state for detonation products and their incorporation into the Tiger code}},\ }in\ \href@noop {} {\emph {\bibinfo {booktitle} {6$^{\,\text{th}}$ Symp. (Int.) on Detonation}}}\ (\bibinfo  {publisher} {Office of Naval Research},\ \bibinfo {year} {1976})\ pp.\ \bibinfo {pages} {162--172}\BibitemShut {NoStop}%
\bibitem [{\citenamefont {Fried}\ and\ \citenamefont {Souers}(1996)}]{FriedSouers1996}%
  \BibitemOpen
  \bibfield  {author} {\bibinfo {author} {\bibfnamefont {L.~E.}\ \bibnamefont {Fried}}\ and\ \bibinfo {author} {\bibfnamefont {P.~C.}\ \bibnamefont {Souers}},\ }\bibfield  {title} {\bibinfo {title} {{BKWC: An empirical BKW parametrization based on cylinder test data}},\ }\href {https://doi.org/https://doi.org/10.1002/prep.19960210411} {\bibfield  {journal} {\bibinfo  {journal} {Propellants, Explosives, Pyrotechnics}\ }\textbf {\bibinfo {volume} {21}},\ \bibinfo {pages} {215} (\bibinfo {year} {1996})}\BibitemShut {NoStop}%
\bibitem [{\citenamefont {Vidal}\ and\ \citenamefont {Khasainov}(1999{\natexlab{a}})}]{Vidal1999a}%
  \BibitemOpen
  \bibfield  {author} {\bibinfo {author} {\bibfnamefont {P.}~\bibnamefont {Vidal}}\ and\ \bibinfo {author} {\bibfnamefont {B.~A.}\ \bibnamefont {Khasainov}},\ }\bibfield  {title} {\bibinfo {title} {A necessary condition for adiabatic explosion behind a shock of arbitrary dynamics},\ }\href {https://doi.org/https://doi.org/10.1016/S1287-4620(99)80016-7} {\bibfield  {journal} {\bibinfo  {journal} {C. R. Acad. Sci. Paris - Series IIB}\ }\textbf {\bibinfo {volume} {327(1)}},\ \bibinfo {pages} {95} (\bibinfo {year} {1999}{\natexlab{a}})}\BibitemShut {NoStop}%
\bibitem [{\citenamefont {Vidal}\ and\ \citenamefont {Khasainov}(1999{\natexlab{b}})}]{Vidal1999b}%
  \BibitemOpen
  \bibfield  {author} {\bibinfo {author} {\bibfnamefont {P.}~\bibnamefont {Vidal}}\ and\ \bibinfo {author} {\bibfnamefont {B.~A.}\ \bibnamefont {Khasainov}},\ }\bibfield  {title} {\bibinfo {title} {{Analysis of critical dynamics for shock-induced adiabatic explosions by means of the Cauchy problem for the shock transformation}},\ }\href {https://doi.org/https://doi.org/10.1007/s001930050165} {\bibfield  {journal} {\bibinfo  {journal} {Shock Waves}\ }\textbf {\bibinfo {volume} {9(4)}},\ \bibinfo {pages} {273} (\bibinfo {year} {1999}{\natexlab{b}})}\BibitemShut {NoStop}%
\end{thebibliography}%

\end{document}